\documentclass{ptephy_v1}

\usepackage{graphics} 
\usepackage{subfig} 
\usepackage{bigints}
\usepackage{siunitx}
\usepackage{color}

\newcommand{\QE}{Q\!E}

\makeatletter
\newcommand{\figcaption}[1]{\def\@captype{figure}\caption{#1}}
\newcommand{\tblcaption}[1]{\def\@captype{table}\caption{#1}}
\makeatother

\graphicspath{{fig/}}

\begin{document}

\title{Expression of the angular dependence of\\the quantum efficiency for a thin multi-alkali photocathode and its optical properties}


\author{Kodai Matsuoka}
\affil{Kobayashi-Maskawa Institute, Nagoya University, Nagoya 464-8602, Japan\email{matsuoka@hepl.phys.nagoya-u.ac.jp}}


\begin{abstract}%
The dependence of the quantum efficiency on the angle and polarization of the incident photon needs to be formulated for a precise description of the response of photomultiplier tubes.
A simplified one-step model of photoelectron emission was derived from Spicer's three-step model, and it enabled the formulation of the dependence of the quantum efficiency in the visible range for thin multi-alkali (NaKSbCs) photocathodes.
The expression of the quantum efficiency was proved by a measurement of the photocurrent for linearly polarized light at various incident angles.
Meanwhile, the measurement revealed the complex refractive indices and thicknesses both of the stratified photocathode and antireflection coating.
It is indicated that the angular dependence of the quantum efficiency is dictated by the optical properties of the photocathode, which are discussed in detail on the basis of the obtained parameters.
\end{abstract}

\subjectindex{H02, H15}

\maketitle

\section{Introduction} \label{sec:intro}
The photocathode is a key component of photomultiplier tubes (PMTs).
It dominates the PMT performance in terms of the sensitivity to photons.
As PMTs are used in a variety of fields, photocathode properties are major subjects of research.
One of the most important properties is the efficiency of photoelectron emission, or the quantum efficiency (QE).
The QE depends not only on the wavelength but also on the incident angle and polarization of the light.
Such dependence is of great interest especially for Cherenkov detectors because the Cherenkov light is emitted along the definite Cherenkov angle and is naturally 100\% linearly polarized~\cite{Cherenkov,Tamm}.
Rigorous expression of the dependence of the QE is demanded to precisely estimate or calibrate the detector performance~\cite{IACT,IceCube,TOP,dualCAL}.
Furthermore, a detailed understanding of the QE could help to find a recipe for enhancement of the QE.

The process of the photoelectron emission can be described by Spicer's three-step model~\cite{ThreeStepModel}: (1) photoexcitation of an electron in the photocathode, (2) transport of the excited electron to the vacuum surface and (3) escape of the electron over the surface barrier into the vacuum.
Therefore the QE is dictated by the optical properties of the PMT, the physical properties of the photocathode material such as the band structure and the electron scattering mechanisms during transportation, and the quality of the photocathode such as the surface characteristics. 
Each of these has to be well described for expression of the dependence of the QE.
Some theoretical expressions or practical models of photoelectron emission were proposed for metal~\cite{photoemit_Berglund,photoemit_Dowell,photoemit_Jensen_metal} and semiconductor~\cite{ThreeStepModel,photoemit_Jensen_semi,photoemit_Kane} under assumption of infinite thickness of the photocathode, where it was not necessary to treat the optical properties of the photocathode in detail because the photon absorption follows a simple exponential function of the depth (Lambert-Beer law).
For thin photocathodes, which are widely used for PMTs, however, the optical properties are of more complication, and there are yet no expressions of photoelectron emission which have been theoretically or experimentally verified.

The optical properties can be expressed by means of the complex refractive index and the thickness of the photocathode.
These parameters can be derived from the standard techniques of reflectance measurement at various angles or of ellipsometry.
Such measurements were reported in Refs.~\cite{index_Dolizy,index_Johnson,index_Lang,index_Moorhead,index_Lay1,index_Lay2,index_Shibamura} for various photocathode types but only at a few wavelengths.
Measurements over the whole visible range were done in Refs.~\cite{photoemit_multialkali,index_Hallensleben,index_Motta,index_Harmer} for bi/multi-alkali photocathodes.
Among them, Refs.~\cite{index_Lay1,index_Lay2} made a comparison between the measured angular dependence of the photocurrent and the predicted one of the absorptance from the measured index and thickness of the photocathode, and found some unexplained discrepancies.
Another attempt to derive the photocathode index and thickness can be made by photocurrent measurement at various angles.
This method was adopted in Ref.~\cite{index_Jones}, but the theoretical function applied in the reference did not fit the data precisely in terms of the angular dependence of the photocurrent.
A discrepancy on the angular dependence was also found between the measurements of the photocurrent and the absorptance by Ref.~\cite{index_Chyba}.
The biggest concern in those comparisons is that some interpretations are needed to connect the photocurrent to the optical absorptance.
Hence the dependence of the QE on the angle and polarization is yet to be investigated.
In addition available data of photocathode indices are not satisfactory for the whole visible range.

In this paper a simplified expression of the QE for thin photocathodes is proposed based on Spicer's three-step model.
It is verified by the measurement of the dependence of the QE on the angle and polarization.
The dependence of the QE was examined directly by the photocurrent measurement for a thin multi-alkali (NaKSbCs) photocathode of the transmission mode, fabricated in a square-shaped micro-channel-plate (MCP) PMT R10754-07-M16(N)~\cite{MCP-PMT} made by HAMAMATSU PHOTONICS K.K.
The photocurrent was measured at wavelengths from 320 to 680~nm and at incident angles from 0 to 80$^\circ$ both for $s$- and $p$-polarizations.
It has to be noted that this PMT has a confidential antireflection coating, of which optical properties are totally unknown.
Hence the optical parameters of the stratified photocathode and antireflection coating are derived at the same time in this work.
Based on the measured parameters, in the latter part of this paper, the optical properties of the photocathode are discussed in connection with effects on the QE.

\section{Expression of the quantum efficiency} \label{sec:theory}
In this section the simplified expression of the QE is derived for thin multi-alkali photocathodes.
Based on the three-step model, the QE can be factorized into the probabilities $A_{\rm pc}$, $P_{\rm transport}$ and $P_{\rm escape}$, where $A_{\rm pc}$ is the absorptance of the photocathode or the fraction of the absorbed light intensity to the incident one, and $P_{\rm transport}$ and $P_{\rm escape}$ are the probabilities that the electron reaches the vacuum surface and escapes into the vacuum, respectively.
These three factors are functions of the photon wavelength because $P_{\rm transport}$ and $P_{\rm escape}$ depend on the energy of the electron, which is determined by the photoexcitation process being subject to the energy of the absorbed photon.
$P_{\rm transport}$ and $P_{\rm escape}$ also depend on the depth $z$ where the photon is absorbed due to a loss of the electron energy during transportation.
That demands a differential function of $A_{\rm pc}$, represented as $A'_{\rm pc}(z)$, which describes the absorption distribution along $z$.
Therefore the QE at a wavelength $\lambda$ is expressed in the form
\begin{equation} \label{eq:QE}
\QE(\lambda) = \int_0^d A'_{\rm pc} (z, \lambda) P_{\rm transport} (z, \lambda) P_{\rm escape} (z, \lambda) {\rm d}z,
\end{equation}
where $d$ is the thickness of the photocathode.
It also has to be expressed as a function of the angle $\theta$ and the polarization $\epsilon$ ($=$ $s$ or $p$) of the incident light.
In the following, each factor of this equation is described by elementary functions with several parameters based on theories.

\subsection{Optical model} \label{sec:optical_model}

\begin{figure}[bt]
 \centering
 \includegraphics[width=0.9\textwidth,keepaspectratio]{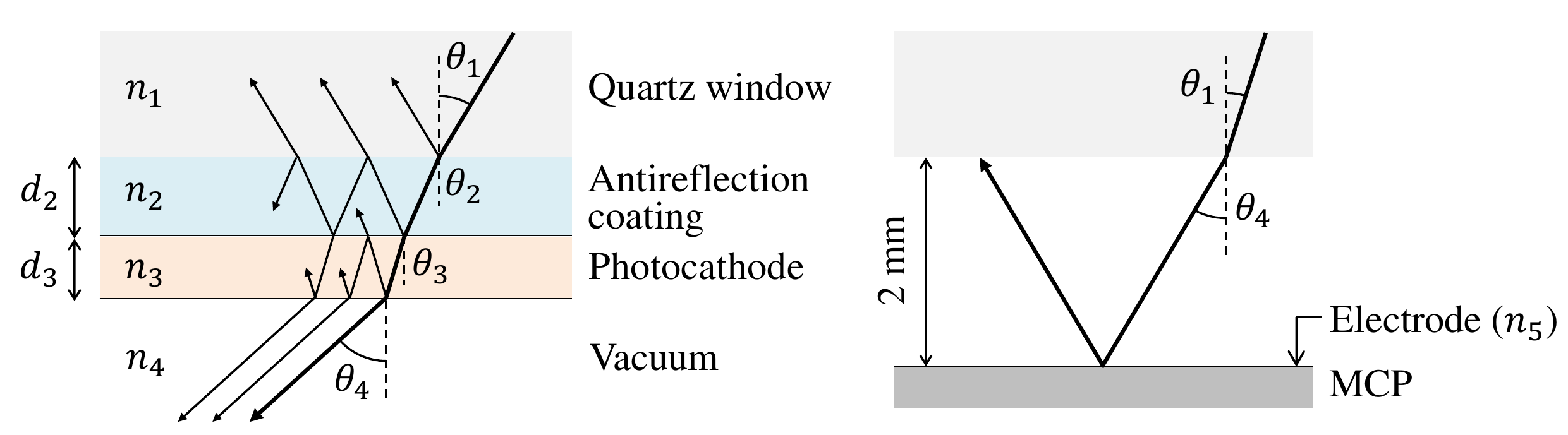}
 \caption{(Left) Diagram of light passing through a double thin layer of the antireflection coating and the photocathode.
 Multiple reflections in the thin layers are also drawn to some extent.
 (Right) Diagram of light transmitted through the photocathode and reflected on the MCP electrode.}
 \label{fig:diagram_optics}
\end{figure}

The absorptance $A_{\rm pc}$ and its derivation $A'_{\rm pc}$ are fully described by the optics.
A specific configuration of the PMT shown in Fig.~\ref{fig:diagram_optics} is considered here\footnote{In most cases light comes from the outside of the window, and one should also consider the reflection and refraction at the outer surface of the window.}:
the light passes from the quartz window (layer~1 with refractive index $n_1$) through a double thin layer of the antireflection coating (layer~2 with $n_2$) and the photocathode (layer~3 with $n_3$) into the vacuum (layer~4 with $n_4=1$).
The amplitude reflection and transmission coefficients at the interface from layer $i$ to $j=i+1$ ($r_{ij}$ and $t_{ij}$, respectively, and $i=1$, 2, 3) are described by Fresnel equations:
\begin{equation} \label{eq:Fresnel}
\begin{array}{lll}
r_{ij} = \cfrac{n_i \cos\theta_i - n_j \cos\theta_j}{n_i \cos\theta_i + n_j \cos\theta_j},&
t_{ij} = \cfrac{2n_i \cos\theta_i}{n_i \cos\theta_i + n_j \cos\theta_j}&
{\rm ({\it s}\mathchar`-polarization)},\\
r_{ij} = \cfrac{n_j \cos\theta_i - n_i \cos\theta_j}{n_j \cos\theta_i + n_i \cos\theta_j},&
t_{ij} = \cfrac{2n_i \cos\theta_i}{n_j \cos\theta_i + n_i \cos\theta_j}&
{\rm ({\it p}\mathchar`-polarization)},
\end{array}
\end{equation}
where $\theta_i$ ($\theta_j$) is the angle between the incident (refracted) light and the normal of the interface, which follows Snell's law:
\begin{equation} \label{eq:Snell}
 n_i \sin\theta_i = n_j \sin\theta_j.
\end{equation}
It should be noted that Eq.~(\ref{eq:Snell}) holds for any combination of $i$ and $j>i$.
When layer $l$ absorbs the light or has stimulated emission, $n_l$ and $\theta_l$ are complex numbers.
Taking into account the multiple reflections on both surfaces of the photocathode, the amplitude reflection and transmission coefficients for the photocathode layer ($\hat{r}_{3}$ and $\hat{t}_{3}$, respectively) are described using Eq.~(\ref{eq:Fresnel}) as follows:
\begin{equation}
\begin{split}
\hat{r}_{3} = r_{23} + t_{23}r_{34}t_{32} e^{i\delta_3} \sum_{m=0}^{\infty} (r_{32} r_{34} e^{i\delta_{3}})^m
 = \frac{r_{23} + r_{34} e^{i\delta_3}}{1 + r_{34}r_{23} e^{i\delta_3}},\\
\hat{t}_{3} = t_{23}t_{34} e^{i\delta_3/2} \sum_{m=0}^{\infty} (r_{34} r_{32} e^{i\delta_{3}})^m
 = \frac{t_{23}t_{34} e^{i\delta_3/2}}{1 + r_{34}r_{23} e^{i\delta_3}},\quad\;
\end{split}
\end{equation}
and similarly for the double layer:
\begin{equation}
\hat{r}_{2+3} = \frac{r_{12} + \hat{r}_{3} e^{i\delta_2}}{1 + \hat{r}_{3}r_{12} e^{i\delta_2}},\quad
\hat{t}_{2+3} = \frac{t_{12}\hat{t}_{3} e^{i\delta_2/2}}{1 + \hat{r}_{3}r_{12} e^{i\delta_2}},
\end{equation}
where
\begin{equation}
\delta_l = \frac{4\pi d_l n_l}{\lambda_0} \cos\theta_l \quad (l=2,3)
\end{equation}
is the phase difference, $d_l$ is the thickness of layer $l$ and $\lambda_0$ is the wavelength of the light in the vacuum.
The reflectance $R_{2+3}$ and transmittance $T_{2+3}$ of the double layer are related to the coefficients:
\begin{equation} \label{eq:reflectance}
R_{2+3} = \left| \hat{r}_{2+3} \right| ^2,
\end{equation}
\begin{equation}
T_{2+3} = \left\{
 \begin{array}{ll}
  \cfrac{{\rm Re}(n_4 \cos\theta_4)}{{\rm Re}(n_1 \cos\theta_1)} \left| \hat{t}_{2+3} \right| ^2
  & {\rm({\it s}\mathchar`-polarization)}\\
  \cfrac{{\rm Re}(n_4 \cos\theta_4^*)}{{\rm Re}(n_1 \cos\theta_1^*)} \left| \hat{t}_{2+3} \right| ^2
  & {\rm ({\it p}\mathchar`-polarization)}
 \end{array}
 \right.
\end{equation}
where $\theta^*$ denotes the complex conjugate of $\theta$ (although $\theta_1$ and $\theta_4$ are real in this case) and ${\rm Re}(x)$ denotes the real part of $x$.
Equation~(\ref{eq:reflectance}) is the base of the reflectance measurement for estimation of the photocathode index and thickness mentioned in Sec.~\ref{sec:intro}.
On the other hand the photocurrent measurement is associated with the absorptance of the double layer $A_{2+3}$, which has the relation:
\begin{equation} \label{eq:R+T+A}
R_{2+3}+T_{2+3}+A_{2+3} = 1.
\end{equation}
If there is no absorption in the antireflection coating, $A_{2+3}$ is equal to the absorptance of the photocathode $A_{3}$.

\begin{figure}[tb]
\centerline{
 \hspace{0.6cm}
 \subfloat{\includegraphics[width=0.55\textwidth,keepaspectratio]{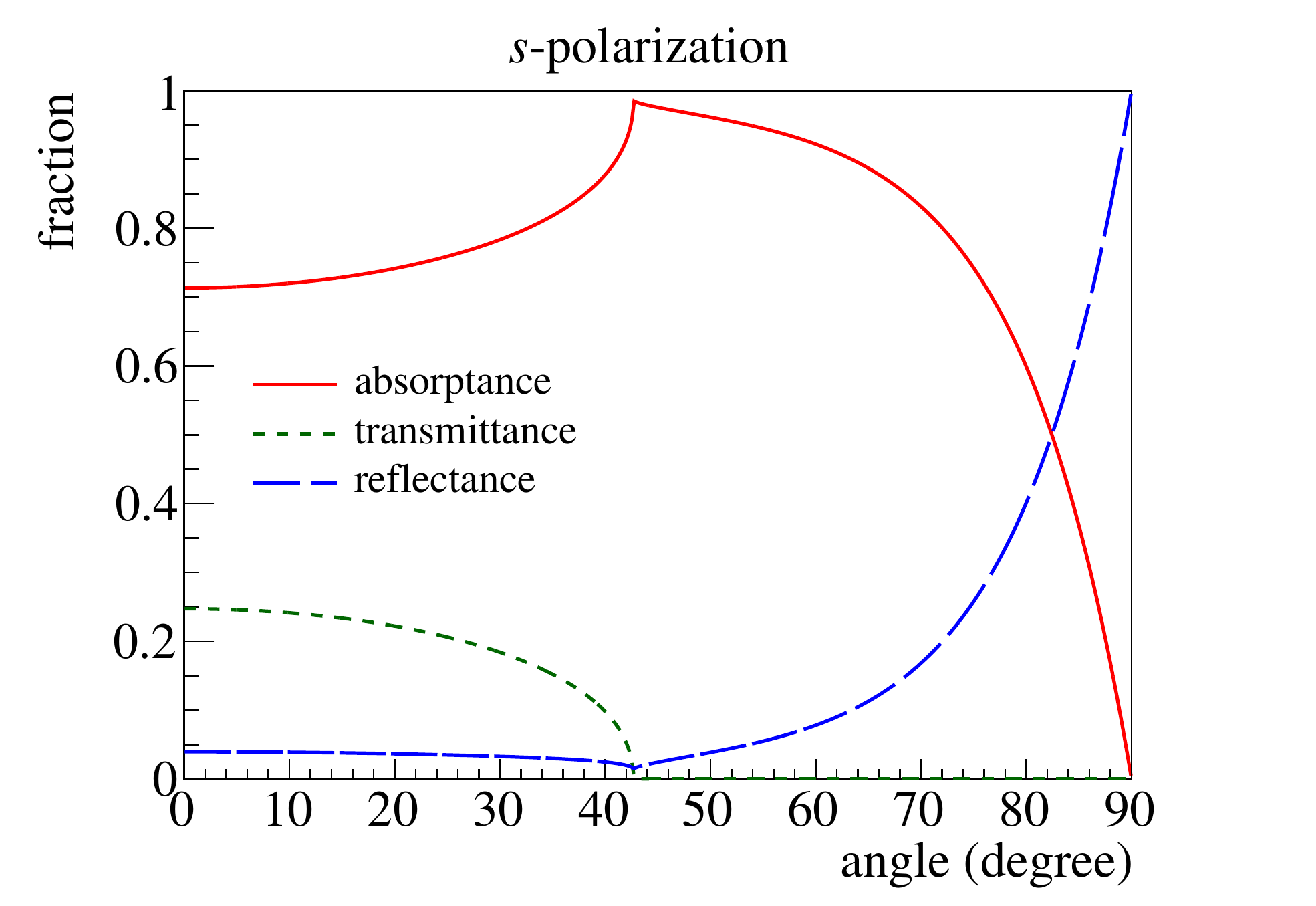}}
 \hspace{-0.6cm}
 \subfloat{\includegraphics[width=0.55\textwidth,keepaspectratio]{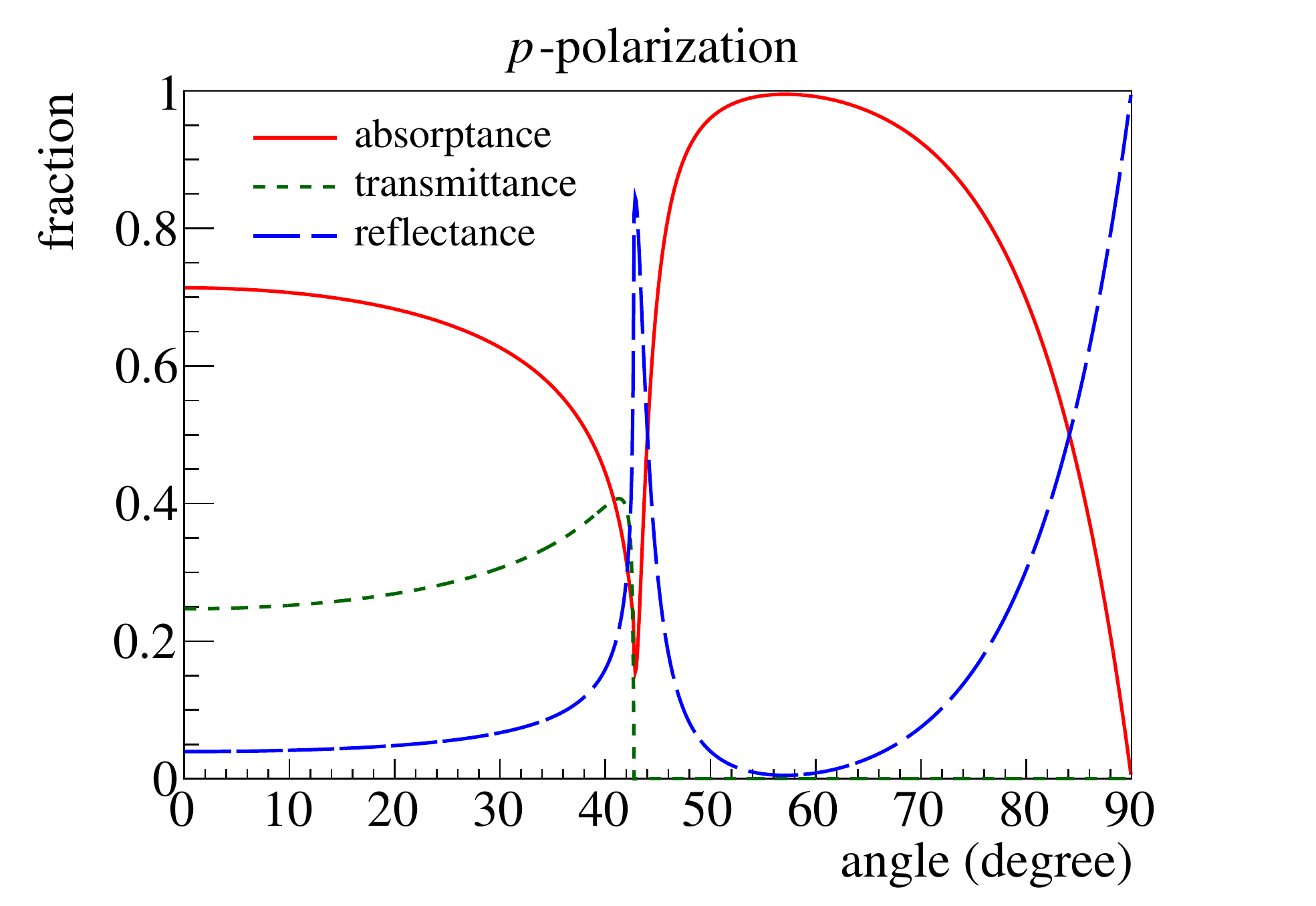}}
}
\caption{Reflectance $R_{2+3}$, transmittance $T_{2+3}$ and absorptance $A_{2+3}$ of the double layer of the antireflection coating ($n_2=2.1$ and $d_2=30$~nm) and the photocathode ($n_3=2.3+3.3i$ and $d_3=10$~nm) at a wavelength $\lambda_0=360$~nm for $s$-polarization (left) and $p$-polarization (right).
The indices of quartz and vacuum are $n_1=1.475$ at 360~nm and $n_4=1$.}
\label{fig:absorptance}
\end{figure}

An example of $R_{2+3}$, $T_{2+3}$ and $A_{2+3}$ as a function of $\theta_1$ is shown in Fig.~\ref{fig:absorptance} for a given set of parameters.
For $s$($p$)-polarization $A_{2+3}$ increases (decreases) along with the incident angle from 0$^\circ$ because at the photocathode-vacuum interface the reflection increases (decreases) and the transmission to the vacuum decreases (increases).
Accordingly $T_{2+3}$ decreases (increases) and falls down to zero at the total reflection angle for the photocathode-vacuum interface.
The incident angle $\theta_1$ corresponding to the total reflection angle is determined by the indices of quartz and vacuum according to Eq.~(\ref{eq:Snell}).
It ranges from 42.4$^\circ$ at 320~nm to 43.4$^\circ$ at 660~nm, where the quartz index varies from 1.48 to 1.46.
Nearly at that angle $A_{2+3}$ peaks for $s$-polarization and drops steeply for $p$-polarization.
$A_{2+3}$ for $p$-polarization then peaks and $R_{2+3}$ drops to nearly zero.
The angle of the minimum $R_{2+3}$ is 57.1$^\circ$, which is slightly different from Brewster's angle of $\arctan (n_2/n_1) = 54.9^\circ$ for the first interface between the window and the antireflection coating due to the reflections from the other interfaces.
When the angle gets close to 90$^\circ$ the reflection at the first interface becomes dominant and thus $A_{2+3}$ declines.

For the sake of precise expression of the absorptances of both stratified layers, the exact expression recently derived~\cite{absorptance} is applied in this work.
Namely the amount of light absorbed per unit length at a depth $z$ from the surface of each layer $l(=2,3)$ ($z=0$ on the light incident surface) is given as
\begin{equation} \label{eq:absorptance_z}
A'_l(z) = a_l \left| v_l \right|^2 e^{-2z {\rm Im}(\tilde{\nu}_l)} + a_l \left| w_l \right|^2 e^{2z {\rm Im}(\tilde{\nu}_l)} + b_l v_l w_l^* e^{2iz {\rm Re}(\tilde{\nu}_l)} + b_l v_l^* w_l e^{-2iz {\rm Re}(\tilde{\nu}_l)},
\end{equation}
\begin{equation*}
\begin{array}{ll}
a_l = b_l = \cfrac{{\rm Im}(\tilde{\nu}_l n_l \cos\theta_l)}{{\rm Re}(n_1 \cos\theta_1)}
& {\rm ({\it s}\mathchar`-polarization)}\\
a_l = \cfrac{2{\rm Im}(\tilde{\nu}_l) {\rm Re}(n_l \cos\theta_l^*)}{{\rm Re}(n_1 \cos\theta_1^*)},\quad
b_l = -\cfrac{2{\rm Re}(\tilde{\nu}_l) {\rm Im}(n_l \cos\theta_l^*)}{{\rm Re}(n_1 \cos\theta_1^*)}
& {\rm ({\it p}\mathchar`-polarization)}
\end{array}
\end{equation*}
where ${\rm Im}(x)$ denotes the imaginary part of $x$, $\tilde{\nu}_l$ is the z-component of the wavenumber in layer $l$, and $v_l$  ($w_l$) is the amplitude of the wave heading forward (backward) on the layer $l$ side of the interface between $l-1$ and $l$:
\begin{equation}
 \tilde{\nu}_l = \frac{2\pi n_l}{\lambda_0} \cos\theta_l,
\end{equation}
\begin{equation}
\begin{split}
v_l = v_{l-1} e^{i\delta_{l-1}/2} t_{l-1,l} + w_l r_{l,l-1}, \;\; \\
w_l = w_{l+1}t_{l+1,l} e^{i\delta_l /2} + v_l r_{l,l+1} e^{i\delta_l}.
\end{split}
\end{equation}
In the case of four layers considered,
\begin{equation}
\begin{array}{lll}
 & v_1 = 1,
 & w_1 = r_{12},\\
 & v_2 = \cfrac{\left( 1-r_{32}r_{34}e^{i\delta_3} \right) t_{12}}{1-r_{21}r_{23}e^{i\delta_2}-r_{32}r_{34}e^{i\delta_3}-r_{21}r_{34}e^{i(\delta_2+\delta_3)}},
 & w_2 = \cfrac{r_{23}+r_{34}e^{i\delta_3}}{1-r_{32}r_{34}e^{i\delta_3}} e^{i\delta_2} v_2,\\
 & v_3 = \cfrac{t_{23}e^{i\delta_2/2}}{1-r_{32}r_{34}e^{i\delta_3}} v_2,
 & w_3 = r_{34}e^{i\delta_3} v_3,\\
 & v_4 = t_{34},
 & w_4 = 0.\\
\end{array}
\end{equation}
The absorptance of layer $l$ is given by integrating Eq.~(\ref{eq:absorptance_z}) over the thickness:
\begin{equation} \label{eq:absorptance}
A_l = \int_0^{d_l} A'_{l}(z) {\rm d}z.
\end{equation}

It is also necessary to consider the transmitted light into the vacuum, which is then reflected on the MCP electrode back to the photocathode as shown in Fig.~\ref{fig:diagram_optics} (right).
Hereafter the illumination from the window/vacuum side is called front/back-illumination.
The back-illumination needs to be taken into account only below the total reflection angle for the photocathode-vacuum interface.
The gap spacing between the photocathode and the MCP is 2~mm, and the back-illumination hits the photocathode at a different position from the front-illumination depending on the incident angle.
The reflectance $R_{\rm el}$ of the electrode with refractive index $n_5$ is calculated using Eq.~(\ref{eq:Fresnel}),
\begin{equation}
 R_{\rm el} = |r_{45}|^2.
\end{equation}
The MCP has micro pores of 10~\si{\micro}m diameter and the open area ratio ($\mathcal{R}_{\rm OA}$) is about 0.6.
Hence the fraction of the reflected light accounts for $(1-\mathcal{R}_{\rm OA}) R_{\rm el}$.
The absorptance of each layer for the back-illumination $A_{\bar{l}}$ and its derivation $A'_{\bar{l}}(z)$ are given in the same manner as described above by reversing the order of the layers (namely $\bar{l}=\bar{1}$, $\bar{2}$, $\bar{3}$, $\bar{4}$ for the vacuum, photocathode, antireflection coating and window, respectively; and the indices $n_{\bar{l}}$ and thicknesses $d_{\bar{l}}$ should be defined correspondingly) and using $\theta_4$ as the incident angle.
The sum of the front- and back-illuminations is the total absorption of the photocathode,
\begin{equation} \label{eq:total_absorption}
 A_{\rm pc} = A_3 + T_{2+3} (1-\mathcal{R}_{\rm OA}) R_{\rm el} A_{\bar{2}}.
\end{equation}
Subsequent reflections of the back-illumination in the vacuum layer are omitted in this work because they should have little contribution to $A_{\rm pc}$.

\subsection{Dispersion model}
In order to describe dispersion functions of refractive indices, the following two models and an empirical equation are applied in this work.

The one for the photocathode and the antireflection coating is the Lorentz dispersion model~\cite{Lorentz}, which represents dielectric response of inter-band transitions by the classical damped harmonic oscillator.
As Ref.~\cite{photoemit_multialkali} infers that the photon absorption in NaKSbCs is probably of direct transition, the Lorentz model is expected to describe the dispersion well.
In this model, the complex dielectric function of the frequency $\omega$ can be expressed as
\begin{equation}
 \varepsilon (\omega) = \varepsilon_\infty + \frac{(\varepsilon_s - \varepsilon_\infty)\omega_t^2}{\omega_t^2 - \omega^2 - i\varGamma_0 \omega},
\end{equation}
where $\varepsilon_\infty$ is the high frequency limit of $\varepsilon$, $\varepsilon_s = \varepsilon_\infty + \omega_p^2 / \omega_t^2$ is the static dielectric function at $\omega = 0$, $\omega_p$ is the plasma frequency, $\omega_t$ is the resonant frequency of the oscillator corresponding to the absorption peak, and $\varGamma_0$ is the damping factor corresponding to the full width at half maximum of the peak.
The real and imaginary parts of $\varepsilon$ are described as follows:
\begin{equation} \label{eq:epsilon}
 {\rm Re}(\varepsilon) = 1 + \frac{(\varepsilon_s - 1) \omega_t^2 \left( \omega_t^2 - \omega^2 \right)}{\left( \omega_t^2 - \omega^2 \right)^2 + \varGamma_0^2 \omega^2},\quad
 {\rm Im}(\varepsilon) = \frac{(\varepsilon_s - 1) \omega_t^2 \varGamma_0 \omega}{\left( \omega_t^2 - \omega^2 \right)^2 + \varGamma_0^2 \omega^2}.
\end{equation}
Here $\varepsilon_\infty$ is fixed to 1 as in general material.
The complex refractive index ($n+ik$) as a function of $\lambda (= 2\pi c/\omega$, where $c$ is the speed of light) is derived as follows:
\begin{equation} \label{eq:index}
 n(\lambda) = \sqrt{\frac{{\rm Re}(\varepsilon) + |\varepsilon|}{2}},\quad
 k(\lambda) = \sqrt{\frac{-{\rm Re}(\varepsilon) + |\varepsilon|}{2}}.
\end{equation}

The other dispersion model for the electrode is described by the Drude free electron theory~\cite{Drude1,Drude2} as follows:
\begin{equation}
 \varepsilon (\omega) = \varepsilon_\infty - \frac{\omega_p^2}{\omega^2 + i\varGamma_d \omega},
\end{equation}
\begin{equation}
 {\rm Re}(\varepsilon) = 1 - \frac{\omega_p^2}{\omega^2 + \varGamma_d^2},\quad
 {\rm Im}(\varepsilon) = \frac{\omega_p^2 \varGamma_d / \omega}{\omega^2 + \varGamma_d^2},
\end{equation}
where $1/\varGamma_d$ is the relaxation time or DC conductivity and $\varepsilon_\infty$ is fixed to 1.
The parameters of NiFe, namely $\omega_p=14.79$~eV and $\varGamma_d=4.78$~eV~\cite{Horiba_TN09}, are applied in this work.

Sellmeier equation~\cite{Sellmeier} is used for the refractive index of quartz:
\begin{equation}
  n^2-1 = \frac{B_1 \lambda^2}{\lambda^2 - C_1} + \frac{B_2 \lambda^2}{\lambda^2 - C_2} + \frac{B_3 \lambda^2}{\lambda^2 - C_3},
\end{equation}
where $B_i$ and $C_i$ ($i=$ 1, 2, 3) are the dispersion constants listed below for $\lambda$ in units of \si{\micro}m at 20\si{\degreeCelsius}~\cite{Suprasil}:
\begin{equation}
\begin{array}{l}
 B_1 = 0.473115591, \quad C_1 = \phantom{0}0.0129957170,\\
 B_2 = 0.631038719, \quad C_2 = \phantom{0}0.00412809220,\\
 B_3 = 0.906404498, \quad C_3 = 98.7685322.\\
\end{array}
\end{equation}

\subsection{Model of photoelectron emission} \label{sec:model}
The other factors in Eq.~(\ref{eq:QE}) are considered in the case of thin semiconductors and visible light in the following.
The loss of the electron during transportation depends on the inelastic and elastic scattering lengths.
The electron-phonon scattering can be regarded as the elastic one because the energy loss by the scattering is quite small~\cite{photoemit_alkali}.
Scattering with electrons in the valence band can take place only if the electron in the conduction band has a sufficient energy to produce an electron pair over the band gap.
The threshold energy was found to be more than twice the band gap above the bottom of the conduction band~\cite{photoelectronic_Sommer}.
It was measured for (Cs)Na$_3$KSb photocathodes to be 3.0~eV~\cite{scatter_ee}, or 4.0~eV above the edge of the valence band, which corresponds to 310~nm wavelength of the photon absorbed.
Scattering with impurities or defects could take place,
but its probability should be small if the photocathode thickness is sufficiently less than the scattering lengths for the impurities and defects.
Therefore, the loss of the electron by the scatterings is expected to be small, and hence
\begin{equation}
 P_{\rm transport} \approx 1.
\end{equation}

When the electron reaches the vacuum surface, it can escape to the vacuum if it has a momentum of which the normal component with respect to the vacuum surface is greater than a critical value~\cite{photoemit_Berglund}.
Namely the electron trajectory is required to fall within the escape cone of the apex angle $2\theta_c$ defined by $\cos\theta_c = \sqrt{\varPhi/E_e} > \sqrt{\varPhi/h\nu}$, where $\varPhi$ is the work function, $E_e (>\varPhi)$ is the energy of the electron at the vacuum surface and $h\nu (>E_e)$ is the energy of the absorbed photon.
This requirement could be a significant reduction factor for the QE when the electron went straight from the initial position to the vacuum surface without any scatterings.
The acceptance of the electron escape in this case would be $(1-\cos\theta_c)/2 < 0.18$ for $\varPhi=1.55$~eV~\cite{ThreeStepModel} and $\lambda=320$~nm.
It means that the QE at 320~nm or longer wavelengths would be limited to 0.18 at most.
However the elastic electron-phonon scattering can enhance the escape probability because it gives the electron many chances to reach the vacuum surface at different angles as in the case of random walk~\cite{photoemit_review}.
Once the electron falls within the escape cone, the escape probability can be regarded as unity~\cite{photoemit_Jensen_semi}.
As far as only electrons above the vacuum level are counted, $P_{\rm escape}$ therefore could be unity.
The proportion of the electrons above the vacuum level is described by the photoexcitation probability $P_{\rm excite}(\lambda)$ which is equal to the ratio of the probability of transitions above the vacuum level to the one of all possible transitions by absorbing a photon of the wavelength $\lambda$.
Thus
\begin{equation}
  P_{\rm escape}(\lambda) = P_{\rm excite}(\lambda).
\end{equation}
When one takes account of the energy loss by the scatterings, $P_{\rm excite}$ can be interpreted as the ratio of the probability of transitions above the vacuum level plus the amount of energy loss to the one of all possible transitions.

Consequently the three-step model could be reduced to only one step for thin photocathodes: photoexcitation of an electron above a certain energy level needed for the escape.
This presumption shall be examined in Sec.~\ref{sec:experiment}.
$P_{\rm excite}$ depends on the work function and normally increases monotonically as the photon energy increases.
To exactly describe the $P_{\rm excite}$ function, one also has to consider the electron density of states which governs the transition probability as studied in Refs.~\cite{band_calc_Ettema,band_calc_Murtaza,band_XPS}.
However it is beyond the scope of this work, and $P_{\rm excite}$ is taken as a parameter to be determined independently at each wavelength.

For thick photocathodes where the approximations above do not hold, $P_{\rm transport}$ and $P_{\rm escape}$ also depend on $z$.
It was mentioned that $P_{\rm transport}$ and $P_{\rm escape}$ can be lumped together and
\begin{equation} \label{eq:PP_thick}
 P_{\rm transport} (z, \lambda) P_{\rm escape} (z, \lambda) = \left\{
 \begin{array}{ll}
  G(\lambda) e^{-(d-z)/L_e} & (\mathrm{front\mathchar`-illumination}) \\
  G(\lambda) e^{-z/L_e}     & (\mathrm{back\mathchar`-illumination})
 \end{array}
 \right.
\end{equation}
is a good approximation~\cite{ThreeStepModel}, where $G(\lambda)$ is an undefined function of $\lambda$ and $L_e$ is the escape length of the excited electron in the photocathode.
This approximation shall also be investigated in Sec.~\ref{sec:experiment}.

\subsection{QE function}
In summary the QE for a thin photocathode is expressed by a function
\begin{equation} \label{eq:QEfunction}
 \QE(\lambda,\theta,\epsilon) = P_{\rm excite} (\lambda) A_{\rm pc} (\lambda,\theta,\epsilon),
\end{equation}
\begin{equation} \label{eq:QEabsorptance}
 A_{\rm pc} (\lambda,\theta,\epsilon) = 
  \int_{0}^{d_3} A'_3(z,\lambda,\theta,\epsilon) {\rm d}z
  + T_{2+3}(\lambda,\theta,\epsilon) (1-\mathcal{R}_{\rm OA}) R_{\rm el}(\lambda,\theta,\epsilon) \int_{0}^{d_{\bar{2}}} A'_{\bar{2}}(z,\lambda,\theta,\epsilon) {\rm d}z.
\end{equation}
The parameters to be determined are $P_{\rm excite}$ for each $\lambda$, and respective $d$, $\varepsilon_s$, $\omega_t$ and $\varGamma_0$ of the photocathode and the antireflection coating.
$\mathcal{R}_{\rm OA}$ is supposed to be 0.6.
As a consequence of the one-step model of photoelectron emission, the dependence of the QE on the angle $\theta$ and the polarization $\epsilon$ is dictated only by the optical properties.

If the photocathode thickness is comparable to the escape length, the dependence of $P_{\rm transport} P_{\rm escape}$ on $z$ expressed in Eq.~(\ref{eq:PP_thick}) has to be taken into account in the integral:
\begin{equation} \label{eq:QEfunction_loss}
\begin{split}
 \QE(\lambda,\theta,\epsilon) = G(\lambda) e^{\beta} \left[
   a |v|^2 \int_{0}^{d} e^{(-2{\rm Im}(\tilde{\nu})+\alpha)z} {\rm d}z
  + a |w|^2 \int_{0}^{d} e^{(2{\rm Im}(\tilde{\nu})+\alpha)z} {\rm d}z \right. \quad\quad\quad \\
 \left.
  +2b {\rm Re}(vw^*) \int_{0}^{d} \cos(2{\rm Re}(\tilde{\nu})z) e^{\alpha z} {\rm d}z
  -2b {\rm Im}(vw^*) \int_{0}^{d} \sin(2{\rm Re}(\tilde{\nu})z) e^{\alpha z} {\rm d}z
 \right] \hspace{-1.0cm} \\
 = G(\lambda) e^{\beta} \left[
   a |v|^2 \frac{e^{(-2{\rm Im}(\tilde{\nu})+\alpha)d}-1}{-2{\rm Im}(\tilde{\nu})+\alpha}
  +a |w|^2 \frac{e^{(2{\rm Im}(\tilde{\nu})+\alpha)d}-1}{2{\rm Im}(\tilde{\nu})+\alpha}
 \right. \quad\quad\quad\quad\;\;\: \\
  +2b {\rm Re}(vw^*) \frac{e^{\alpha d}\{ \alpha\cos(2{\rm Re}(\tilde{\nu})d)+2{\rm Re}(\tilde{\nu})\sin(2{\rm Re}(\tilde{\nu})d) \} -\alpha}{4({\rm Re}(\tilde{\nu}))^2+\alpha^2} \quad \\
 \left.
  -2b {\rm Im}(vw^*) \frac{e^{\alpha d}\{ \alpha\sin(2{\rm Re}(\tilde{\nu})d)-2{\rm Re}(\tilde{\nu})\cos(2{\rm Re}(\tilde{\nu})d) \} +2{\rm Re}(\tilde{\nu})}{4({\rm Re}(\tilde{\nu}))^2+\alpha^2}
 \right] , \hspace{-0.9cm}
\end{split}
\end{equation}
\begin{equation*}
\begin{array}{lll}
 \alpha = \cfrac{1}{L_e}, & \beta = -\cfrac{d}{L_e} & {\rm (front\mathchar`-illumination)}\\
 \alpha = -\cfrac{1}{L_e}, & \beta = 0 & {\rm (back\mathchar`-illumination)}
\end{array}
\end{equation*}
Here the second term of Eq.~(\ref{eq:QEabsorptance}) for the back-illumination is not explicitly written and the subscript $l$ is omitted.
As $L_e$ gets larger, $G$ becomes closer to $P_{\rm excite}$, and 
Eq.~(\ref{eq:QEfunction_loss}) with $L_e=\infty$ is identical to Eq.~(\ref{eq:QEfunction}).

\section{Experimental verification} \label{sec:experiment}

This section describes the measurement of the dependence of the QE on the angle and polarization.
The QE function derived in the previous section shall be verified experimentally.

\subsection{Measurement setup}
\begin{figure}[bt]
 \centering
 \includegraphics[width=\textwidth,keepaspectratio]{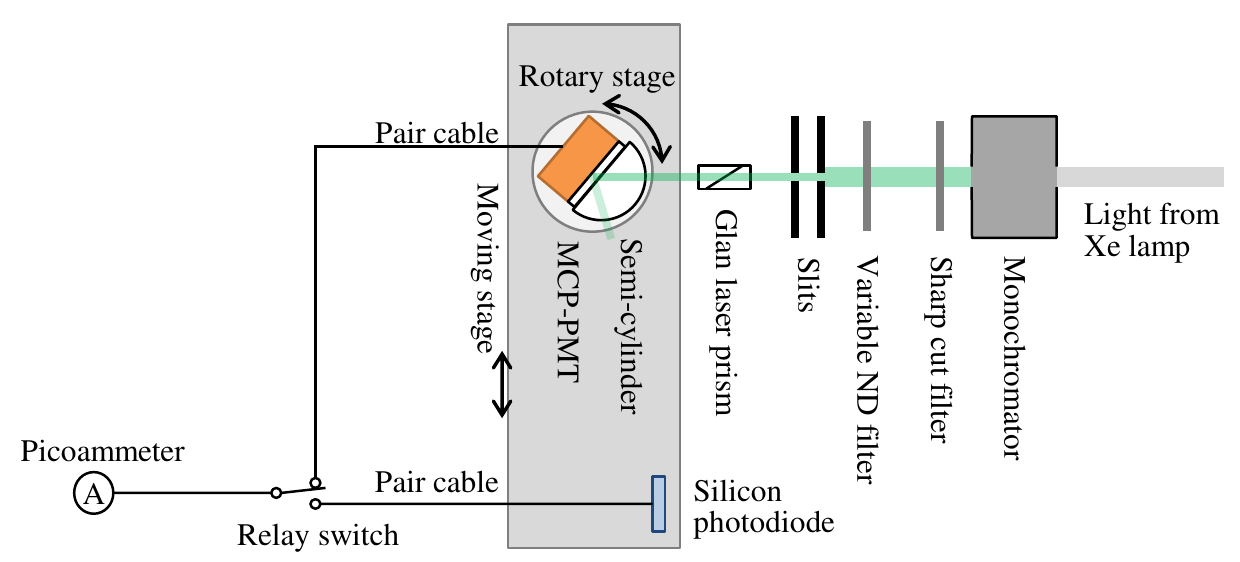}
 \caption{Schematic top view of the setup of the QE measurement.}
 \label{fig:setup}
\end{figure}

The QE was measured as the photocurrent on the PMT photocathode under illumination of linearly polarized light.
The data were taken at 81~incident angles (2$^\circ$ intervals from $-80^\circ$ to 80$^\circ$)  and at 18~wavelengths (20-nm intervals from 320 to 660~nm) for both polarizations in the setup shown in Fig.~\ref{fig:setup}.
The incident angle was selected by rotating the PMT on a rotary stage while fixing the optical system.
The normal incident angle was defined as 0$^\circ$.
The light source was a xenon lamp, and the wavelength was selected by a monochromator.
Two types of sharp cut filters were placed in a switchable holder and one of them or a blank was selected according to the wavelength in order to cut out stray light from the monochromator.
Following a round continuously variable neutral density (ND) filter and two slits, a Glan laser prism was placed.
It polarized the light with an extinction ratio less than $5\times 10^{-6}$.
The photocathode was illuminated through a semi-cylinder whose radius of the curvature was 15~mm and through the 1.5~mm thick PMT window.
The semi-cylinder and the window were both made of synthetic quartz and were coupled together via index-matching optical oil made of fused silica.
Therefore the light passed from the air to the photocathode without refraction at any orientations of the PMT rotation as the PMT was rotated about the axis of the semi-cylinder.
At 0$^\circ$ of the incident angle, the light spotted at the center of the photocathode and the size of the light spot was about 1~mm in diameter.
Since the photocathode was located off from the rotation center, the light spot shifted and ovalized on the photocathode at oblique angles.
At $\pm80^\circ$, the profile of the light spot was nearly on the verge of the photocathode, of which the size was $23\times23$~mm$^2$.
The photoinduced current on the photocathode $I_{\rm PMT}$ was measured as the mean of ten readings by a picoammeter without any amplification.
The light intensity was adjusted by the ND filter so that $I_{\rm PMT}$ did not exceed 10~nA to avoid saturation due to the space charge effect in the photocathode.
The counter-current of the photoelectrons was got out to the MCP electrode by applying +200~V on it.
A silicon photodiode was used to measure the light intensity.
It was mounted on a moving stage together with the PMT rotary stage and took turns measuring the light.
The photocurrent of the photodiode $I_{\rm PD}$ was measured by the same picoammeter (60~times each in this instance) as a relay switch selected either of the photodevices.
Therefore $I_{\rm PMT} \cdot \QE_{\rm PD}/I_{\rm PD}$ amounts to the PMT QE, where the absolute QE spectrum of the photodiode $\QE_{\rm PD}(\lambda)$ was calibrated in advance with a precision of 0.5-1.7\% depending on the wavelength.
In this measurement the PMT QE was defined as the efficiency of detecting the incident light onto the quartz semi-cylinder, and about 4\% loss by the reflection at the air-quartz interface was counted as the inefficiency.

A misalignment of the centers of the semi-cylinder and the rotary stage could systematically deviate the incident angle on the photocathode due to the refraction at the surface of the semi-cylinder.
The measurement for both rotation directions (from 0$^\circ$ to $\pm 80^\circ$) was intended for canceling out the error of the misalignment as well as one of the offset angle.
In the analysis the QEs at the same absolute angle were averaged, and the difference between the QEs, which was typically $\Delta\QE/\QE \approx 1$\%, was considered as the systematic error.
The misalignment of the centers along the light axis however cannot be compensated by averaging the QEs, and it was counted in the error on the incident angle.
The uncertainty of the alignment was 0.25~mm and the error on the incident angle was estimated to be 0-0.7$^\circ$ depending on the angle.
The other errors on the angle were 0.3$^\circ$ due to the spread of the light spot and 0.2$^\circ$ for the positioning accuracy of the rotary stage.

The degree of polarization could be another major source of the systematic error on the QE.
It was determined mostly by the alignment of the polarization axis of the Glan laser prism relative to the plane of incidence for the photocathode.
The uncertainty of the alignment was estimated to be $0.25^\circ$, and the systematic error on the QE was typically 0.1\%, depending on the difference in the QEs between $s$- and $p$-polarizations at the same incident angle.
The statistical error in this measurement was typically 0.2\%, which was quoted from the standard deviation of the ten readings of the picoammeter.
The accuracy of the picoammeter, 0.2\% of the reading, was also added in the error.
The errors on $\QE_{\rm PD}$ and $I_{\rm PD}$, in total 0.5-2.5\% depending on the wavelength, were treated as the systematic error on the normalization of the PMT QE because they do not distort the angular dependence of the QE.

Compared to the standard technique of reflectance measurement, this photocurrent measurement has some advantages.
Another photodevice to measure the reflected light is unnecessary, and therefore the difficulties can be avoided in aligning that photodevice in accordance with the angle of the incident light, in particular around 0$^\circ$ where the incident and reflected light overlaps.
In addition the measurement uncertainties related to that photodevice can be omitted.
Regarding the dependence of the QE on the angle and polarization, it should be measured directly by the photocurrent as it is in general not trivial to interpret the absorptance deduced by the reflectance measurement as the QE.

\subsection{Results} \label{sec:result}
\begin{figure}[t]
\begin{minipage}{0.49\textwidth}
 \hspace{-0.3cm}
 \includegraphics[width=1.12\textwidth,keepaspectratio]{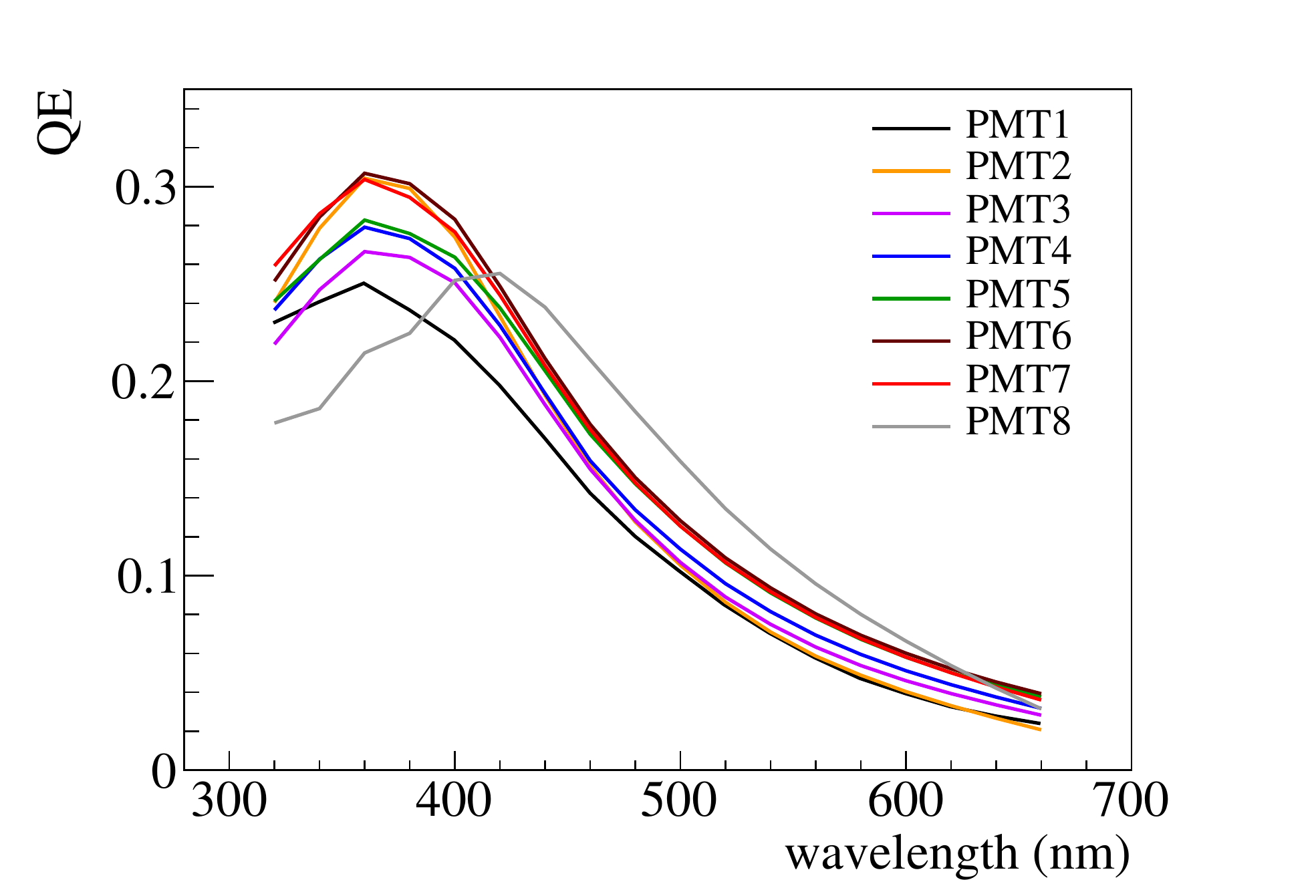}
 \caption{Measured QE spectra at 0$^\circ$.
 The error bars of the data are omitted in this figure.}
 \label{fig:QE_spectrum}
\end{minipage}
\hspace{0.02\textwidth}
\begin{minipage}{0.49\textwidth}
 \hspace{-0.3cm}
 \centering
 \includegraphics[width=0.865\textwidth,keepaspectratio]{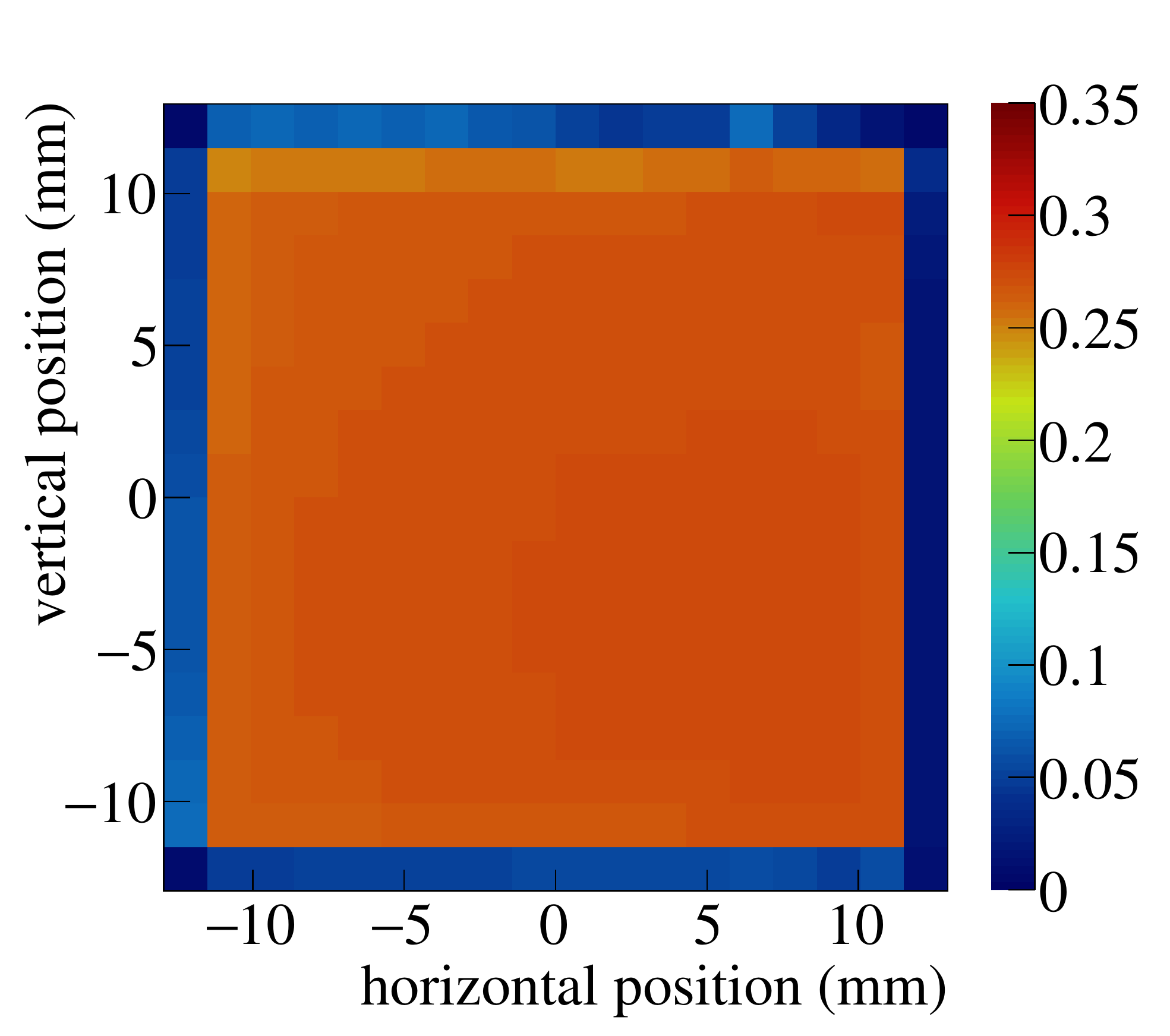}
 \caption{QE distribution of PMT3 at 0$^\circ$ and 360~nm.
 The color represents the QE.}
 \label{fig:QE_uniformity}
\end{minipage}
\end{figure}

Eight PMTs (called hereafter PMT1-8) of the same type of photocathode were measured.
Though all the photocathodes were produced in the same way, some differences in the QE spectra measured at 0$^\circ$ are recognizable in Fig.~\ref{fig:QE_spectrum}.
While PMT1-7 have similar spectra, PMT8 differs from the others:
the former spectra peak around 360~nm and the latter peaks around 420~nm.
The QE at the peak varies PMT-by-PMT from 0.249 to 0.306.
Within the individual PMTs the QE is uniform over the photocathode.
It was confirmed in advance by measuring the QE at $18\times18$ points over the photocathode including the fringe at 0$^\circ$ without the semi-cylinder.
A typical example is shown in Fig.~\ref{fig:QE_uniformity} for PMT3.

\begin{figure}[!h]
 \vspace{1cm}
 \centerline{
  \hspace{0.6cm}
  \subfloat{\includegraphics[width=0.55\textwidth]{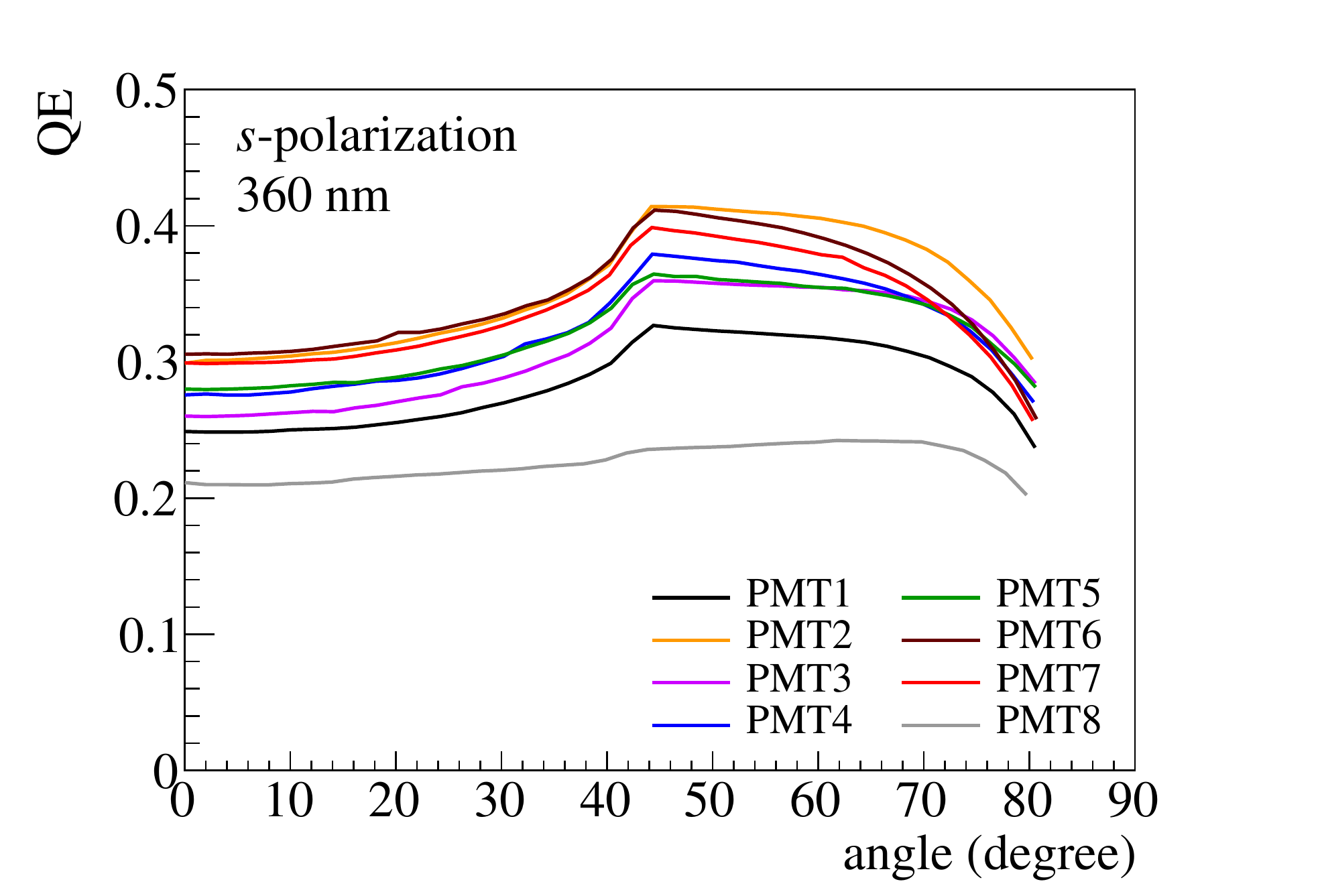}}
  \hspace{-0.6cm}
  \subfloat{\includegraphics[width=0.55\textwidth]{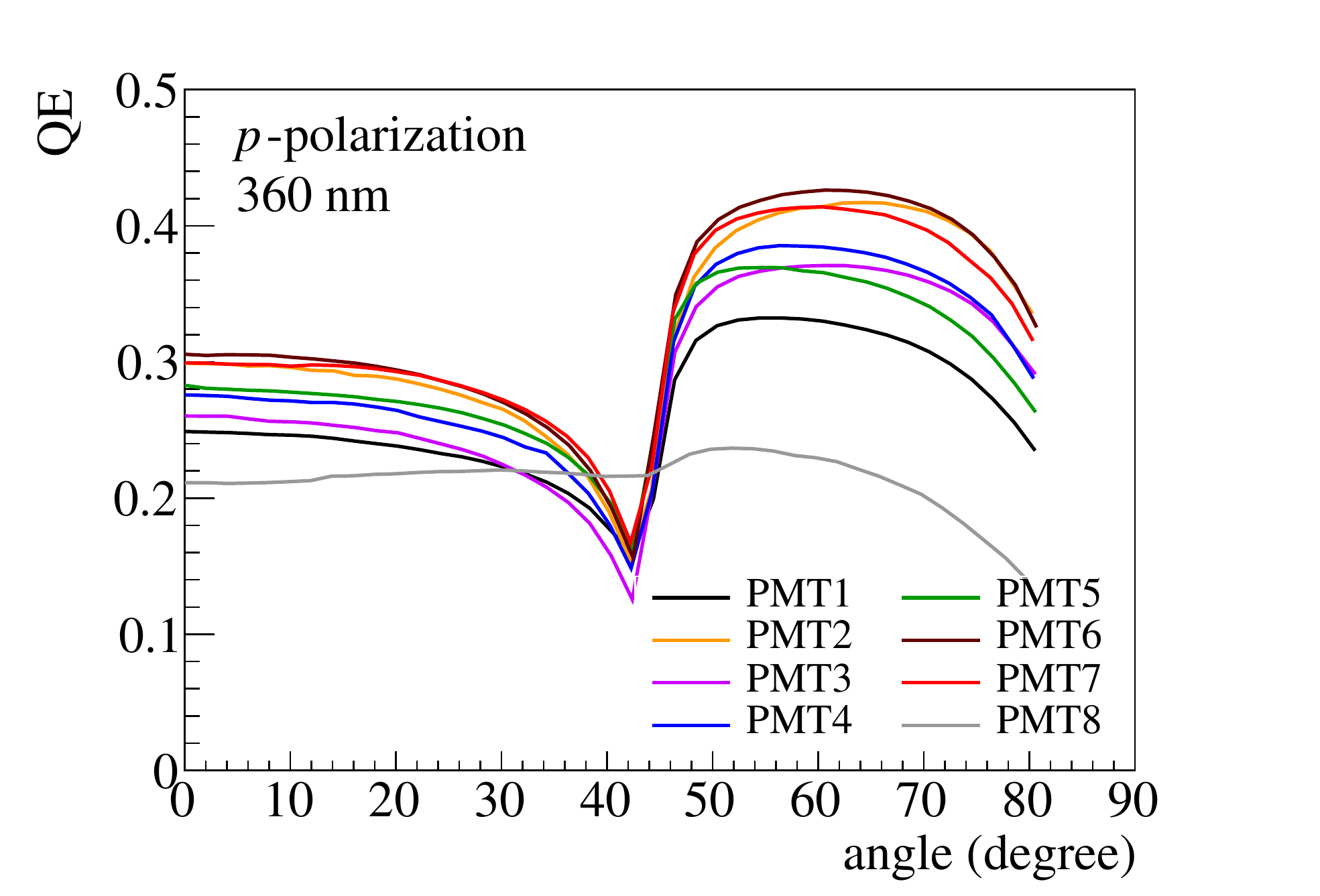}}
 }
 \centerline{
  \hspace{0.6cm}
  \subfloat{\includegraphics[width=0.55\textwidth]{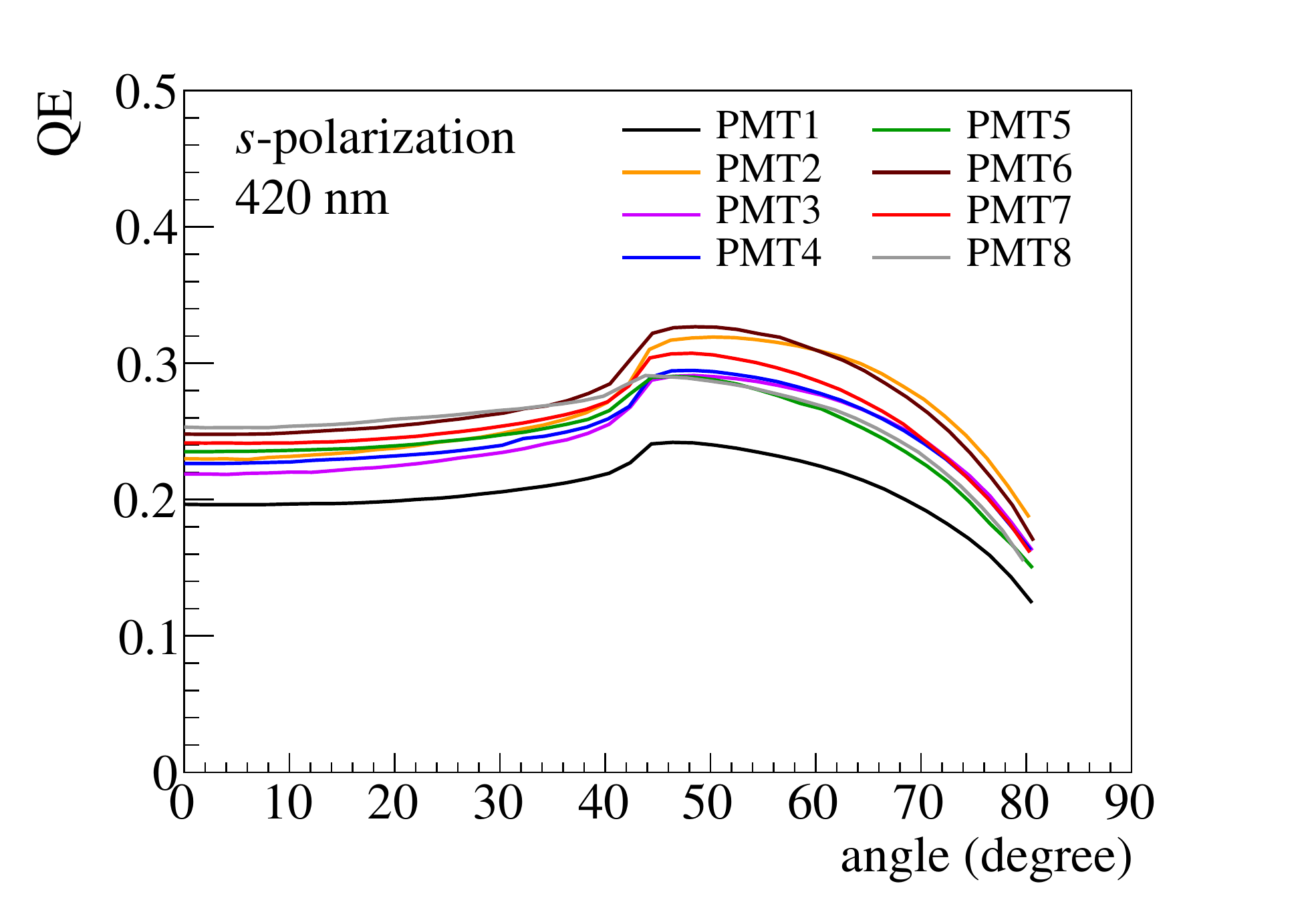}}
  \hspace{-0.6cm}
  \subfloat{\includegraphics[width=0.55\textwidth]{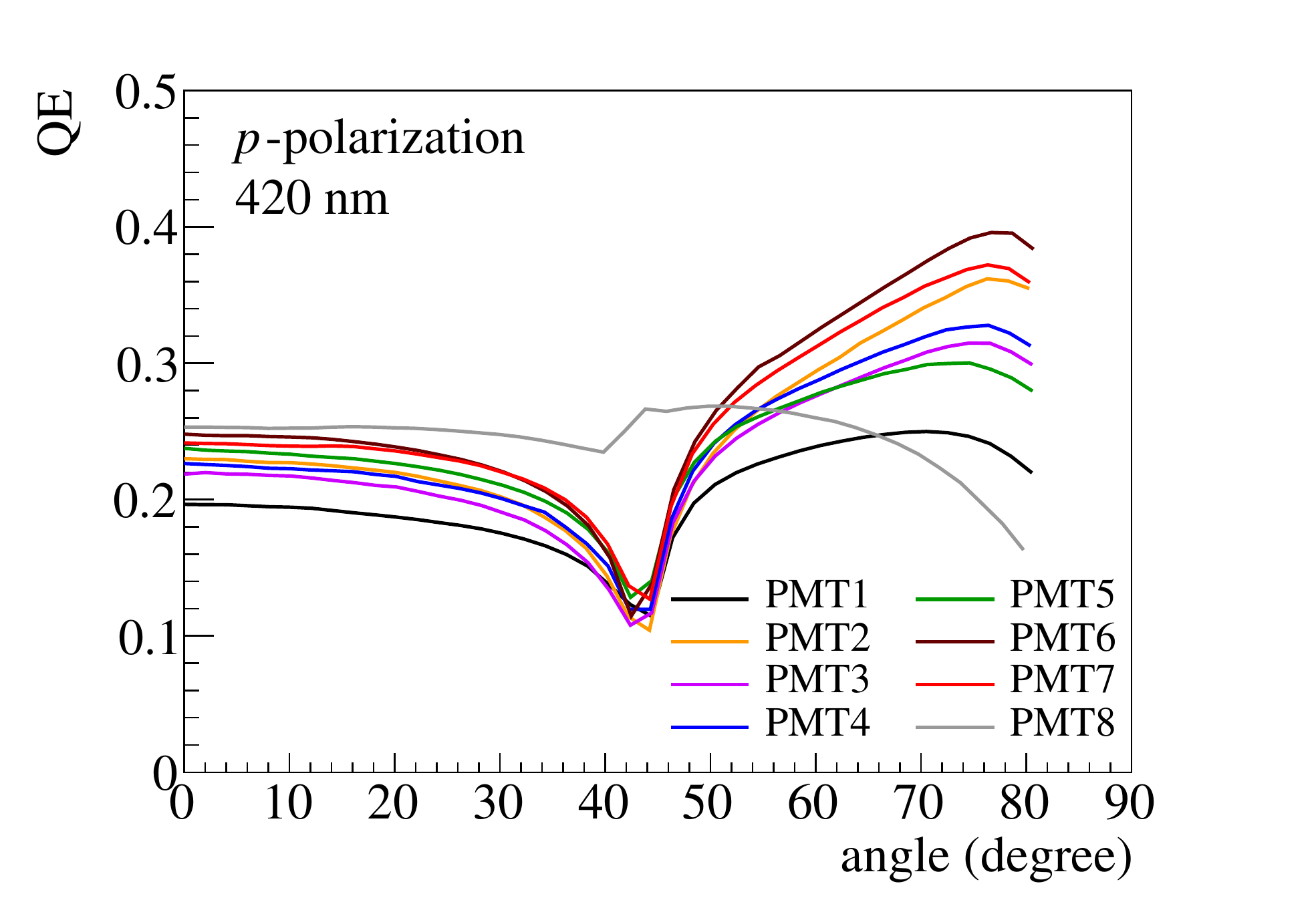}}
 }
 \centerline{
  \hspace{0.6cm}
  \subfloat{\includegraphics[width=0.55\textwidth]{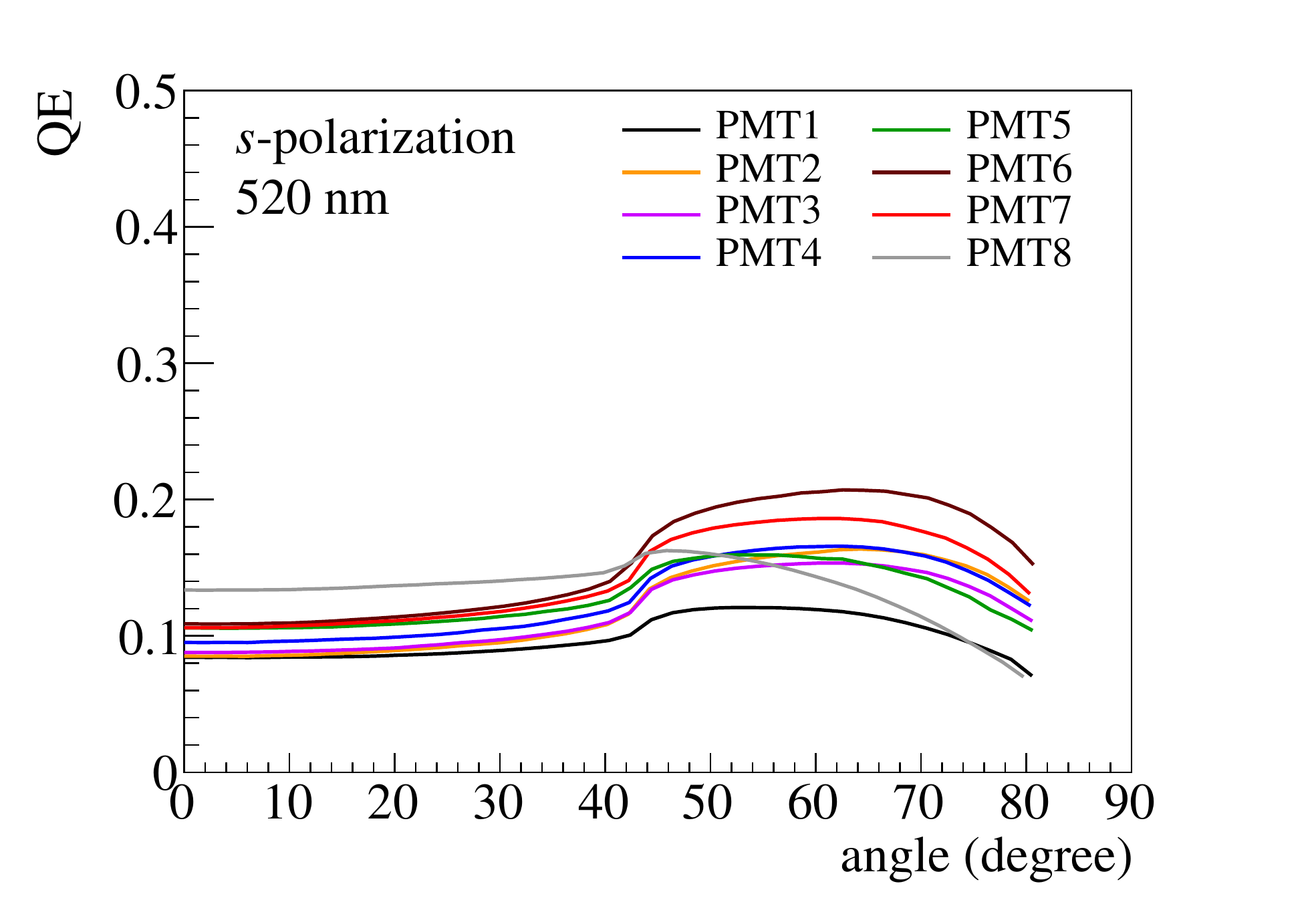}}
  \hspace{-0.6cm}
  \subfloat{\includegraphics[width=0.55\textwidth]{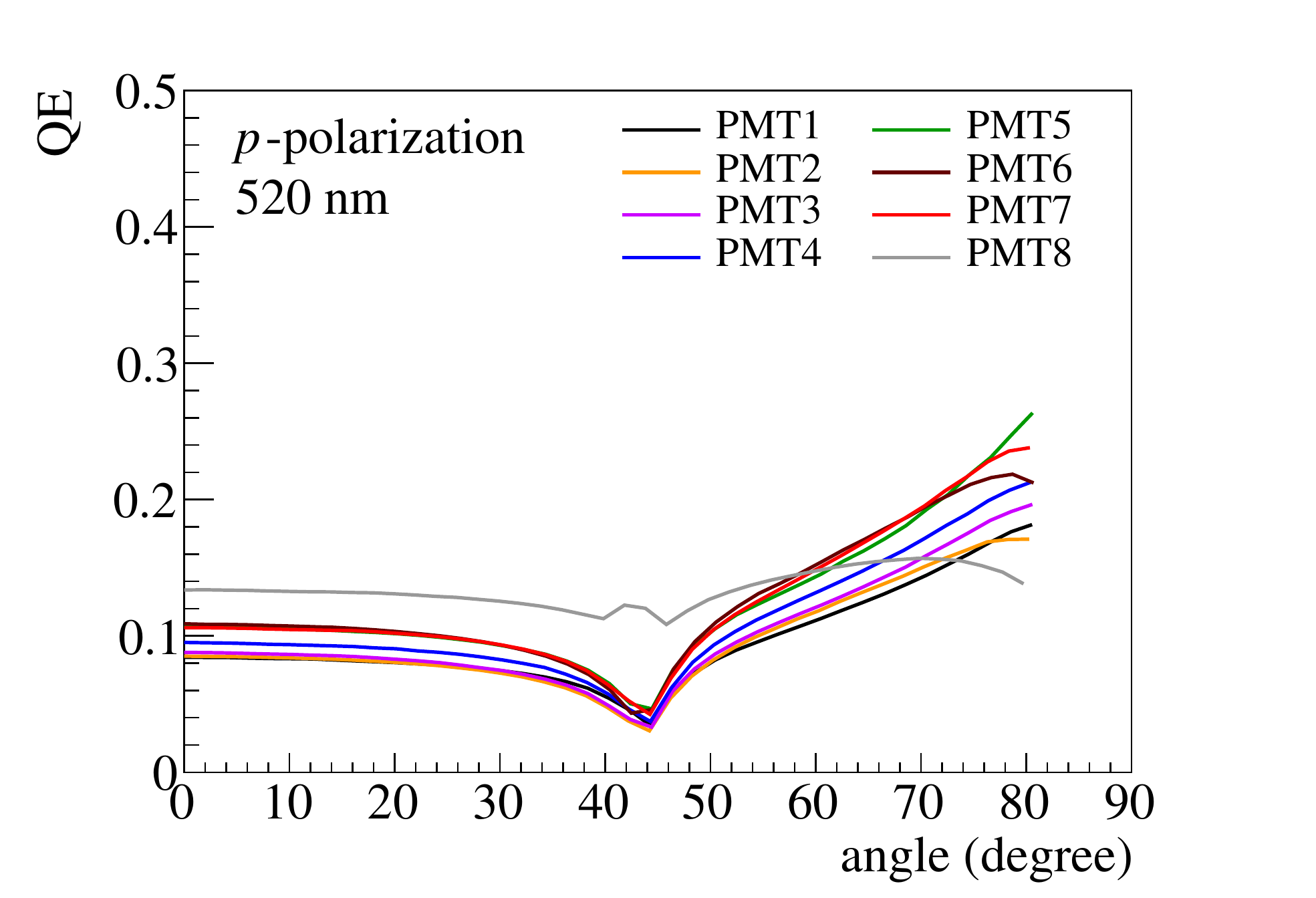}}
 }
 \caption{Measured QE as a function of the incident angle for $s$-polarized (left) and $p$-polarized (right) light.
 The data only at 360~nm (top), 420~nm (middle) and 520~nm (bottom) wavelengths are shown in this figure.
 The error bars of the data are omitted.}
 \label{fig:QE_angle}
 \vspace{2cm}
\end{figure}

The QEs measured as a function of the incident angle are shown in Fig.~\ref{fig:QE_angle}.
The broad feature of the curve is similar for all the PMTs except PMT8.
The prominent rise (valley) of the QE for $s$($p$)-polarization around 43$^\circ$ mentioned in Sec.~\ref{sec:optical_model} cannot be seen for PMT8.

\begin{figure}[p]
 \centering
 \includegraphics[width=1.1\textwidth,keepaspectratio,bb=25 38 572 651]{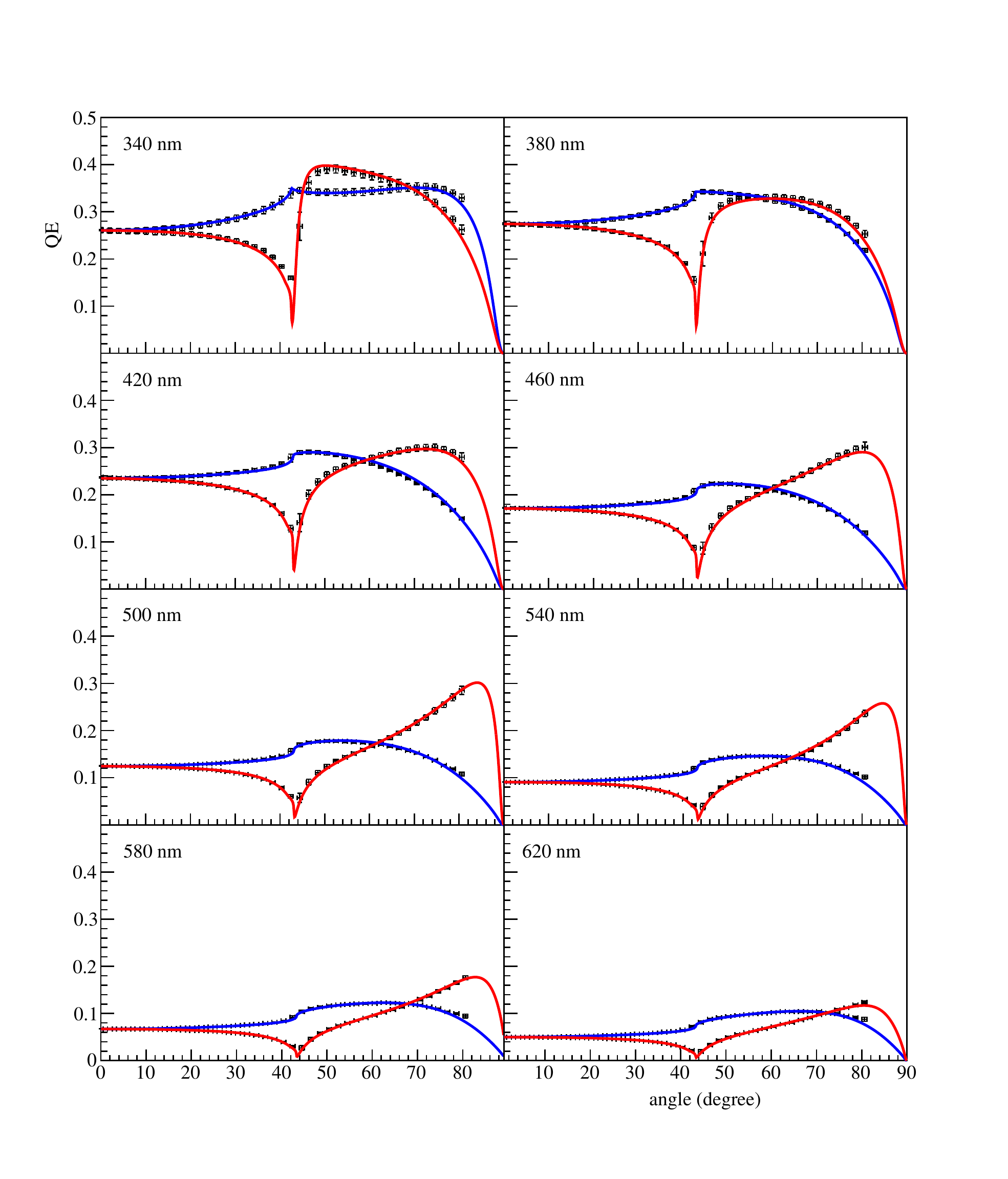}
 \caption{Fitted function (blue lines for $s$-polarization and red ones for $p$-polarization) overlaid on the data (dots with error bars) for PMT5.
 Only those at 340, 380, 420, 460, 500, 540, 580 and 620~nm wavelengths are shown in this figure.}
 \label{fig:QE_fit}
 \vspace{1.5cm}
\end{figure}

The function of Eq.~(\ref{eq:QEfunction}) was fitted to the QE data of each PMT.
The fitting was done simultaneously for both polarizations, all the 18~wavelengths (320-660~nm) and all the 41~angles (0-80$^\circ$) using chi-squared minimization.
Comparisons between the data and the fitted function can be seen for PMT5 as an example in Fig.~\ref{fig:QE_fit}.
The error bars on the data points in this figure include all the statistical and systematical errors.
It is remarkable that the single function fits all the data of the PMT accurately.
For the other PMTs the agreement between the data and the fitted function is similar or better except for PMT8.
The function of Eq.~(\ref{eq:QEfunction_loss}), which has the additional parameter $L_e$, fits better to PMT8.
It was also fitted for the other PMTs, and consistent results with Eq.~(\ref{eq:QEfunction}) were obtained.

\begin{figure}[p]
 \tblcaption{Thickness and the optical parameters of the photocathode and the antireflection coating obtained by the fitting with Eq.~(\ref{eq:QEfunction}) for PMT1-7 and Eq.~(\ref{eq:QEfunction_loss}) for PMT8.
 The reduced $\chi^2$ of the fit is also shown in the last column.
 The last row shows the average of the eight PMTs.}
 \label{tab:parameter}
 \centering
 {
 \begin{tabular}{llllllllll}
  \hline
  & \multicolumn{4}{l}{Photocathode} \\
  PMT & $d$ (nm) & $\varepsilon_s$ & $\omega_t$ (eV) & $\varGamma_0$ (eV) & $L_e$ (nm) \\
  \hline

  1 & 11.2$\pm$2.6 & \phantom{0}8.5$\pm$1.8 & 3.17$\pm$0.11 & 1.22$\pm$0.19 & \phantom{0}-- \\
  2 & \phantom{0}6.4$\pm$2.7 & \phantom{0}9.7$\pm$4.0 & 3.30$\pm$0.14 & 1.23$\pm$0.20 & \phantom{0}-- \\
  3 & \phantom{0}5.5$\pm$2.3 & 10.9$\pm$4.2 & 3.22$\pm$0.13 & 1.10$\pm$0.19 & \phantom{0}-- \\
  4 & \phantom{0}8.2$\pm$1.5 & \phantom{0}8.3$\pm$1.4 & 3.23$\pm$0.07 & 1.08$\pm$0.12 & \phantom{0}-- \\
  5 & \phantom{0}9.9$\pm$2.9 & \phantom{0}8.2$\pm$2.1 & 3.24$\pm$0.10 & 0.98$\pm$0.14 & \phantom{0}-- \\
  6 & \phantom{0}6.6$\pm$3.9 & \phantom{0}8.7$\pm$4.8 & 3.25$\pm$0.10 & 0.98$\pm$0.15 & \phantom{0}-- \\
  7 & \phantom{0}8.5$\pm$0.9 & \phantom{0}7.8$\pm$0.8 & 3.25$\pm$0.05 & 0.99$\pm$0.10 & \phantom{0}-- \\
  8 & 34.1$\pm$3.1 & \phantom{0}6.8$\pm$0.6 & 3.32$\pm$0.05 & 1.04$\pm$0.06 & 27.1$\pm$2.5 \\
  Avg. & \multicolumn{1}{c}{NA} & \phantom{0}7.4$\pm$0.4 & 3.26$\pm$0.03 & 1.05$\pm$0.04 & \multicolumn{1}{c}{NA} \\
  \hline
  \\
  \hline
  & \multicolumn{4}{l}{Antireflection coating} & \\
  PMT & $d$ (nm)  & $\varepsilon_s$ & $\omega_t$ (eV) & $\varGamma_0$ (eV) & $\chi^2$/ndf\phantom{0} \\
  \hline
  1 & 36.8$\pm$5.7  & 4.3$\pm$0.6 & 17.2$\pm$28.4 & \phantom{0}3.4$\pm$13.6 & 0.91 \\
  2 & 38.6$\pm$8.0  & 4.0$\pm$0.5 & 19.7$\pm$24.3 & \phantom{0}5.9$\pm$11.7 & 0.71 \\
  3 & 40.6$\pm$6.6  & 4.1$\pm$0.5 & 16.4$\pm$12.4 & \phantom{0}4.4$\pm$\phantom{0}7.4 & 0.67 \\
  4 & 36.9$\pm$5.5  & 4.2$\pm$0.4 & 26.6$\pm$16.9 & 10.6$\pm$13.5 & 0.85 \\
  5 & 37.6$\pm$5.9 & 4.5$\pm$0.6 & 24.7$\pm$25.3 & \phantom{0}7.3$\pm$15.8 & 0.91 \\
  6 & 30.5$\pm$8.4 & 4.5$\pm$0.8 & 12.5$\pm$10.3 & \phantom{0}3.6$\pm$\phantom{0}5.4 & 0.56 \\
  7 & 31.7$\pm$4.3 & 4.7$\pm$0.5 & 29.5$\pm$29.1 & 14.5$\pm$12.5 & 0.91 \\
  8 & 48.3$\pm$6.9  & 3.6$\pm$0.3 & 11.7$\pm$13.5 & \phantom{0}1.7$\pm$\phantom{0}1.4 & 1.22 \\
  Avg. & 36.9$\pm$2.1 & 4.1$\pm$0.2 & 16.7$\pm$\phantom{0}5.7 & \phantom{0}2.3$\pm$\phantom{0}1.3 & \phantom{0}-- \\
  \hline
 \end{tabular}
 }

 \vspace{1.6cm}

 \includegraphics[width=0.55\textwidth]{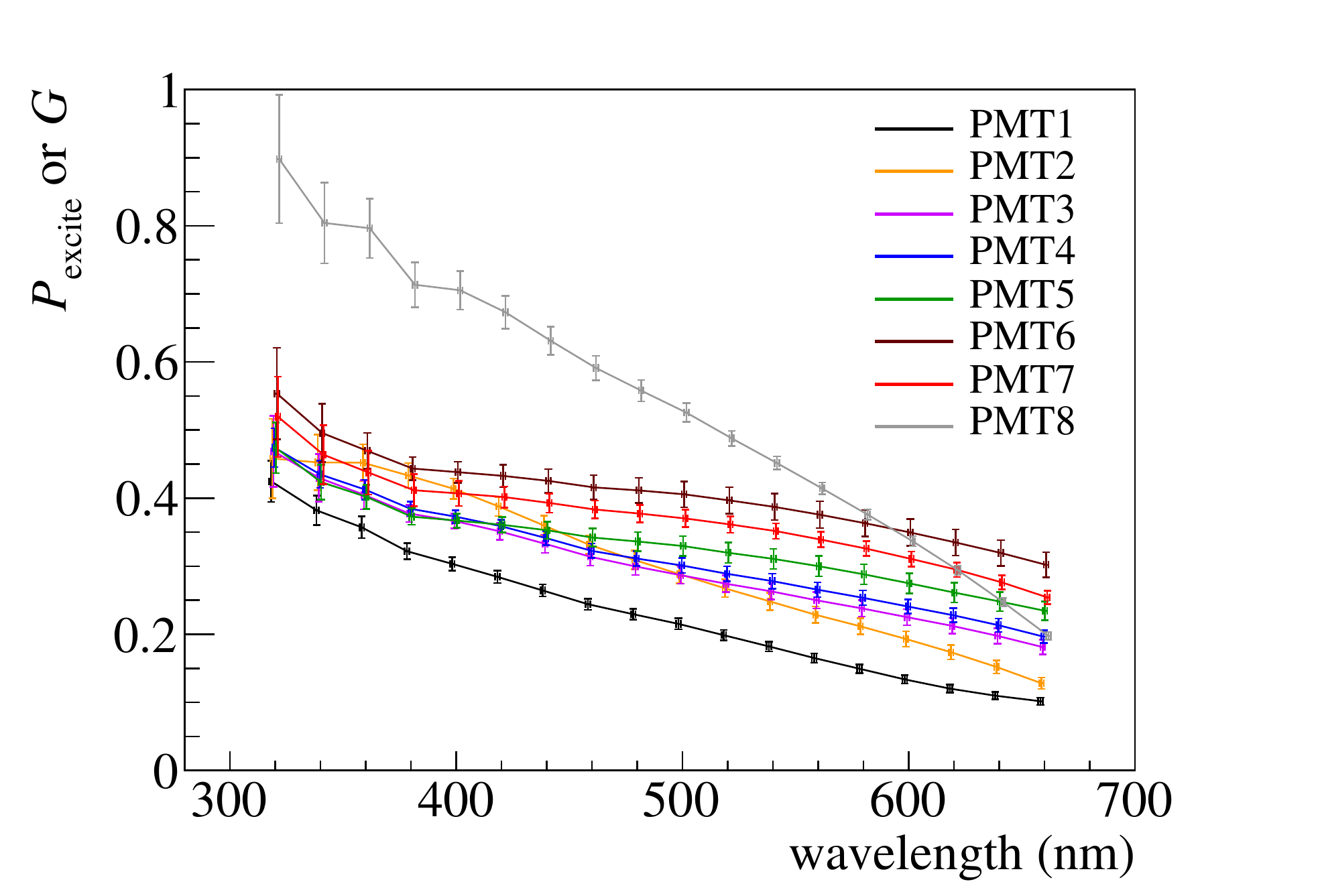}
 \figcaption{$P_{\rm excite}$ or $G$ for PMT8 obtained by the fitting.}
 \label{fig:Pexcite}
\end{figure}

The best fit parameters of Eq.~(\ref{eq:QEfunction}) for PMT1-7 and those of Eq.~(\ref{eq:QEfunction_loss}) for PMT8 are shown in Table~\ref{tab:parameter} and Fig.~\ref{fig:Pexcite}.
No significant differences in the thickness and the optical parameters are identified for the antireflection coating.
The errors on $\omega_t$ and $\varGamma_0$ of the antireflection coating are sizable.
That is because the measurement was done in the wavelength range of 320-660~nm (or 1.9-3.9~eV) and had little sensitivity to $\omega_t$ and $\varGamma_0$.
Nevertheless the refractive index of the antireflection coating in that wavelength range should be reliable since it is derived mainly from $\varepsilon_s$.
The photocathode has a variation of the thickness.
Especially the one of PMT8 is thicker than the others.
It distinguishes the photocathode response as being different from the others as seen in Figs.~\ref{fig:QE_spectrum} and \ref{fig:QE_angle};
a thicker photocathode is more sensitive at longer wavelengths, and has less QE variation along with the angle.
It is notable that the photocathode of PMT8 is thicker than $L_e$, which requires incorporating the scattering loss into the QE function as discussed in Sec.~\ref{sec:model}.
On the other hand, for the thin photocathodes of PMT1-7, the approximation by the one-step model holds well.
The optical parameters of the photocathodes are consistent with each other while $P_{\rm excite}$ differs PMT-by-PMT.
$P_{\rm excite}$ and $G$ seem to be a monotonically increasing function of the photon energy.

The reduced $\chi^2$ of the fit is also listed in Table~\ref{tab:parameter}.
Since the number of degrees of freedom is as large as 1450 or 1449 with $L_e$, a reduced $\chi^2$ which deviates from 1 by about 0.1 or more is statistically improbable.
The small reduced $\chi^2$ for PMT1-7 could be due to overestimation of the systematic errors.
Nevertheless the good agreement between the data and the fitted function indicates validity of the models.
The Lorentz dispersion model is appropriate for this type of photocathodes.
The one-step model of the photoelectron emission is applicable for photocathodes much thinner than the escape length.
For thicker photocathodes the approximation of $P_{\rm transport}P_{\rm escape}$ in Eq.~(\ref{eq:PP_thick}) holds well.
A little deviation of the fitted function from the data was found around $42^\circ$ only in the wavelengths not longer than 340~nm for $p$-polarization as shown in Fig.~\ref{fig:QE_fit}.
It was found in every PMT systematically.
It could be an indication of the effect of the electron-electron scattering, which we have neglected for 310~nm or longer wavelengths as discussed in Sec.~\ref{sec:model}.
The large reduced $\chi^2$ for PMT8 is also attributed mainly to that deviation.
It is reasonable that a larger deviation by the scattering loss is seen in a thicker photocathode.

\subsection{Effects of the scattering loss} \label{sec:scattering-loss}

\begin{figure}[t]
\centerline{
 \hspace{0.6cm}
 \subfloat{\includegraphics[width=0.55\textwidth,keepaspectratio]{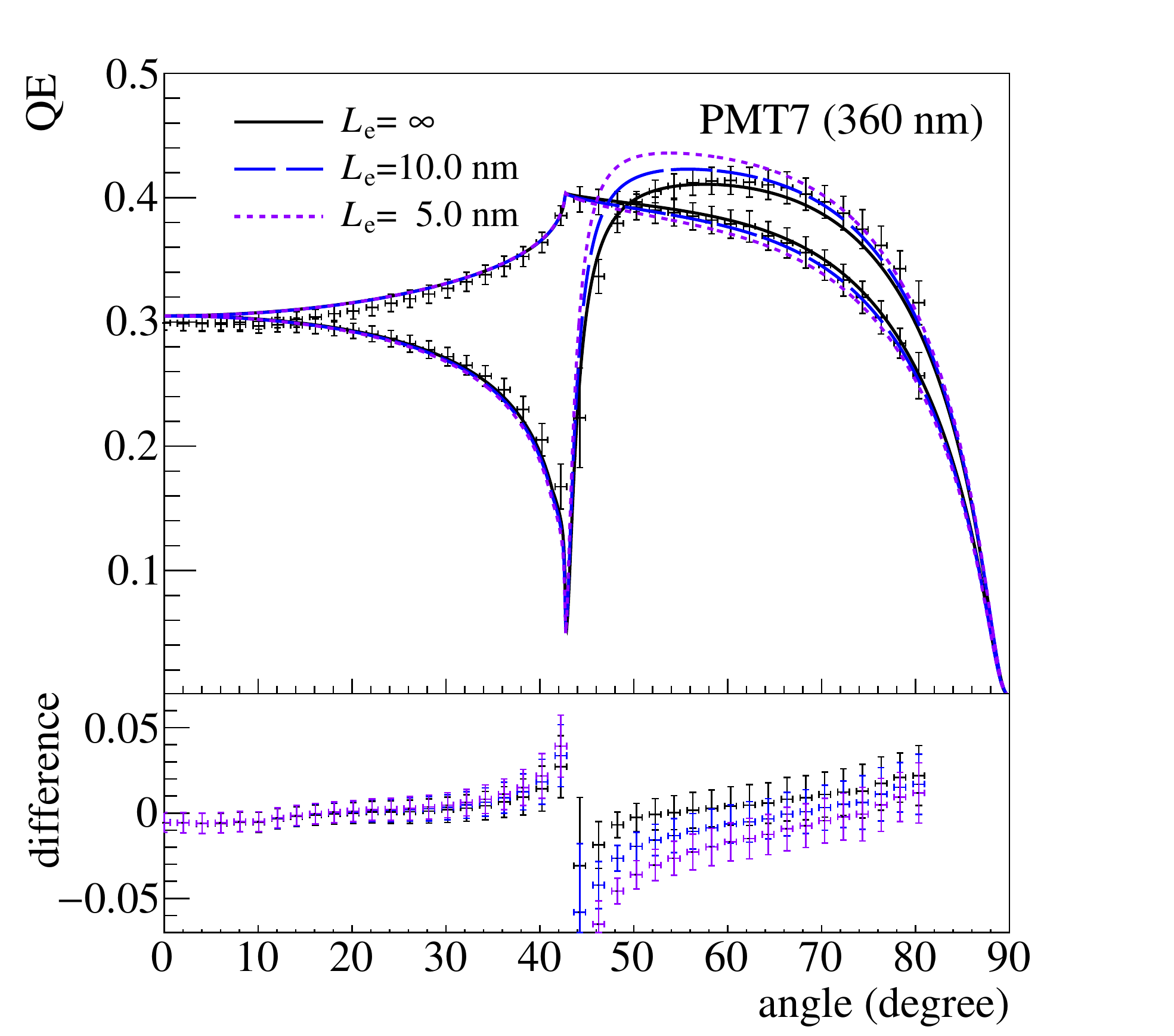}}
 \hspace{-0.6cm}
 \subfloat{\includegraphics[width=0.55\textwidth,keepaspectratio]{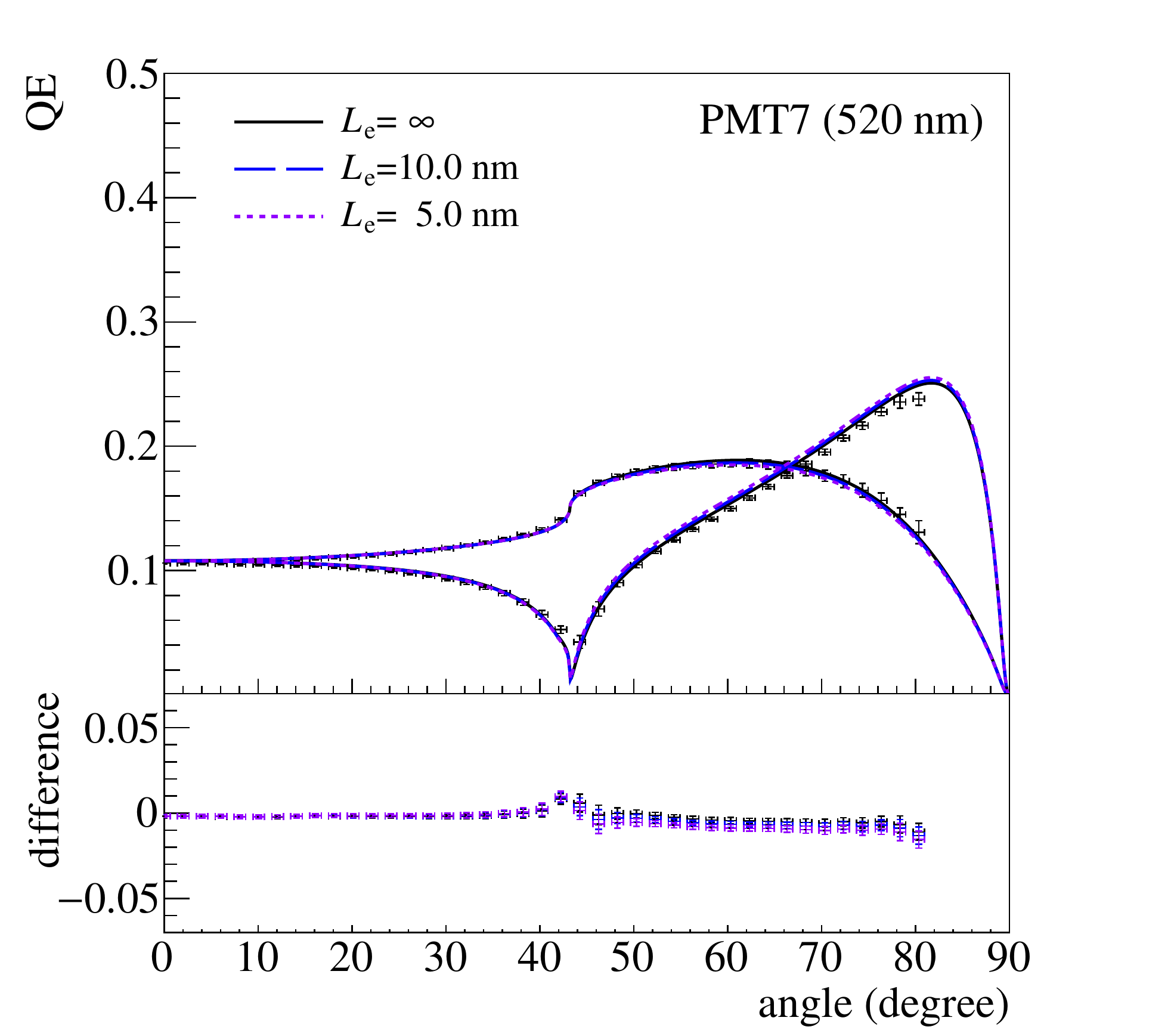}}
}
\centerline{
 \hspace{0.6cm}
 \subfloat{\includegraphics[width=0.55\textwidth,keepaspectratio]{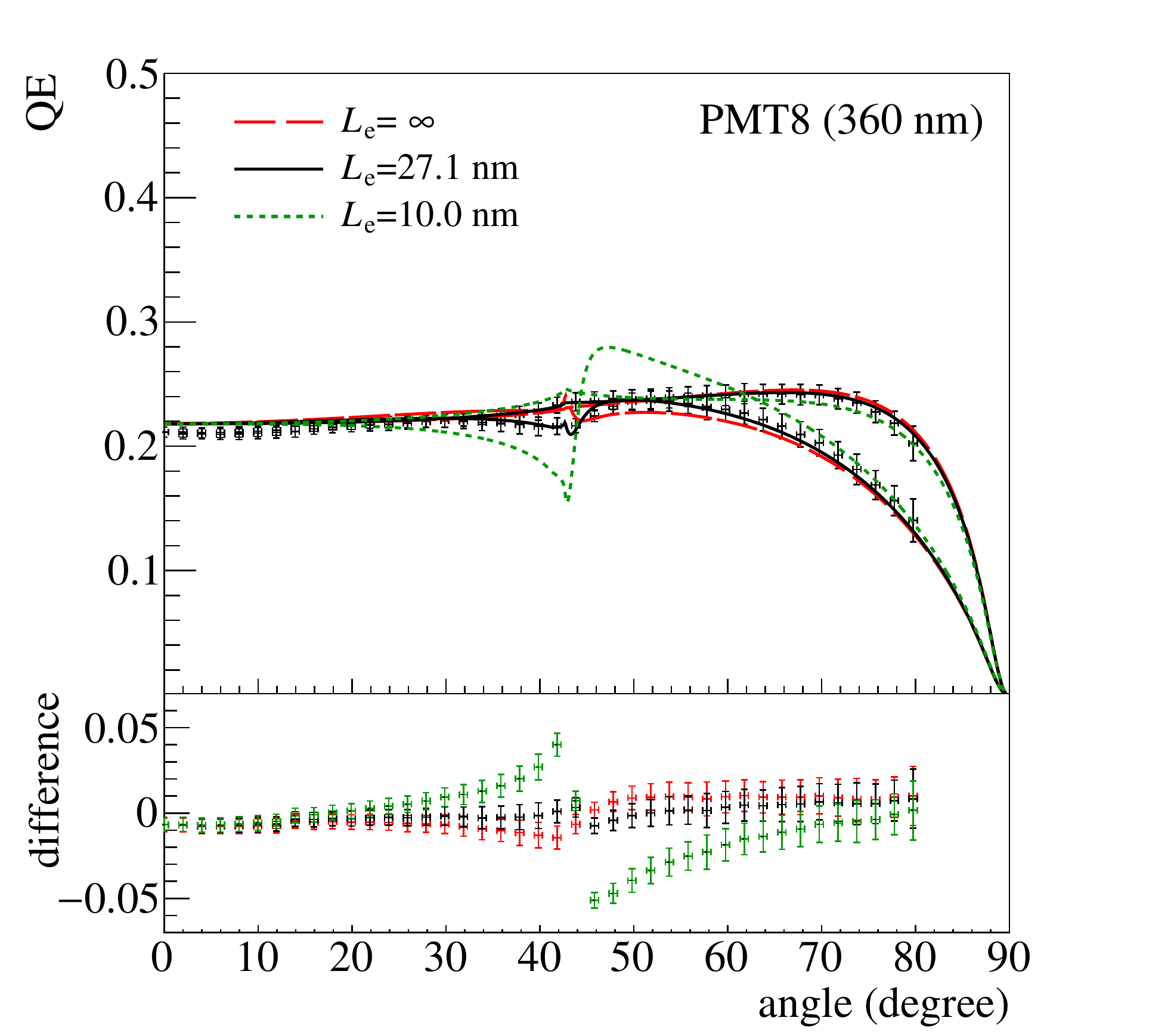}}
 \hspace{-0.6cm}
 \subfloat{\includegraphics[width=0.55\textwidth,keepaspectratio]{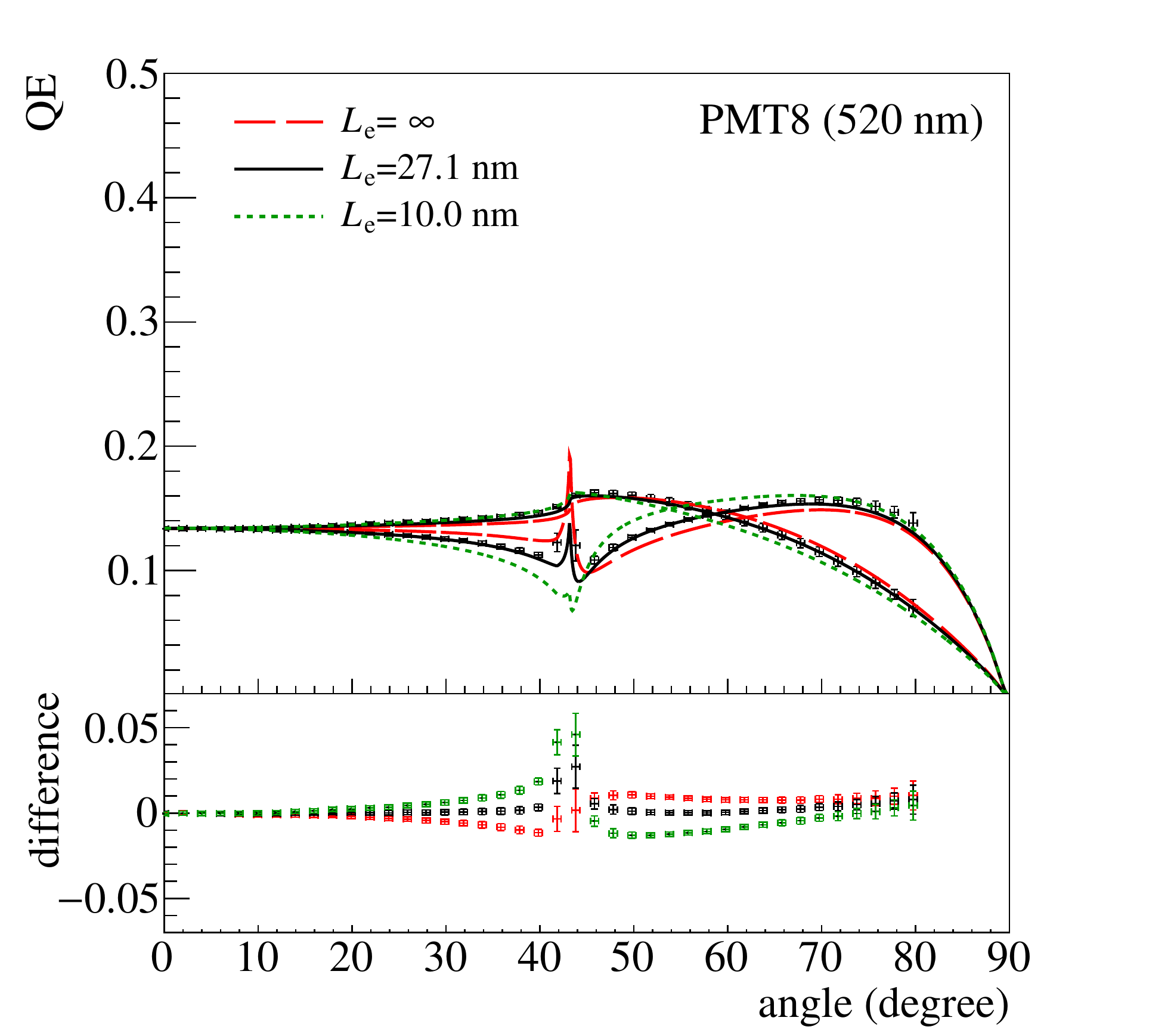}}
}
 \caption{Expected angular dependence of the QE at 360~nm (left) and 520~nm (right) with different escape lengths $L_e$ for PMT7 (upper) and PMT8 (lower).
 The QE curves both for $s$- and $p$-polarizations are plotted in the same upper panels.
 The dots with error bars represent the data.
 The differences of the data from each curve are shown only for $p$-polarization in the lower panels.}
 \label{fig:QE_loss}
\end{figure}

\begin{figure}[t]
\centerline{
 \hspace{0.6cm}
 \subfloat{\includegraphics[width=0.55\textwidth,keepaspectratio]{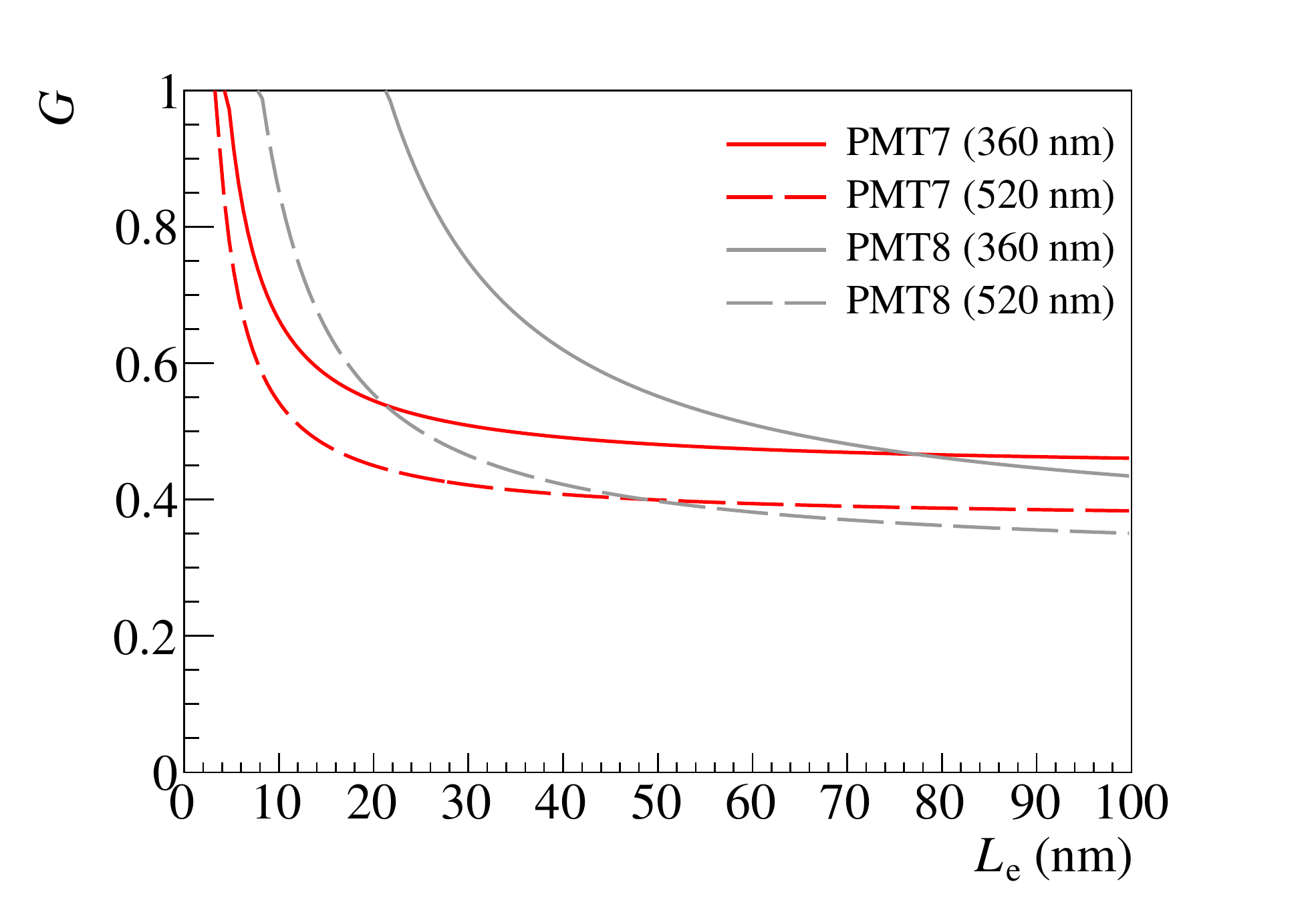}}
 \hspace{-0.6cm}
 \subfloat{\includegraphics[width=0.55\textwidth,keepaspectratio]{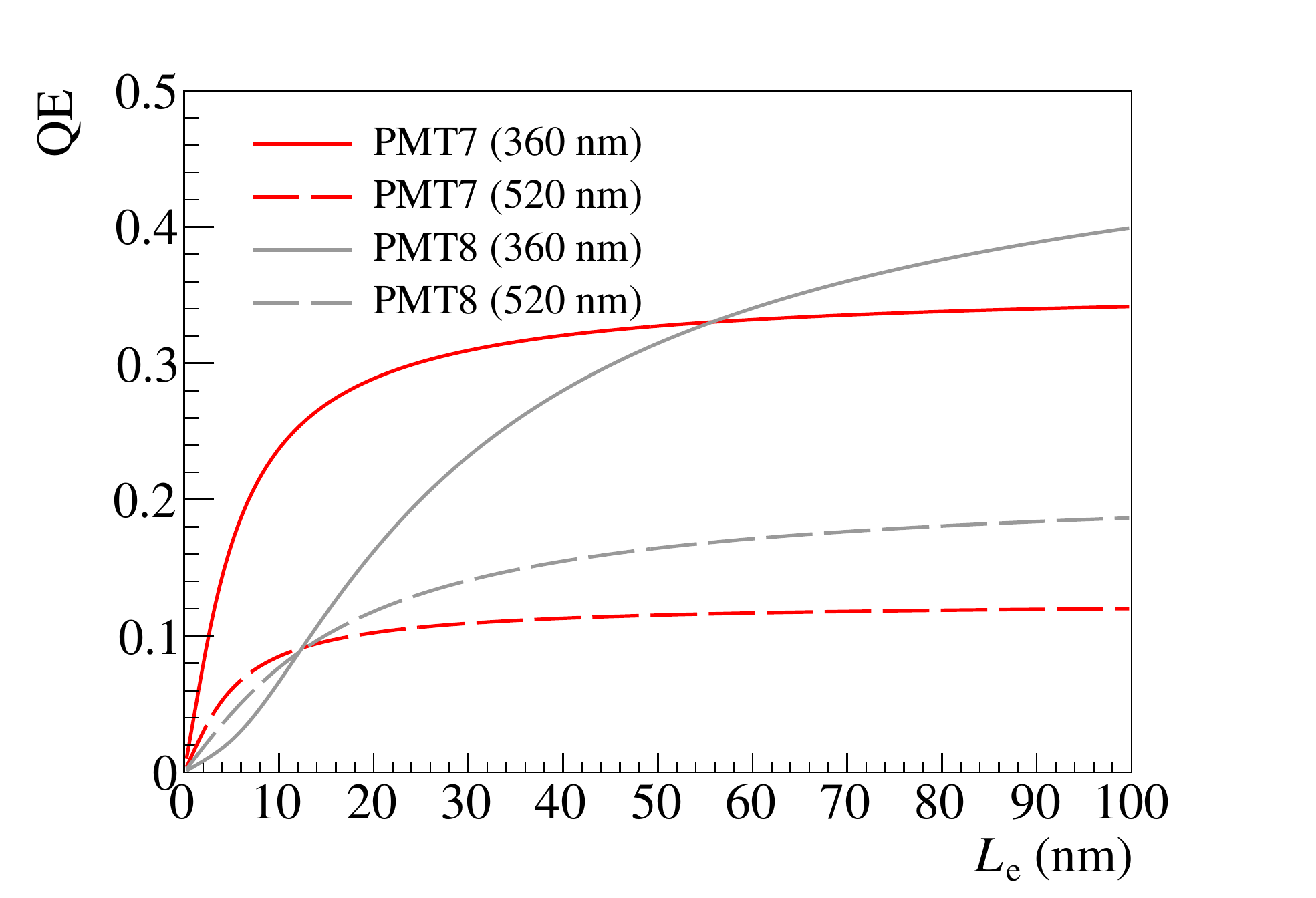}}
}
 \caption{(Left) $G$ which makes the QE function normalized to the data at 0$^\circ$ depending on the escape length.
 (Right) Variation of the QE at 0$^\circ$ as a function of the escape length when $G$ is fixed to the one for $L_e$ $= 27.1$~nm.
 $G$ for PMT7 is determined to have the same QE at $L_e=27.1$~nm as the one of the data at 0$^\circ$.
 Those for PMT7 and 8 at 360 and 520~nm wavelengths are shown in these plots.}
 \label{fig:QE_loss_Le}
\end{figure}

In order to compare the angular dependence of the QE with different $L_e$'s, Eq.~(\ref{eq:QEfunction_loss}) is plotted for PMT7 and 8 in Fig.~\ref{fig:QE_loss}, where the optical parameters are fixed to those obtained by the fitting but $G$ is adjusted to normalize the QE to the one of the data at $\theta = 0^\circ$.
$G$ to be applied as a function of $L_e$ is shown in Fig.~\ref{fig:QE_loss_Le} (left).
The effect of $L_e$ on the angular dependence of the QE is found to be little for $s$-polarization, which indicates that the optical parameters are strongly constrained by the angular dependence for $s$-polarization independently of $L_e$.
The one for $p$-polarization is rather visible, but the angular dependence does not start to differ clearly from the data down to $L_e\approx10$~nm.
That corroborates the consequence of the one-step model that the angular dependence is dictated only by the optical properties for thin photocathodes.
It also indicates that $L_e$ for PMT1-7 is larger than the photocathode thickness.
Actually $L_e$'s obtained by the fitting with Eq.~(\ref{eq:QEfunction_loss}) for PMT1-7 are 19.9-32.3~nm though the errors are sizable.
Moreover a distinct lower limit on $L_e$ can be obtained from Fig.~\ref{fig:QE_loss_Le} (left) as $G < 1$ by definition: for example, $L_e > 4.6$~nm (PMT7) or 21.4~nm (PMT8) at 360~nm wavelength.
It is consistent with the measured $L_e$ reported in Ref.~\cite{escapelength_Dolizy}.

According to Eq.~(\ref{eq:QEfunction_loss}), the QE largely depends on $L_e$ when $L_e$ is roughly equal to or less than the photocathode thickness as shown in Fig.~\ref{fig:QE_loss_Le} (right), where $\theta = 0^\circ$ and the parameters other than $L_e$ are fixed.
With $L_e=10$~nm, for example, the QE at 0$^\circ$ and 360~nm wavelength is reduced to 78\% of the one with $L_e=$ 27.1~nm for PMT7.
Since the effect of the scattering loss is implicitly incorporated into $P_{\rm excite}$ in the one-step model, one of the causes for the PMT-by-PMT difference of $P_{\rm excite}$ therefore could be a difference of $L_e$ especially due to the scattering with impurities or defects.

\section{Optical properties of the photocathode}
As denoted in Eq.~(\ref{eq:QEfunction}) the optical properties of the photocathode as well as those of the antireflection coating are of great importance in the QE and its angular dependence.
In the following they are discussed using the optical parameters obtained in the previous section.

\subsection{Refractive index and absorptance of the photocathode}
\begin{figure}[t]
\centerline{
 \hspace{0.6cm}
 \subfloat{\includegraphics[width=0.55\textwidth,keepaspectratio]{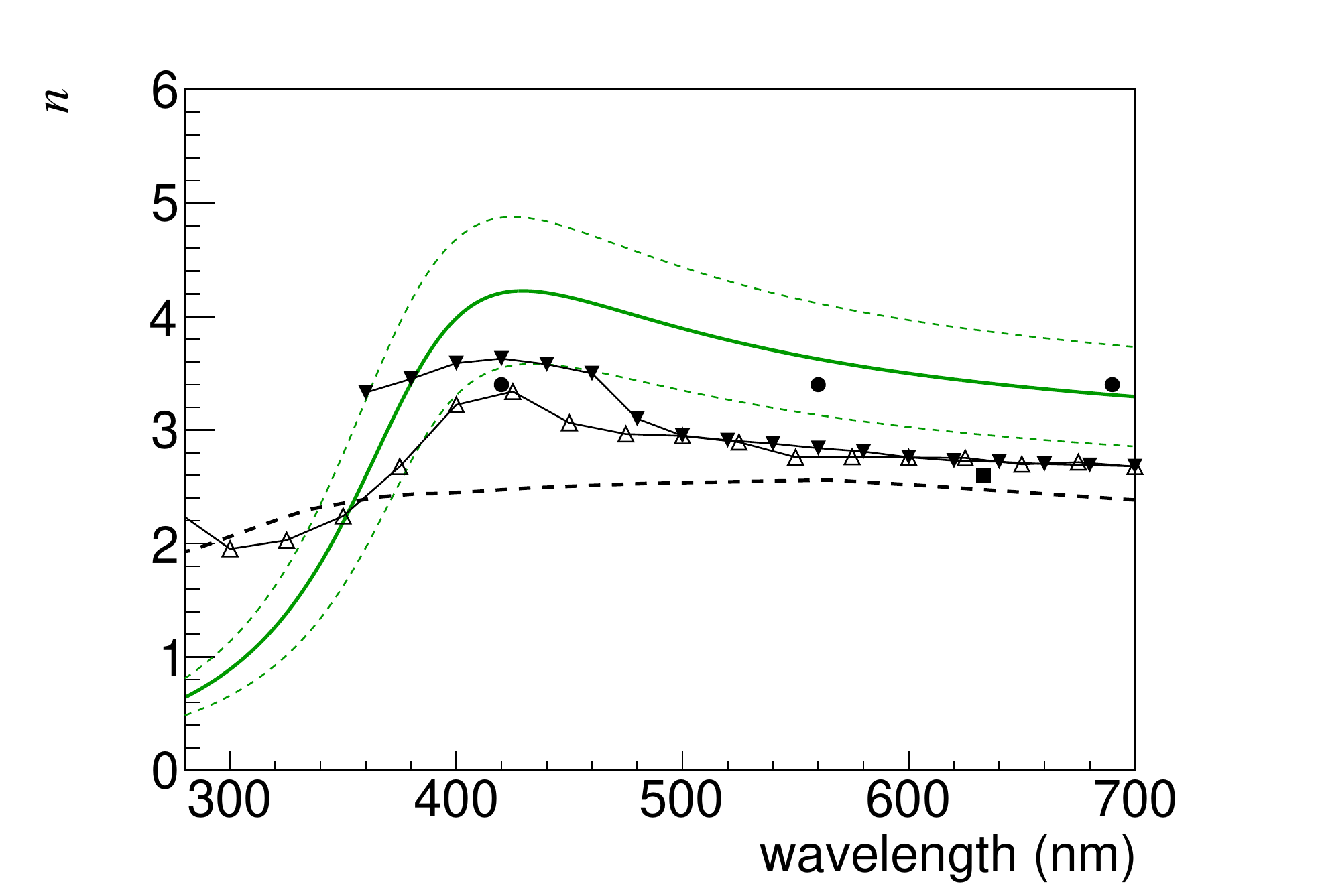}}
 \hspace{-0.6cm}
 \subfloat{\includegraphics[width=0.55\textwidth,keepaspectratio]{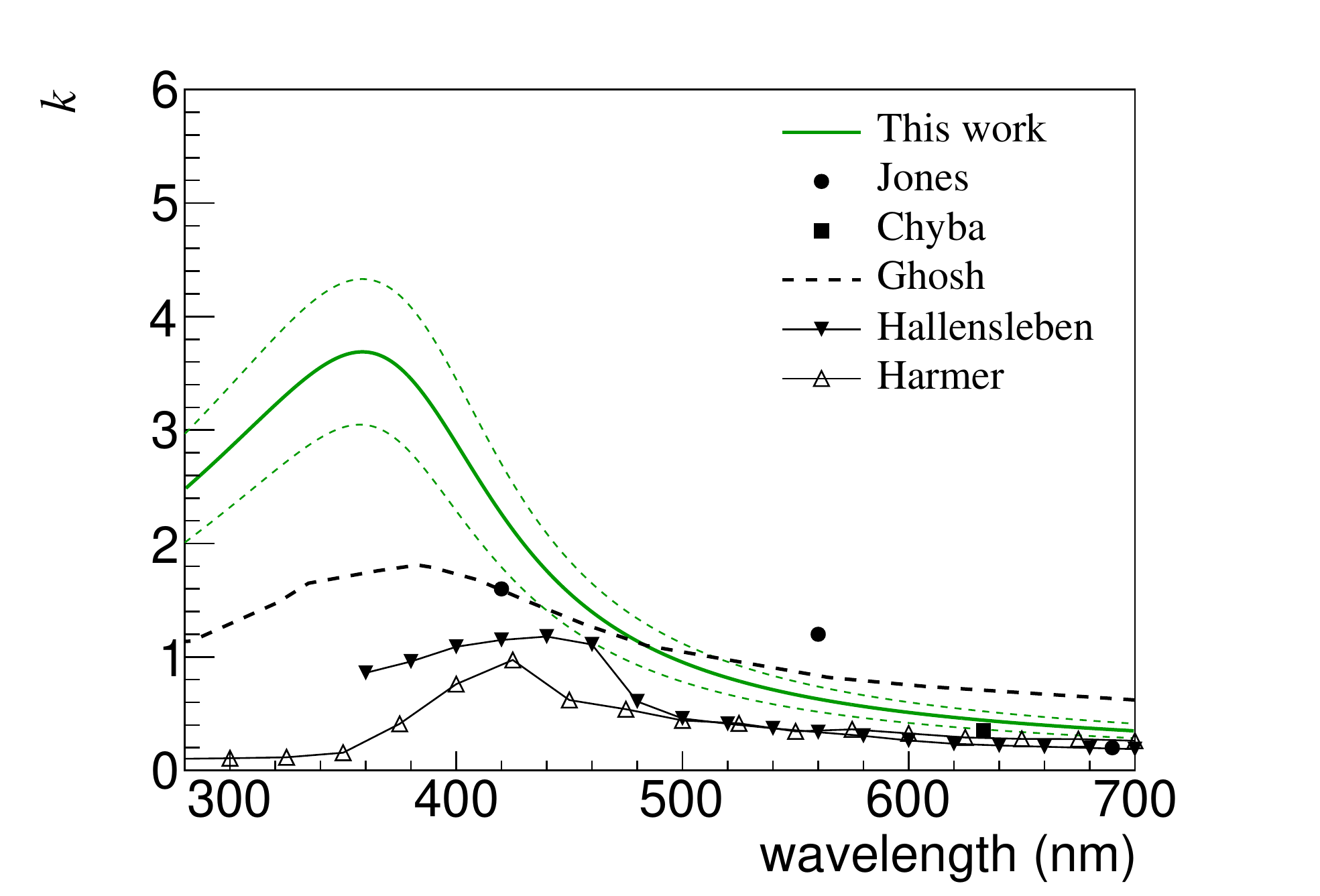}}
}
\caption{Real (left) and imaginary (right) parts of the refractive index of the photocathode for PMT5.
The solid and dotted green lines represent the best fit dispersions and the 1$\sigma$ allowed bands, respectively.
The previously published data for the same type of photocathodes are also plotted for comparison.}
\label{fig:index_pc}
\end{figure}

\begin{figure}[t]
\centerline{
 \hspace{0.6cm}
 \subfloat{\includegraphics[width=0.55\textwidth,keepaspectratio]{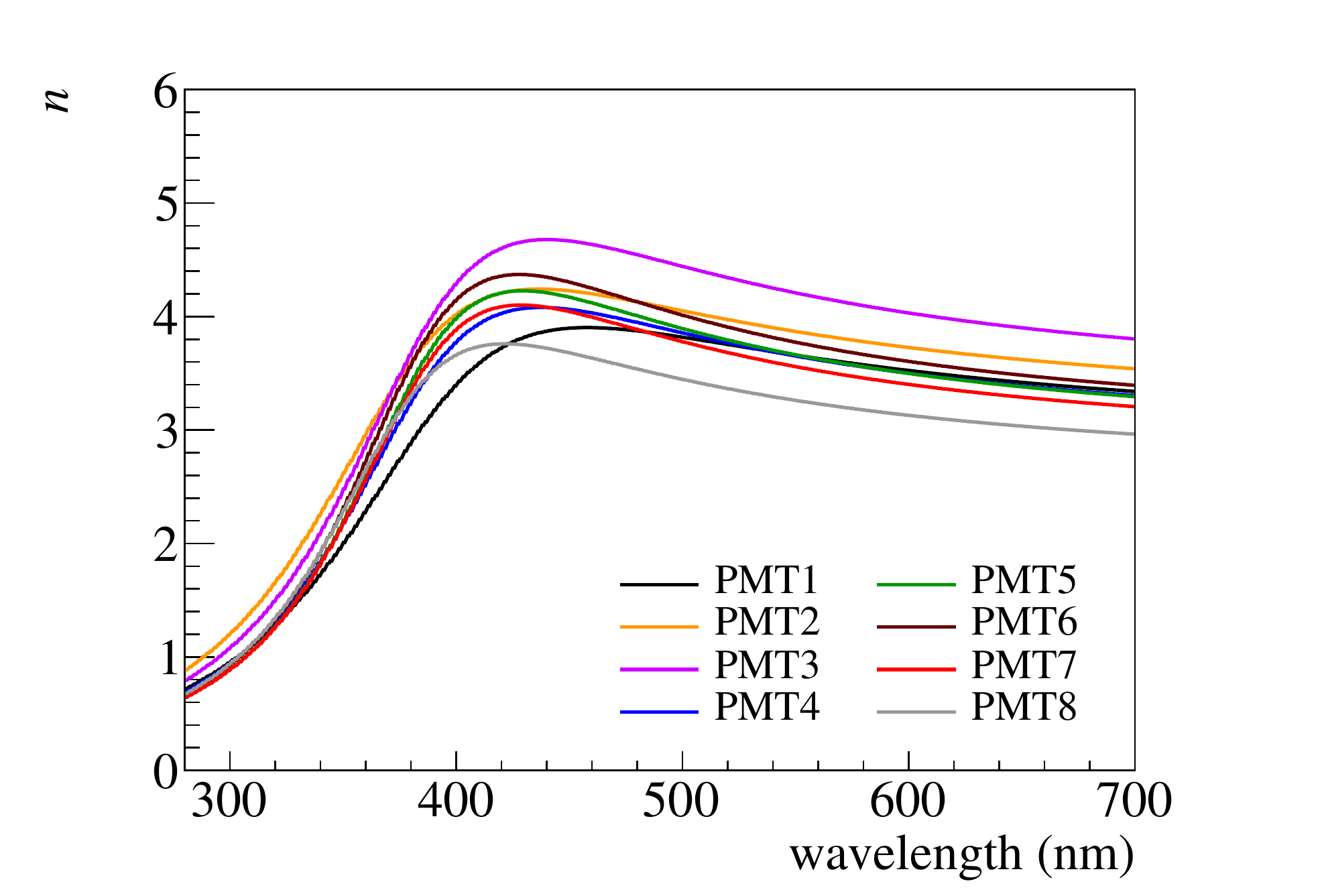}}
 \hspace{-0.6cm}
 \subfloat{\includegraphics[width=0.55\textwidth,keepaspectratio]{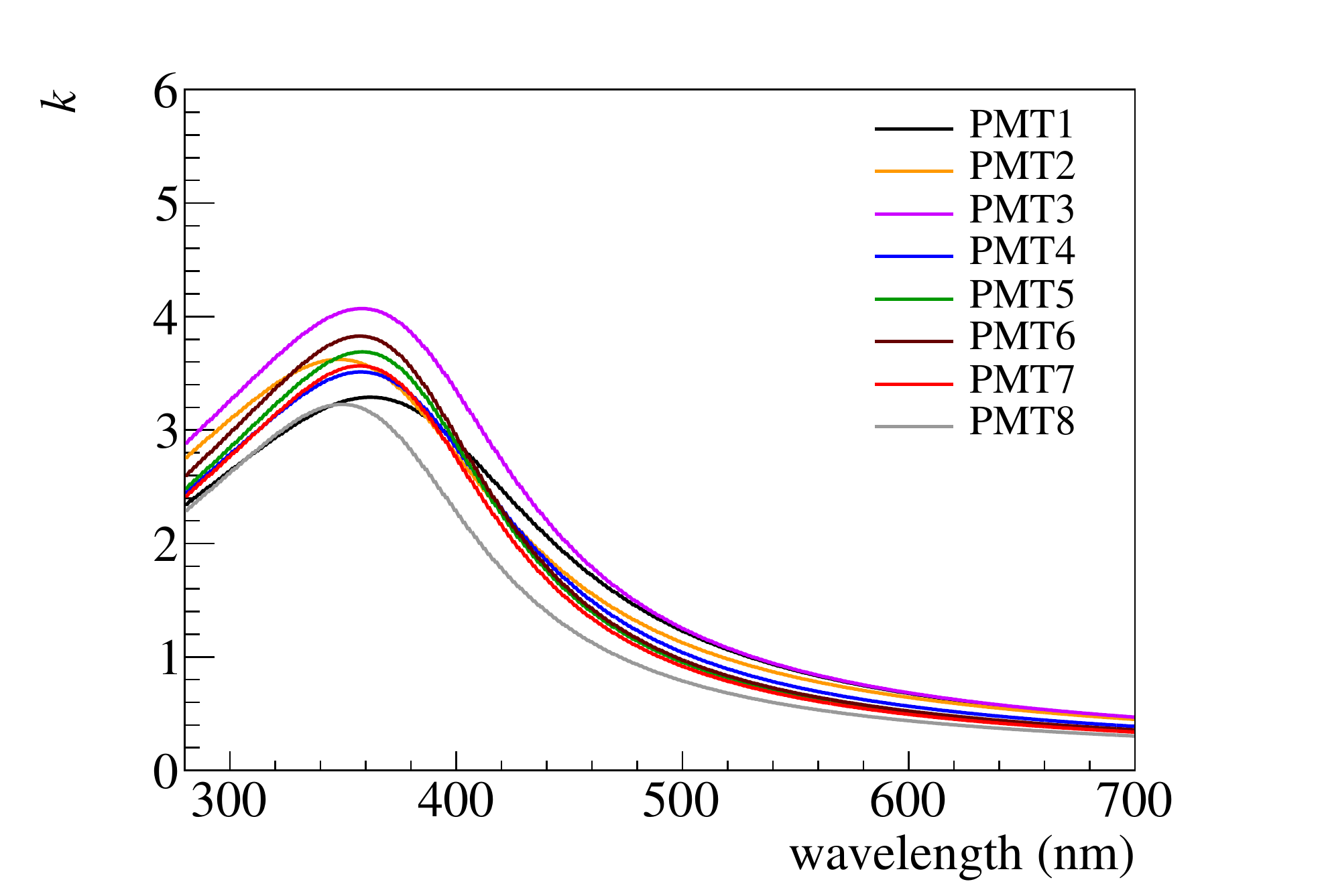}}
}
\caption{Real (left) and imaginary (right) parts of the refractive index of the photocathode.
Only the best fit dispersions are plotted in this figure.}
\label{fig:index_pc_all}
\end{figure}

Figure~\ref{fig:index_pc} shows the refractive index ($n+ik$) of PMT5 photocathode derived from Eqs.~(\ref{eq:epsilon}) and (\ref{eq:index}) using the best fit parameters in Table~\ref{tab:parameter}.
The peak of $n$ and $k$ is around 420~nm and 360~nm, respectively.
The spectrum of $k$ is similar to the one of the QE shown in Fig.~\ref{fig:QE_spectrum} as $k$ relates to the absorption.
For comparison Fig.~\ref{fig:index_pc} also includes previously published data for multi-alkali photocathodes composed of the same elements (NaKSbCs) and windows of two different materials: Jones~\cite{index_Jones} and Chyba~\cite{index_Chyba} for borosilicate windows; Ghosh~\cite{photoemit_multialkali}, Hallensleben~\cite{index_Hallensleben} and Harmer~\cite{index_Harmer} for fused silica windows. 
The index measured in this work differs at the shorter wavelengths from Ghosh, Hallensleben and Harmer.
Especially the discrepancy in $k$ with Harmer is quite large, but $k\sim0$ below 350~nm or insensitivity to near-ultraviolet light is rather strange for the multi-alkali photocathode.
The dispersion curves for the other PMTs are plotted in Fig.~\ref{fig:index_pc_all}.
They are consistent within the errors since there are no significant differences in the optical parameters among the PMTs.

\begin{figure}[t]
 \centerline{
  \hspace{0.6cm}
  \subfloat{\includegraphics[width=0.55\textwidth,keepaspectratio]{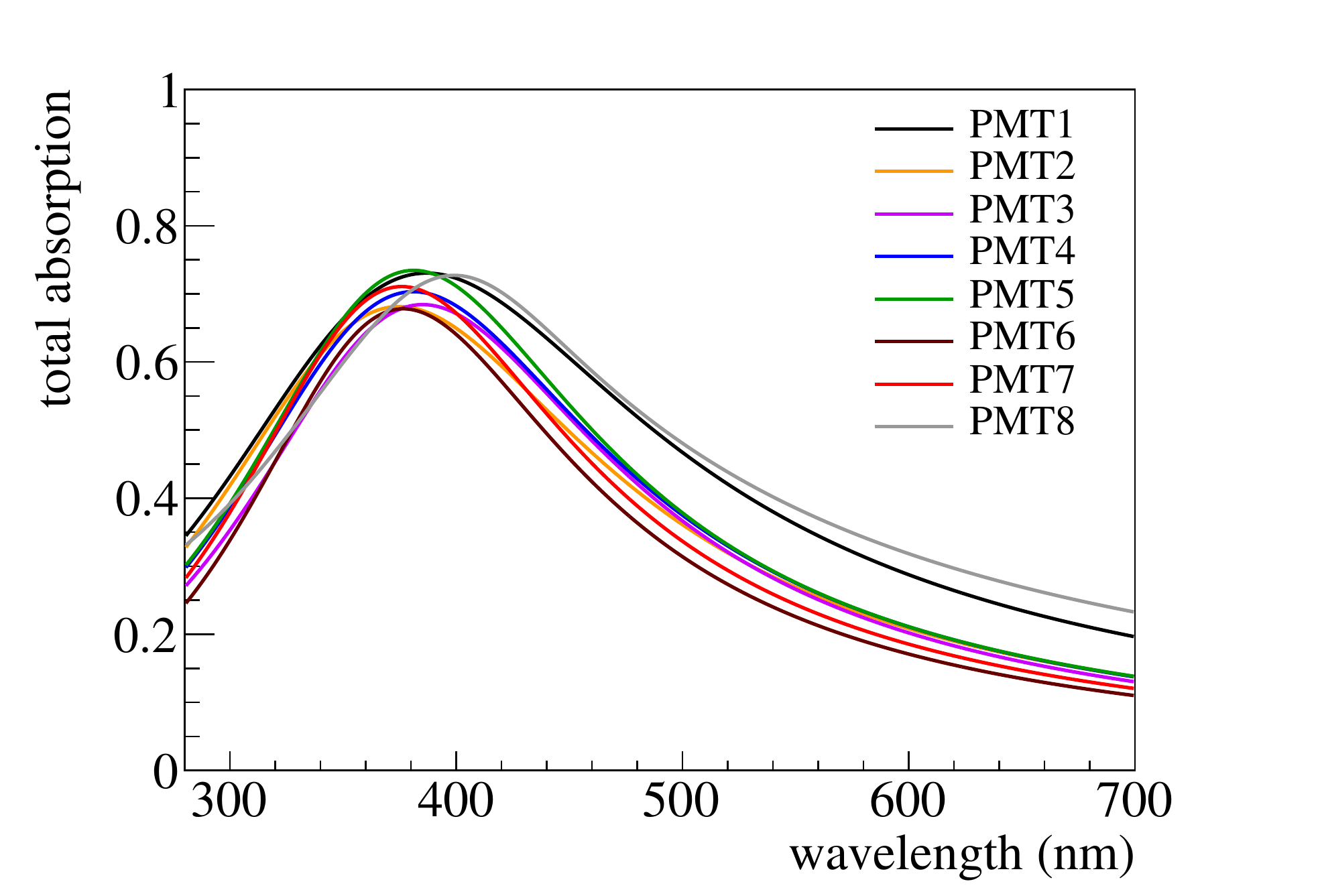}}
  \hspace{-0.6cm}
  \subfloat{\includegraphics[width=0.55\textwidth,keepaspectratio]{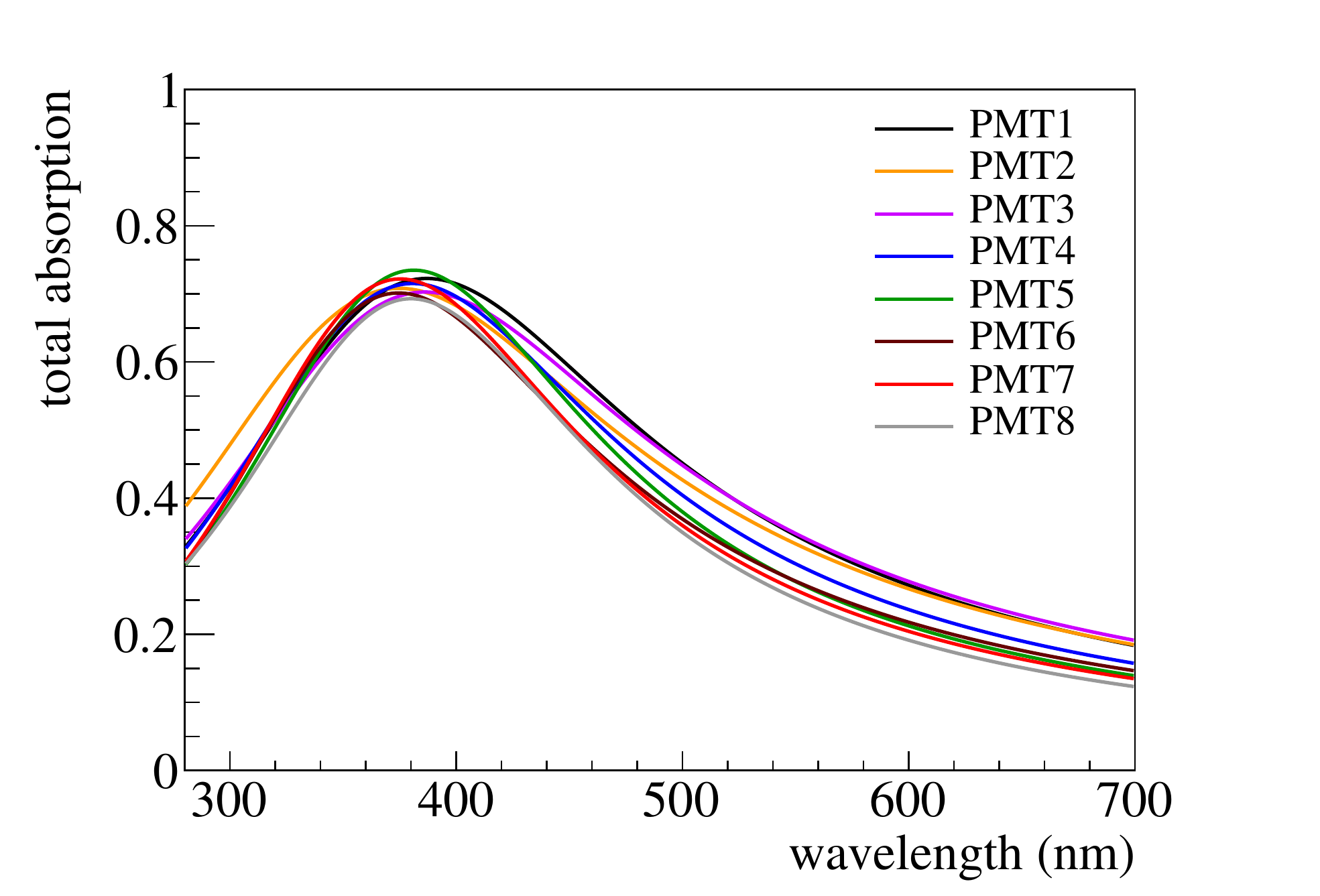}}
 }
 \caption{Total absorption of the photocathode at 0$^\circ$ including the back-illumination.
 (Left) Calculated with the best fit parameters in Table~\ref{tab:parameter}.
 (Right) Calculated with the same parameters other than the photocathode thickness, which is fixed to 10.0~nm for all the photocathodes.}
 \label{fig:absorption}
\end{figure}

\begin{figure}[t]
\centerline{
 \hspace{0.6cm}
 \subfloat{\includegraphics[width=0.55\textwidth,keepaspectratio]{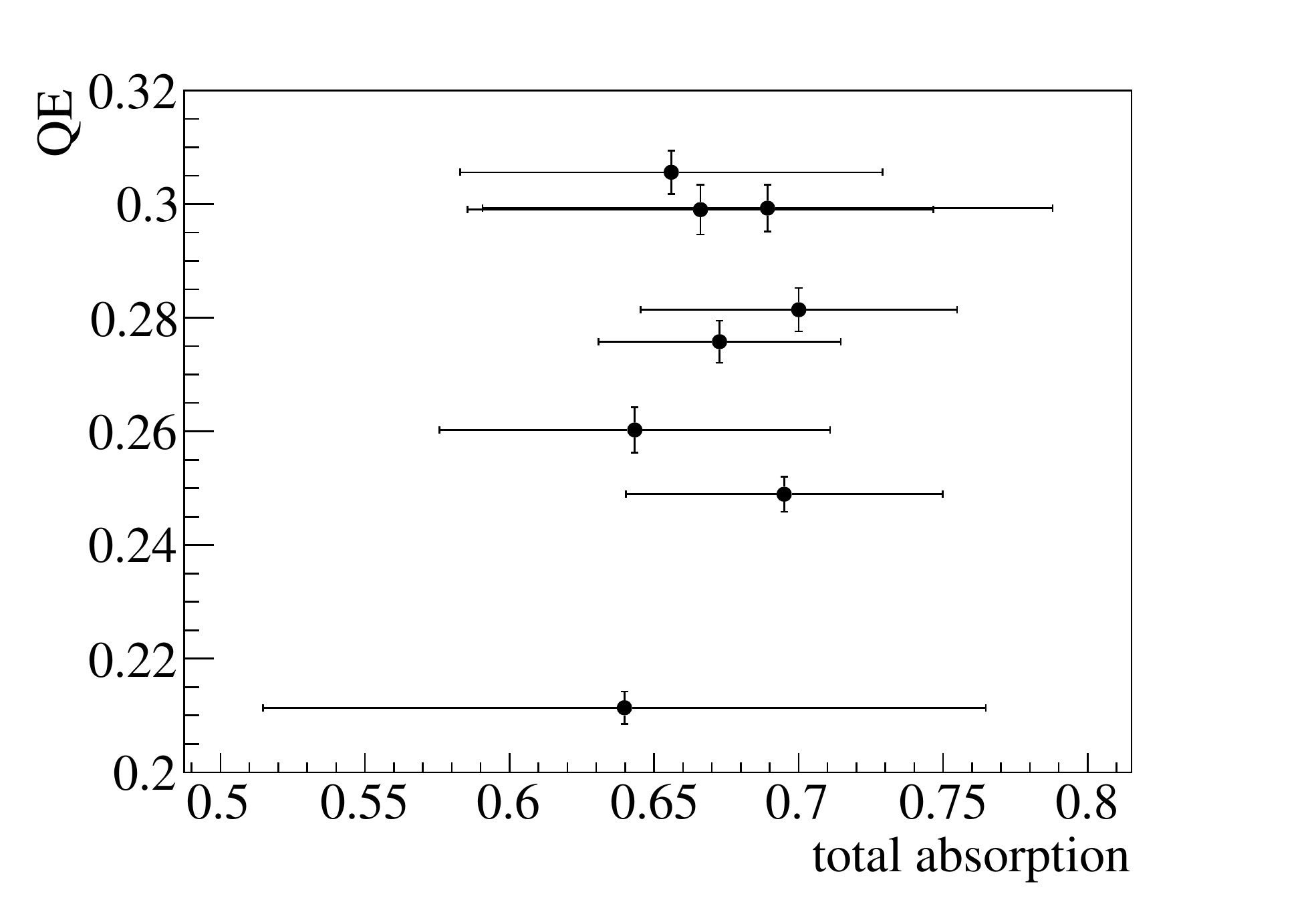}}
 \hspace{-0.6cm}
 \subfloat{\includegraphics[width=0.55\textwidth,keepaspectratio]{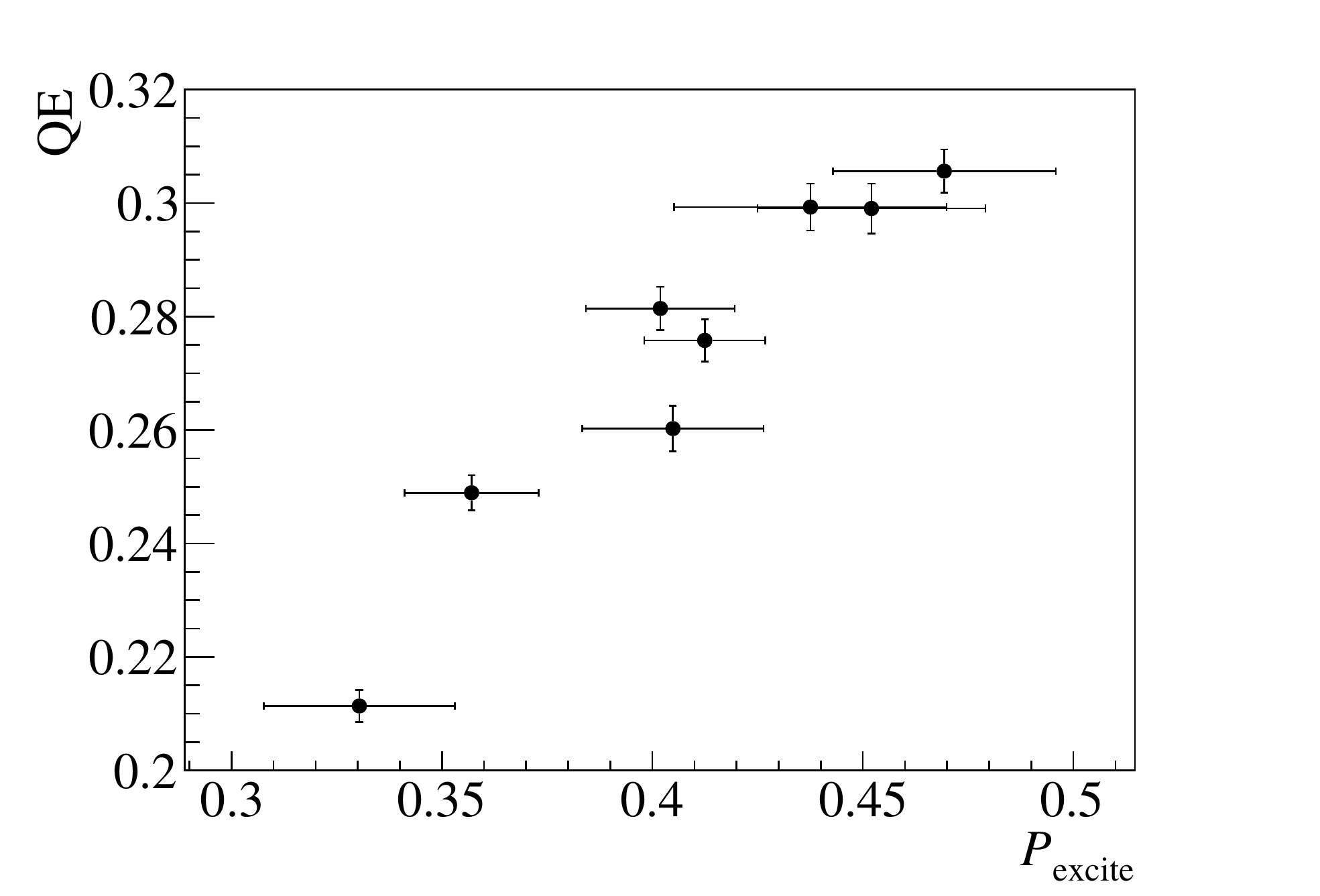}
}}
\caption{Correlation of the QE at 0$^\circ$ with the total absorption (left) and with $P_{\rm excite}$ (right) at 360~nm.
$P_{\rm excite}$ for PMT8 is derived from the QE divided by the total absorption.}
\label{fig:QE_correlation}
\end{figure}

The total absorptions of the photocathodes at 0$^\circ$ calculated using Eq.~(\ref{eq:total_absorption}) are shown in Fig.~\ref{fig:absorption} (left).
They are nearly the same at wavelengths below 400~nm.
The difference above 400~nm comes mainly from the difference of the photocathode thickness.
Given the same thickness of 10.0~nm for all the photocathodes, the absorption spectrum is similar to each other as shown in Fig.~\ref{fig:absorption} (right).
Compared to the QE spectrum in Fig.~\ref{fig:QE_spectrum}, therefore, the variation of the QE can be attributed not to the optical property of the photocathode other than the thickness but to $P_{\rm excite}$.
It can be clearly seen in Fig.~\ref{fig:QE_correlation}, where the QE at 0$^\circ$ is plotted relative to the total absorption or $P_{\rm excite}$ at 360~nm for the eight PMTs.
The QE has a correlation not with the total absorption but with $P_{\rm excite}$, which is subject to the work function and the escape length.

\subsection{Effect of the back-illumination}

\begin{figure}[t]
\begin{minipage}{0.49\textwidth}
 \hspace{-0.3cm}
 \includegraphics[width=1.12\textwidth,keepaspectratio]{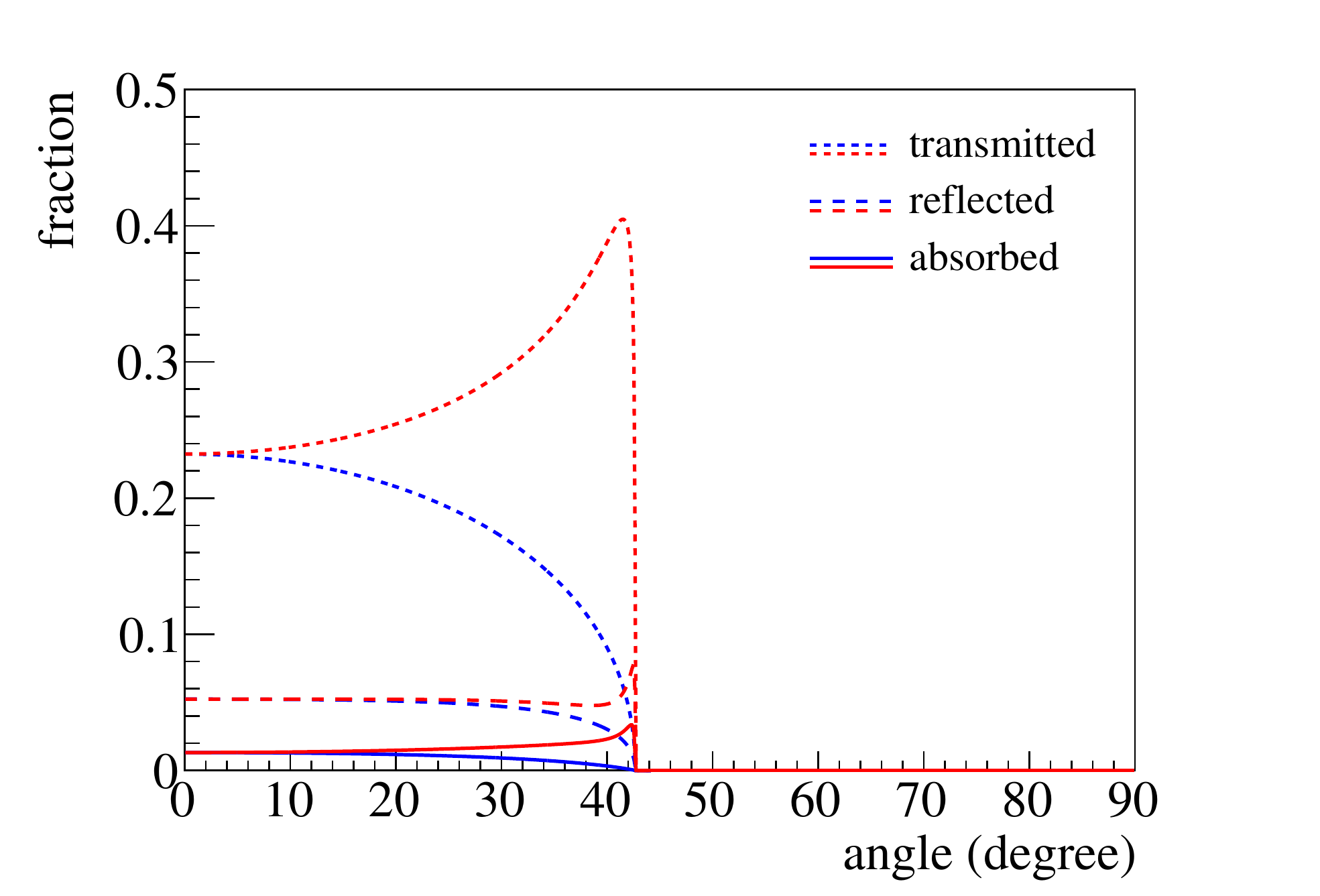}
 \caption{Intensity fraction of 360~nm light transmitted through PMT7 photocathode (dotted line), one of the back-illumination after reflection on the electrode (dashed line) and one of the back-illumination absorbed by the photocathode (solid line) with respect to the initial light.
 The blue (red) lines represent those for $s$($p$)-polarization.}
 \label{fig:fraction_back}
\end{minipage}
\hspace{0.02\textwidth}
\begin{minipage}{0.49\textwidth}
 \hspace{-0.3cm}
 \includegraphics[width=1.12\textwidth,keepaspectratio]{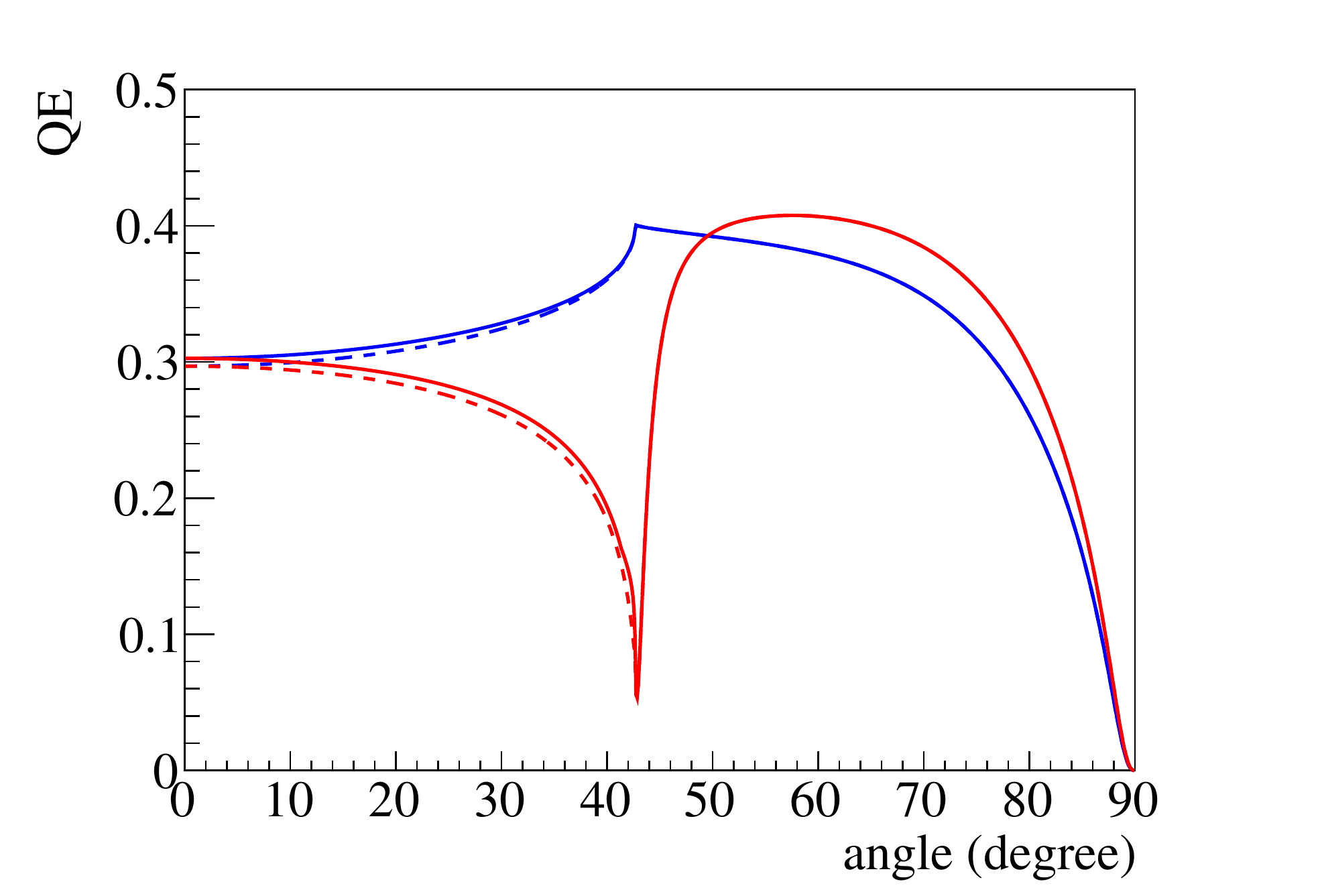}
 \caption{QE at 360~nm as a function of the incident angle calculated with and without the back-illumination (solid and dashed lines, respectively) for PMT7.
 The blue (red) lines represent those for $s$($p$)-polarization.
 \vspace{1.57cm}}
 \label{fig:QE_back}
\end{minipage}
\end{figure}

The QE was measured as a sum of the front- and back-illuminations.
As denoted in Eq.~(\ref{eq:total_absorption}) the back-illumination is reduced by the factors $T_{\rm 2+3}$, $1-\mathcal{R}_{\rm OA}$ and $R_{\rm el}$, and it should have less contribution to the QE.
Figure~\ref{fig:fraction_back} shows $T_{\rm 2+3}$, $T_{\rm 2+3}(1-\mathcal{R}_{\rm OA})R_{\rm el}$ and $T_{\rm 2+3}(1-\mathcal{R}_{\rm OA})R_{\rm el}A_{\bar{2}}$ at 360~nm as a function of the incident angle $\theta_1$.
$T_{\rm 2+3}$ is 0.232 at 0$^\circ$ and increases as the angle for $p$-polarization, while it decreases for $s$-polarization.
Hence the contribution of the back-illumination is larger for $p$-polarization.
The following discussion takes the case of the normal incidence at 360~nm.
The factor $(1-\mathcal{R}_{\rm OA})R_{\rm el}$ reduces the fraction of the back-illumination down to 0.052.
In addition the reflection at the vacuum-photocathode interface for the back-illumination is higher than the one at the quartz-photocathode interface for the front-illumination due to a larger mismatch of the refractive indices.
For example, the reflectance of the photocathode $|\hat{r}|^2$ for the back- and front-illumination is 0.405 and 0.224, respectively, where the antireflection coating is omitted for simplicity and the parameters of PMT7 photocathode are used.
It further reduces the fraction of the absorbed back-illumination down to 0.013, resulting in the QE of 0.006 regarding only the back-illumination.
The effect of the back-illumination is feeble at any angle as shown in Fig.~\ref{fig:QE_back}.

\begin{figure}[t]
\centerline{
 \hspace{0.6cm}
 \subfloat{\includegraphics[width=0.55\textwidth,keepaspectratio]{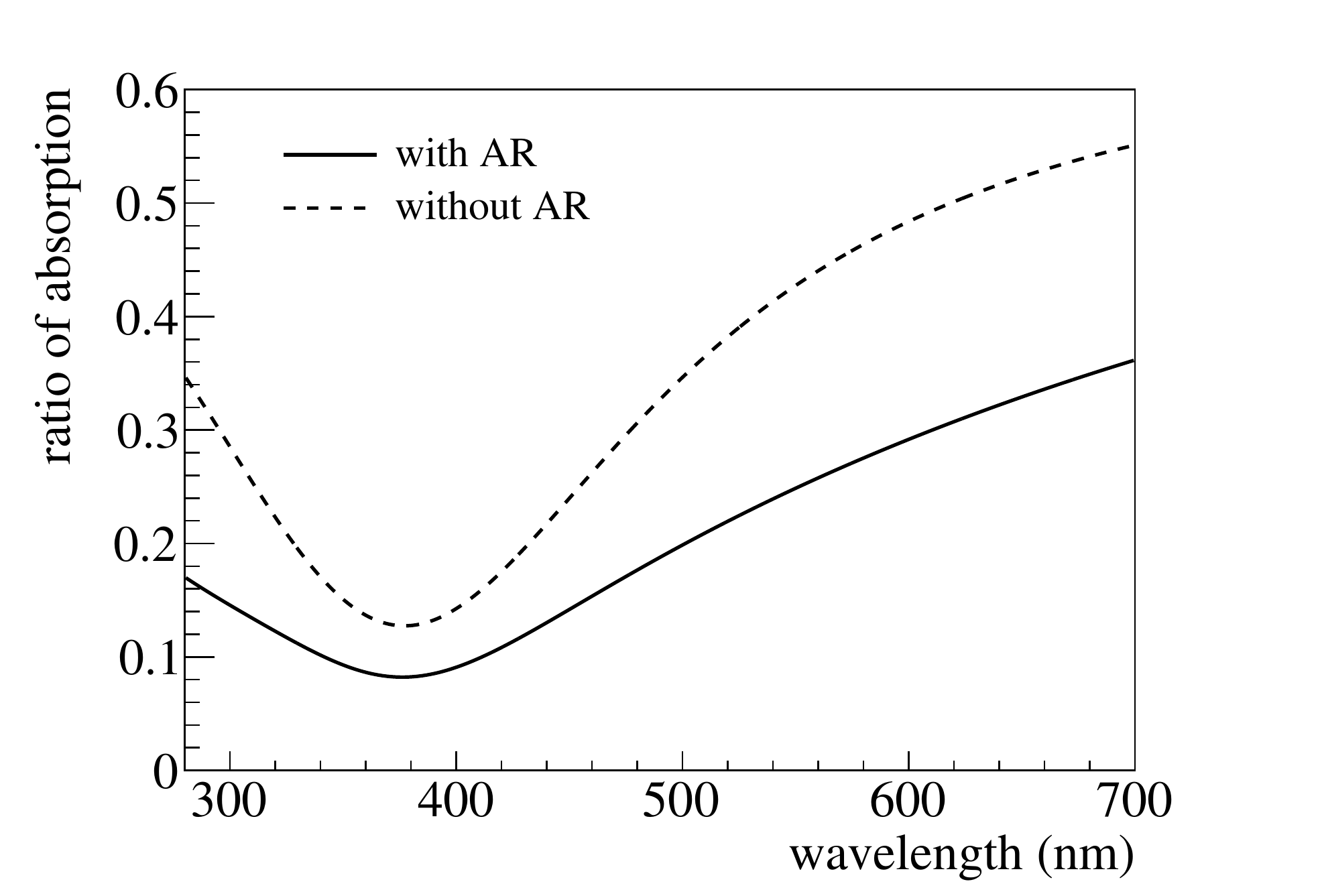}}
 \hspace{-0.6cm}
 \subfloat{\includegraphics[width=0.55\textwidth,keepaspectratio]{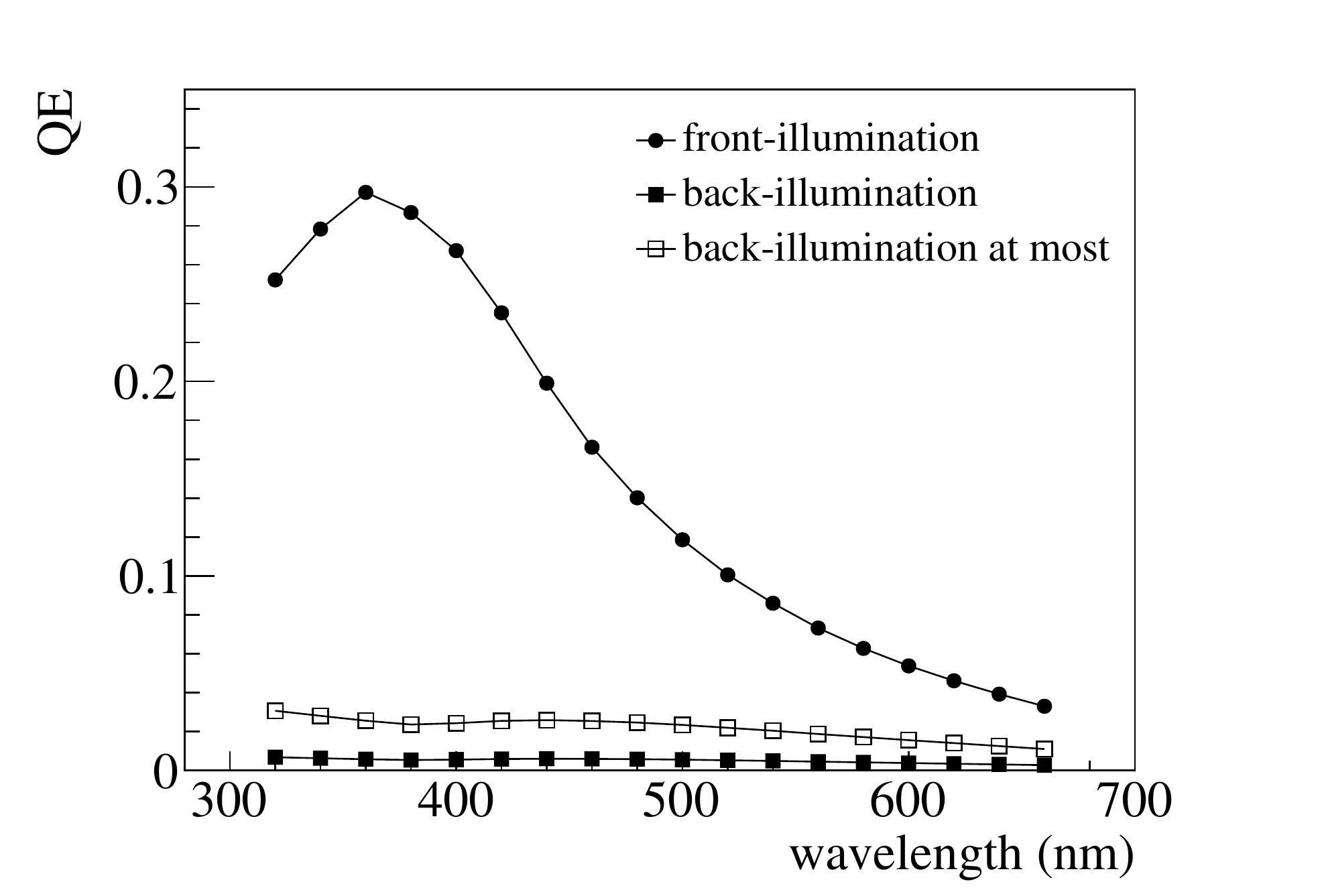}}
}
\caption{(Left) Ratio of the absorbed back- to front-illumination at 0$^\circ$ when  $(1-\mathcal{R}_{\rm OA})R_{\rm el}$ is assumed to be unity.
It is calculated for PMT7 photocathode with and without the antireflection coating (AR).
(Right) Respective contributions of the front- and back-illuminations to the QE at 0$^\circ$ for PMT7 photocathode with the antireflection coating.
Both the actual and the maximum possible back-illumination are plotted.}
\label{fig:QE_back_max}
\end{figure}

Even if a more reflective material is put behind the photocathode to fully reflect the transmitted light, the enhancement of the QE cannot be significant.
The maximum possible contribution of the back-illumination at 0$^\circ$, namely under assumption of $(1-\mathcal{R}_{\rm OA})R_{\rm el}=1$, is shown in Fig.~\ref{fig:QE_back_max}.
The factor of enhancement by the full back-illumination is only 1.09 at 360~nm for this photocathode with the antireflection coating.
It is inversely related to the QE for the front-illumination since the transmittance of the photocathode decreases as the absorptance increases.
Therefore the fraction of the back-illumination becomes larger away from the wavelength of the peak QE, but the net QE increase by the full back-illumination is nearly constant at about 0.01-0.03 regardless of the wavelength.
In the same manner the ratio of the back-illumination becomes larger without the antireflection coating.
The increase of the QE by the antireflection coating is, however, larger than the one by the back-illumination at the peak wavelength, and removing the antireflection coating to increase the back-illumination adversely decreases the QE.

Another attempt is to put the second photocathode of the reflection mode which detects transmitted light through the first photocathode~\cite{JUNO_MCP-PMT}.
For the same reason mentioned above, however, one can expect a little improvement of the QE, while the dark noise rate increases in proportion to the photocathode area.

Since the contribution of the back-illumination is little for the PMTs in this work, only the front-illumination is considered in the following discussions.

\subsection{Effect of the antireflection coating}
\begin{figure}[tb]
\centerline{
 \hspace{0.6cm}
 \subfloat{\includegraphics[width=0.55\textwidth,keepaspectratio]{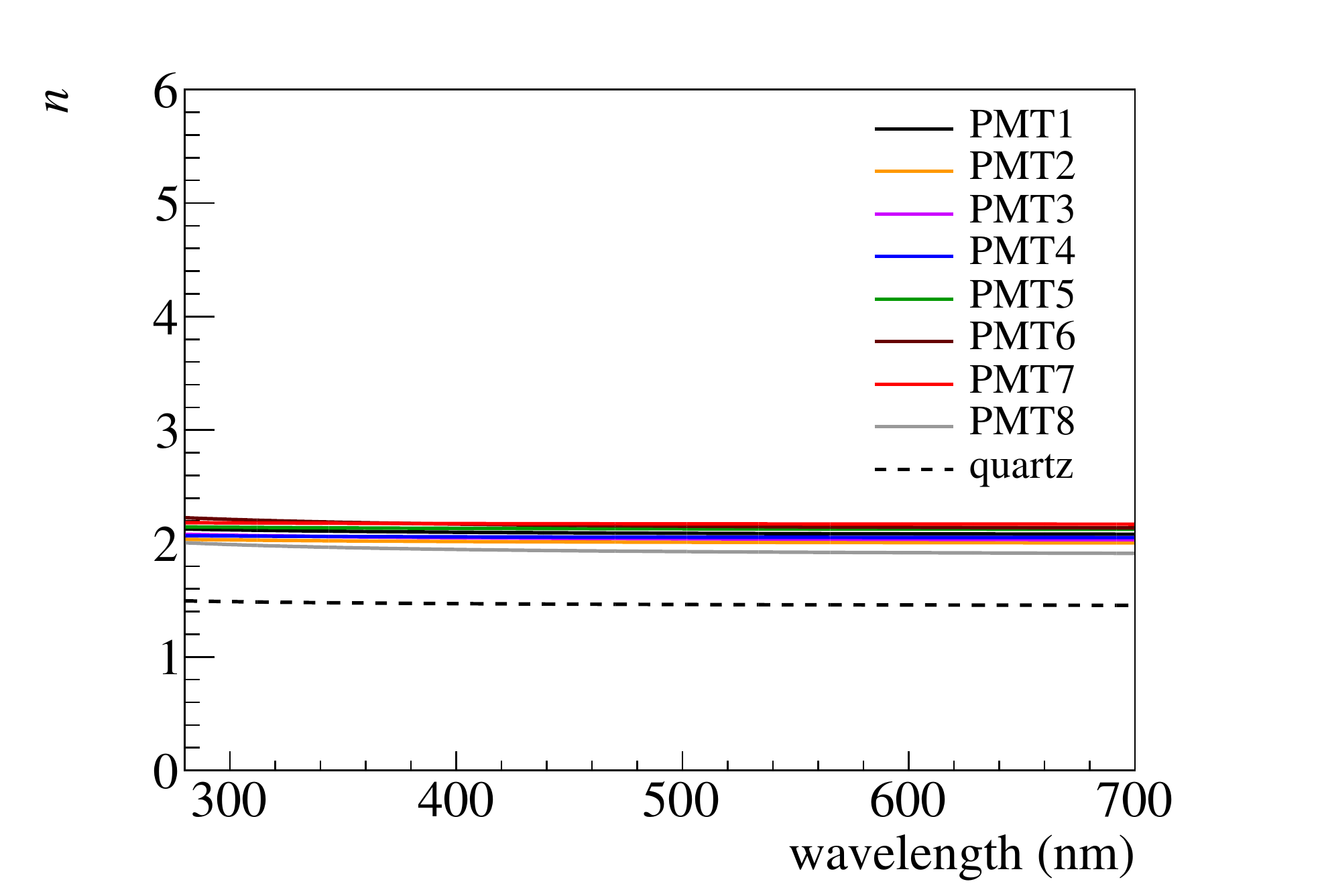}}
 \hspace{-0.6cm}
 \subfloat{\includegraphics[width=0.55\textwidth,keepaspectratio]{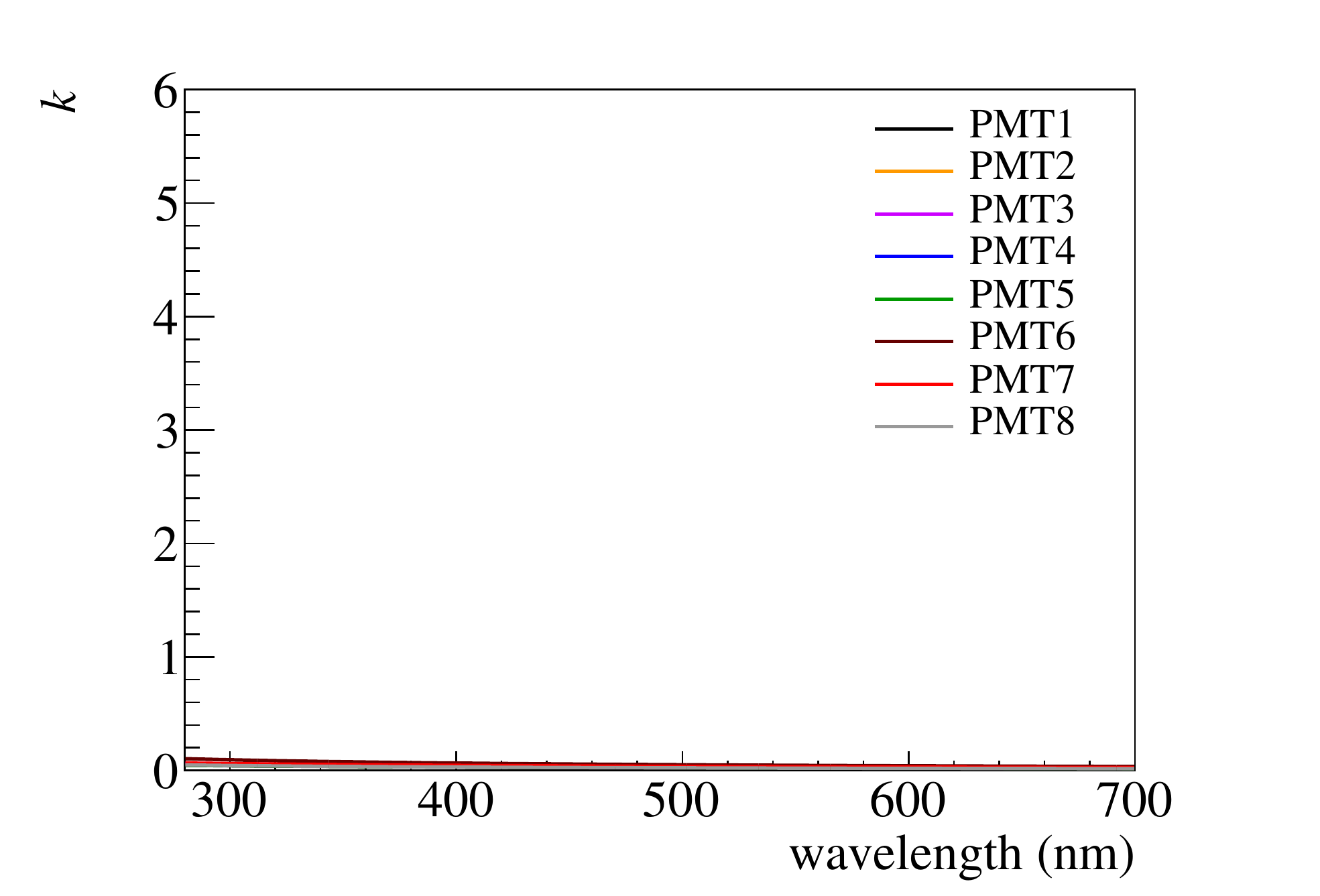}}
}
\caption{Real (left) and imaginary (right) parts of the refractive index of the antireflection coating.
The one of quartz is also shown for comparison.}
\label{fig:index_ar}
\end{figure}

\begin{figure}[tb]
\centerline{
 \hspace{0.6cm}
 \subfloat{\includegraphics[width=0.55\textwidth,keepaspectratio]{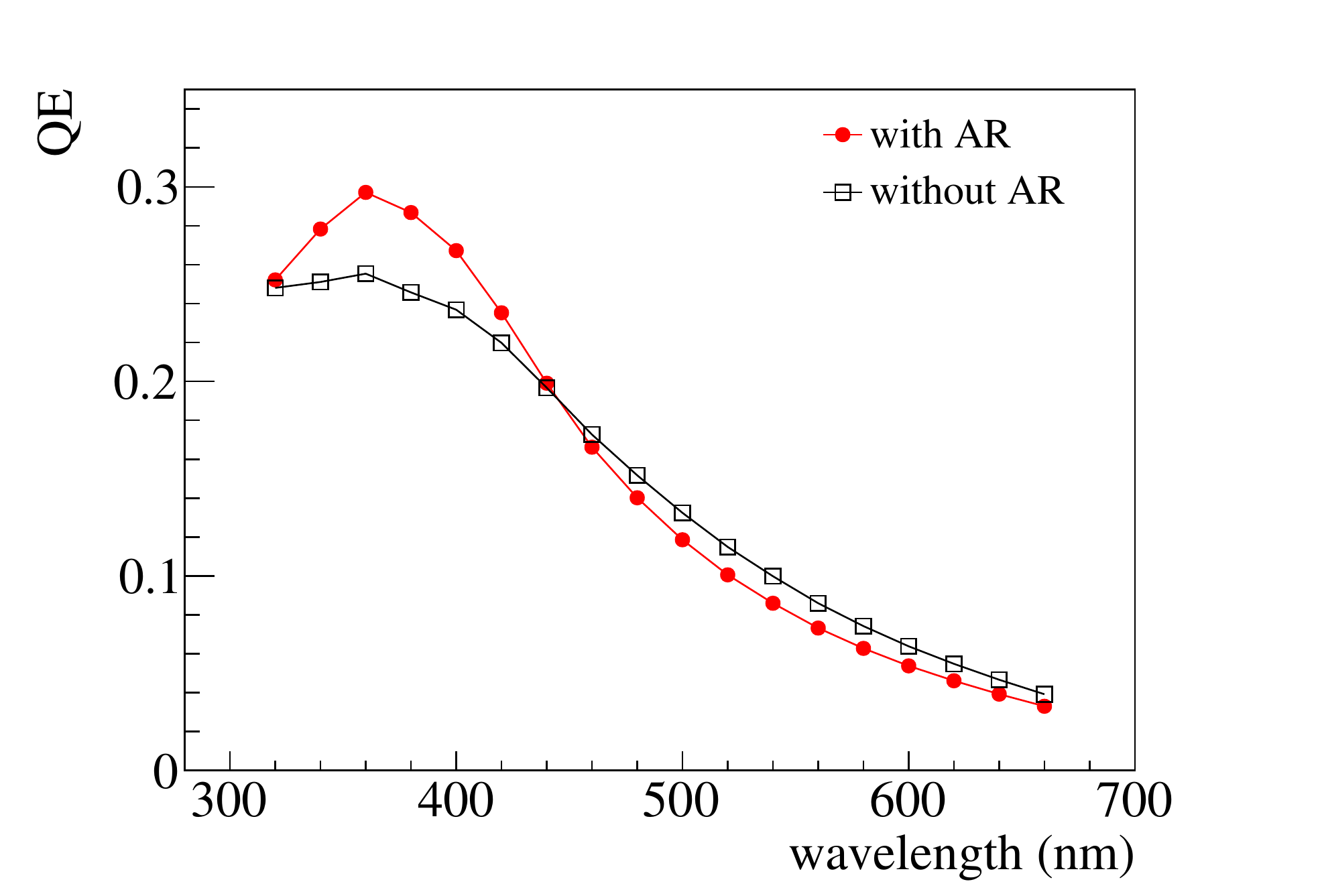}}
 \hspace{-0.6cm}
 \subfloat{\includegraphics[width=0.55\textwidth,keepaspectratio]{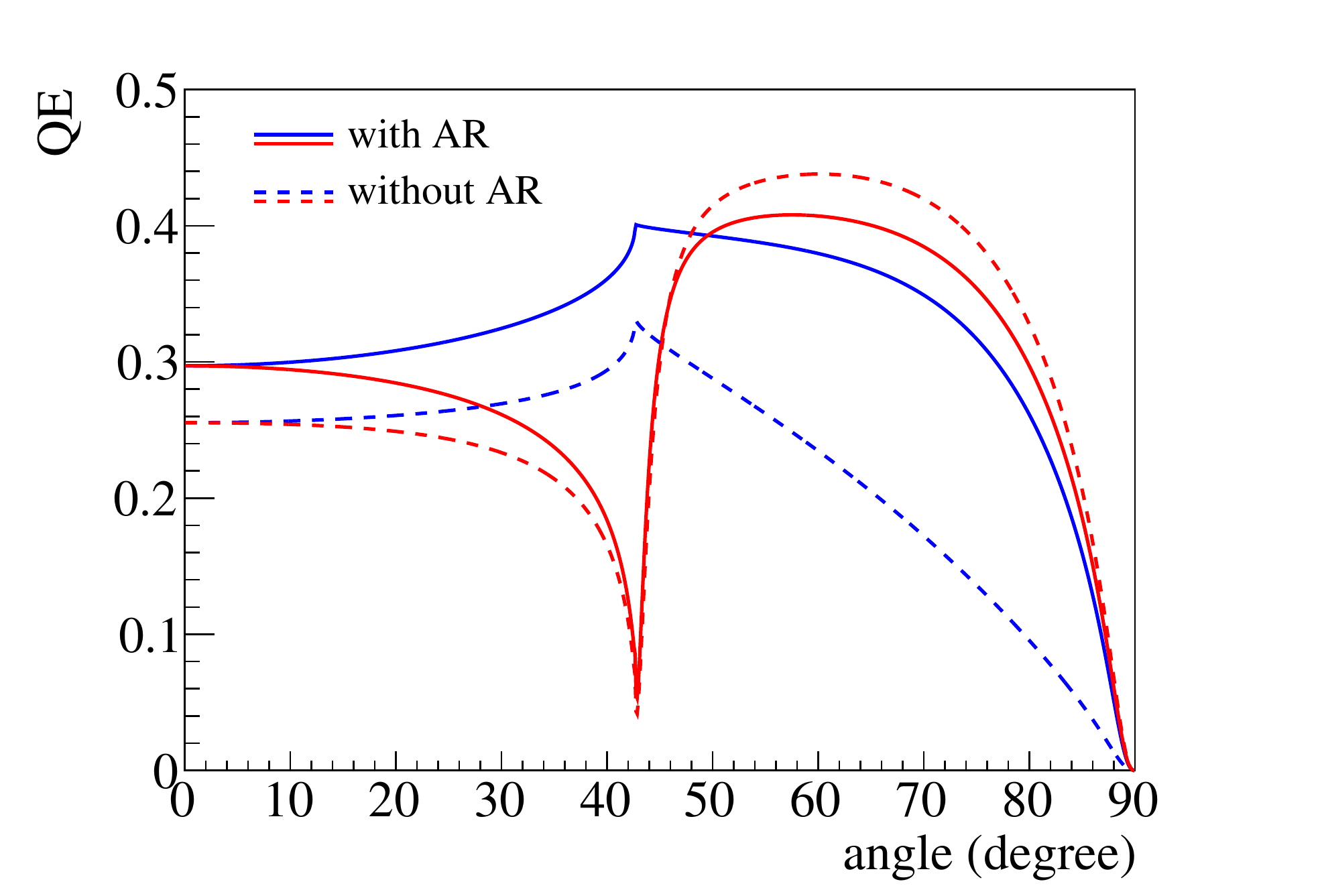}}
}
 \caption{Comparisons of the QE for PMT7 calculated with and without the antireflection coating (AR).
 Only the front-illumination is taken into account.
 (Left) QE at 0$^\circ$ as a function of the wavelength.
 (Right) QE at 360~nm as a function of the incident angle.
 The blue (red) lines represent the QE for $s$($p$)-polarization.}
 \label{fig:QE_ar}
\end{figure}

Figure~\ref{fig:index_ar} shows the refractive index ($n+ik$) of the antireflection coating derived from Eqs.~(\ref{eq:epsilon}) and (\ref{eq:index}) using the best fit parameters in Table~\ref{tab:parameter}.
The dispersion in this range of wavelength is quite small; $n$ is about 2.1 and $k$ is nearly 0.
The thickness of the antireflection coating ranges from 30.5 to 48.3~nm.
These values of the index and the thickness are adequate for destructive interference of the reflected light waves to reduce the reflection.
The effect of the antireflection coating on the QE is manifest as shown in Fig.~\ref{fig:QE_ar}.
The factor of enhancement of the QE is 1.16 at 360~nm and 0$^\circ$.
More significant enhancement can be seen at large incident angles for $s$-polarization.
On the other hand the QE at 460~nm or longer is reduced by the antireflection coating due to constructive interference of the reflected light waves to increase the reflection.
Although the antireflection coating has a nonzero value of $k$ and absorbs the light, it is not the main cause of the reduction. 
The absorptance of the antireflection coating is, for example in the case of PMT7, 0.074 at 360~nm and almost constant at 0.016 above 500~nm. 

\begin{figure}[tb]
\begin{minipage}{0.49\textwidth}
 \hspace{-0.3cm}
 \includegraphics[width=1.12\textwidth,keepaspectratio]{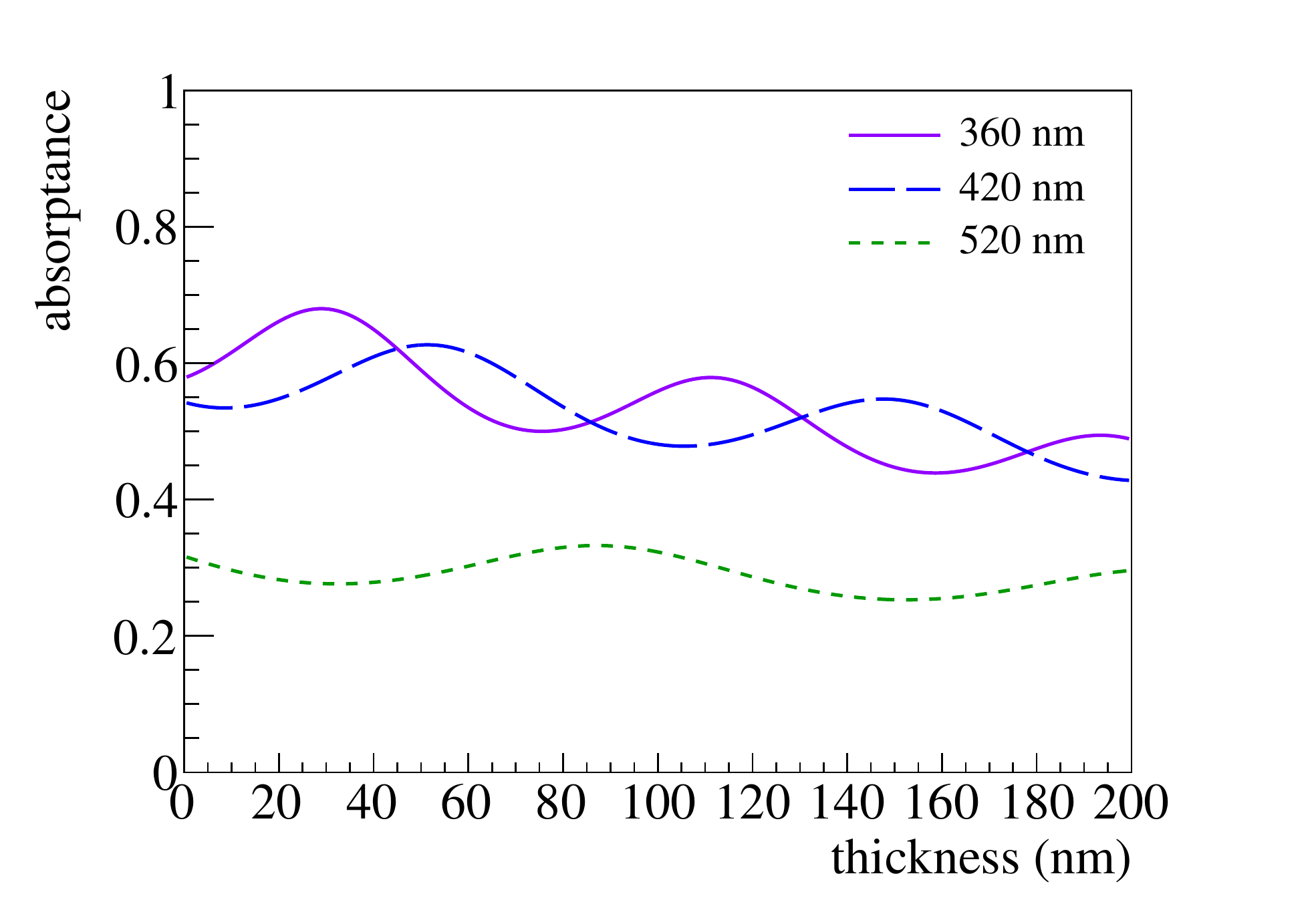}
 \caption{Absorptance of PMT7 photocathode at 0$^\circ$ as a function of the thickness of the antireflection coating.
 There are three curves for 360, 420 and 520~nm wavelengths.
 Only the front-illumination is taken into account.}
 \label{fig:absorptance_ar}
\end{minipage}
\hspace{0.02\textwidth}
\begin{minipage}{0.49\textwidth}
 \hspace{-0.3cm}
 \centering
 \includegraphics[width=1.12\textwidth,keepaspectratio]{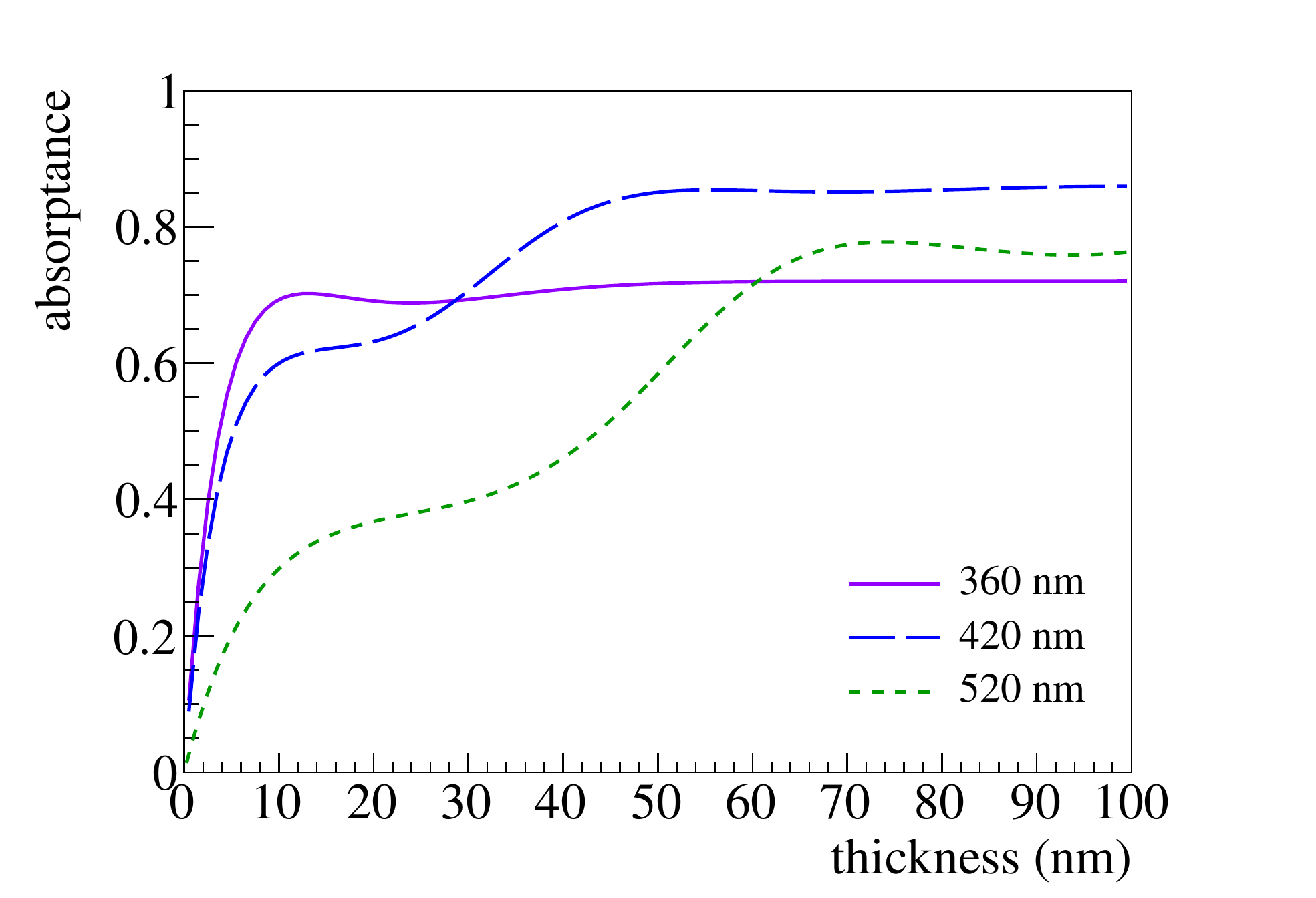}
 \caption{Absorptance of PMT7 photocathode at 0$^\circ$ as a function of the photocathode thickness.
 There are three curves for 360, 420 and 520~nm wavelengths.
 Only the front-illumination is taken into account.}
 \label{fig:absorptance_thickness}
\end{minipage}
\end{figure}

Adjustment of the thickness of the antireflection coating could help to enhance the QE at a desired range of the wavelength.
Given the refractive indices of the window, antireflection coating and photocathode, the best matching thickness of the antireflection coating can be estimated.
Figure~\ref{fig:absorptance_ar} shows the dependence of the absorptance of the photocathode on the thickness of the antireflection coating, where the other parameters than the thickness are the same as PMT7.
The absorptance is maximized at the thickness of 28.9~nm, 51.4~nm and 87.0~nm for the wavelength of 360~nm, 420~nm and 520~nm, respectively.
By adjusting the thickness to these values, the absorptance increases by 0.3\%, 7.6\% and 20.3\%, respectively, from the one at the actual thickness of 31.7~nm.

\subsection{Photon absorption in depth}
Another parameter which could be tuned to enhance the QE is the thickness of the photocathode.
In terms of the absorptance a thicker photocathode is better especially for longer wavelengths.
It is also better in terms of the QE under the condition of the one-step model, where the QE is simply proportional to the absorptance.
Nevertheless for a thick photocathode comparable to the electron escape length, one has to take into account the scattering loss.
The optimal thickness can be foreseen in Fig.~\ref{fig:absorptance_thickness}, where the absorptance is plotted as a function of the photocathode thickness by using Eq.~(\ref{eq:absorptance}).
The absorptance at 10~nm thickness is close to its maximum for 360~nm wavelength, and a thicker photocathode could result in a less QE as only the scattering loss increases.
Therefore the thickness around 10~nm is optimal for the QE at 360~nm.
To enhance the QE at longer wavelengths, for example up to 520~nm, the optimal thickness should be around 70~nm or less.

\begin{figure}[t]
\centerline{
 \hspace{0.6cm}
 \subfloat{\includegraphics[width=0.55\textwidth,keepaspectratio]{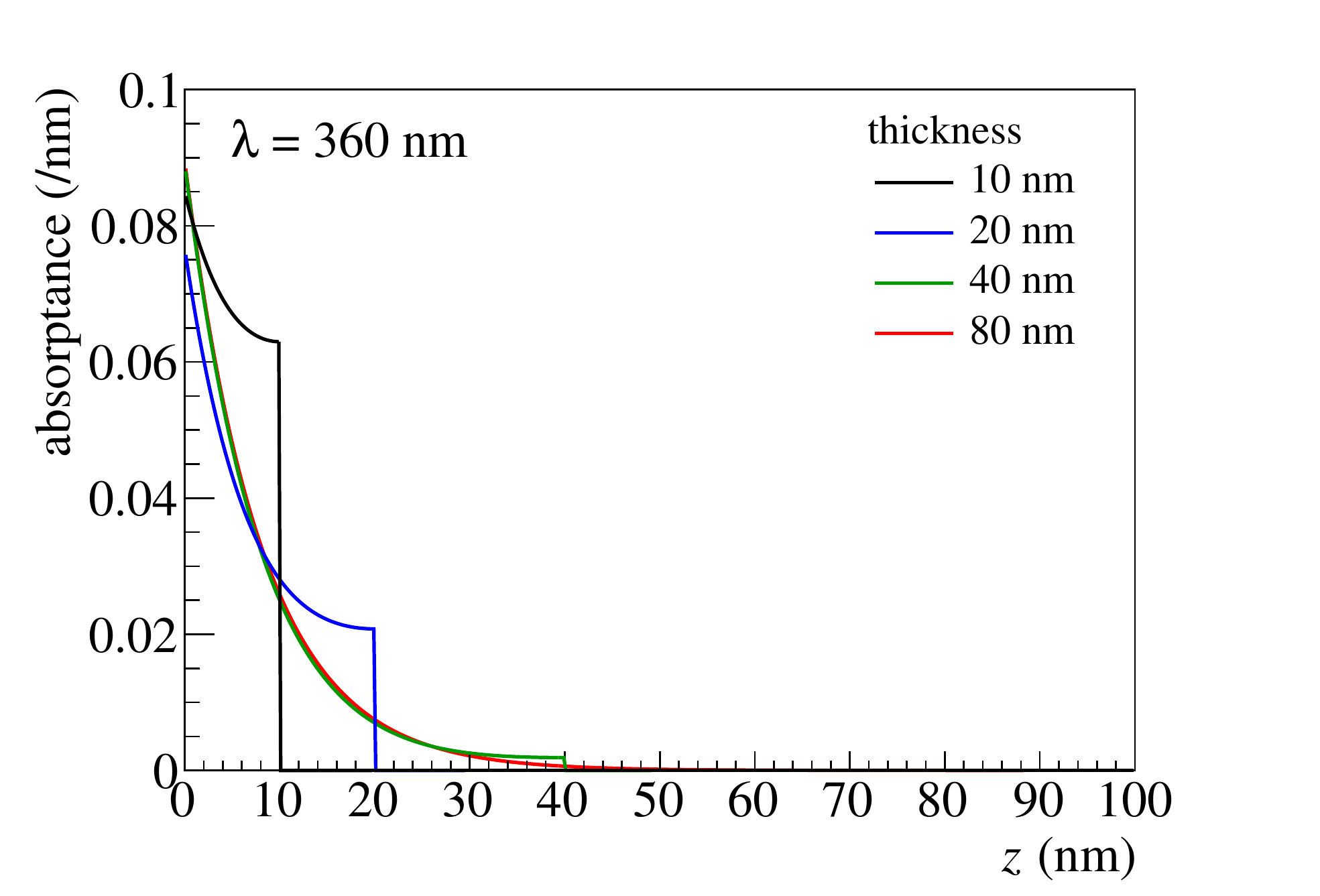}}
 \hspace{-0.6cm}
 \subfloat{\includegraphics[width=0.55\textwidth,keepaspectratio]{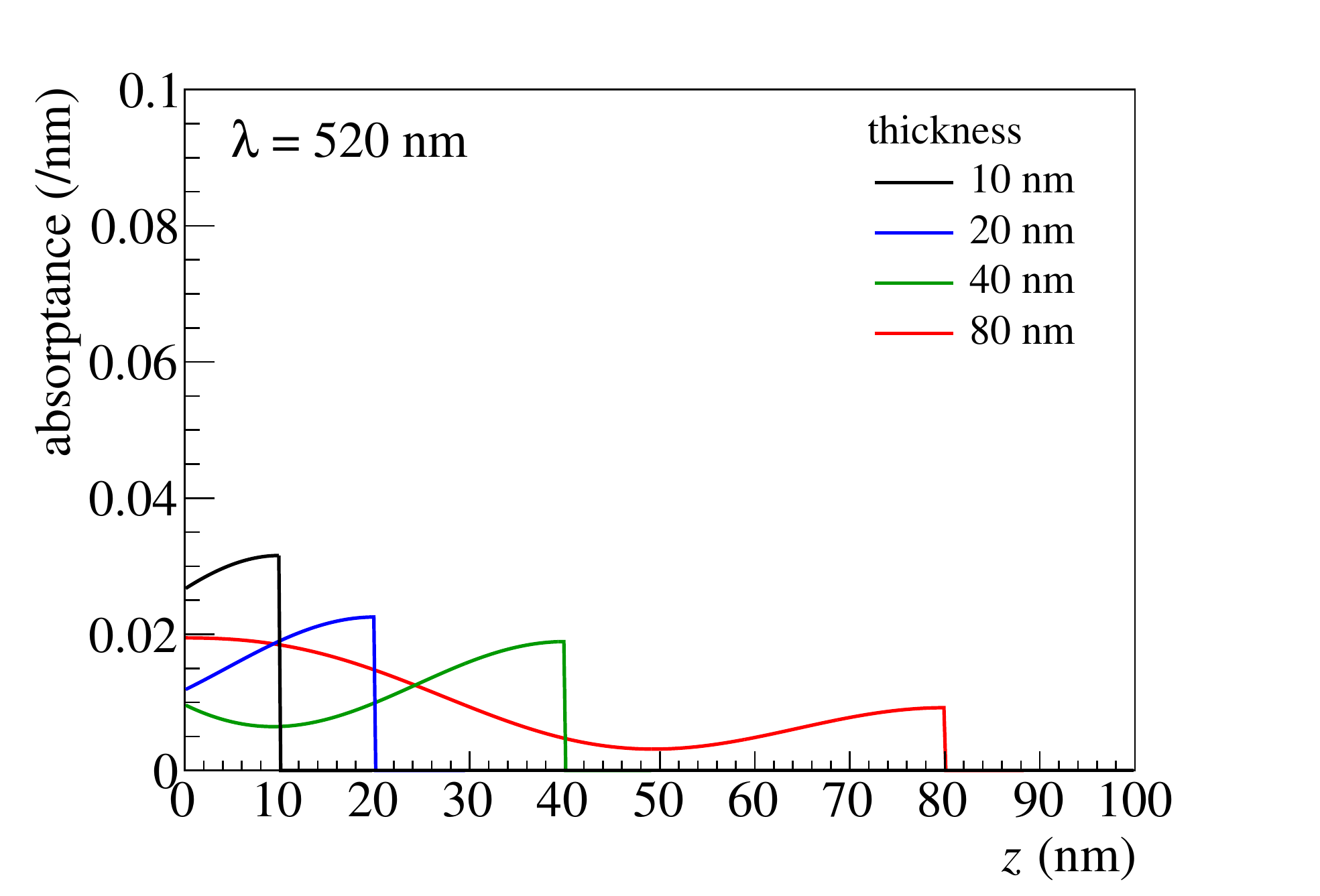}}
}
 \caption{Absorptance per unit length at 360~nm (left) and 520~nm (right) wavelengths as a function of the depth from the photocathode surface.
 There are four curves for different photocathode thicknesses, 10, 20, 40 and 80~nm.
 The refractive indices of the photocathode and antireflection coating for PMT7 are used in these plots.
  Only the front-illumination is taken into account.}
 \label{fig:absorptance_z}
\end{figure}

The absorptance per unit length expressed by Eq.~(\ref{eq:absorptance_z}) is shown in Fig.~\ref{fig:absorptance_z}.
It is different from the often-used Lambert-Beer law, $e^{-\alpha z}$, described by an absorption coefficient $\alpha$.
That is due to the light waves reflected at the interfaces and their interference.
Only when the photocathode is enough thick, the absorptance can be approximated by $e^{-\alpha z}$ as the light wave heading forward is fully absorbed before reflected.
An example can be found in Fig.~\ref{fig:absorptance_z} for 80~nm thickness and 360~nm wavelength.
The assumption of an infinite thickness, which is usually taken, is therefore not appropriate for thin photocathodes.

\subsection{Prospect for enhancement of the QE}
A straightforward way to enhance the QE is to put the photocathode aslant so that the light strikes at 50-80$^\circ$, where the QE is higher than the one at 0$^\circ$ as shown in Fig.~\ref{fig:QE_angle}.
Because it is discussed elsewhere~\cite{QE_angle1,QE_angle2}, other possibilities are mentioned here in terms of enhancement of the intrinsic QE for the same type of photocathodes.
As discussed above there is a little room for improvement of the peak QE by adjusting the thicknesses of the antireflection coating and the photocathode.
The strong correlation of the QE with $P_{\rm excite}$ found in Fig.~\ref{fig:QE_correlation} indicates that one of the dominant factors dictating the QE could be the work function.
Because the low work function of the NaKSbCs photocathode is derived from a Cs-Sb dipole monolayer~\cite{S20_Holtom}, matters to the QE could be, for example, the arrangement of surface atoms~\cite{workfunction_Smoluchowski}, the monolayer coverage~\cite{workfunction_Gyftopoulos}, or the surface roughness~\cite{photocathode_roughness}.
Less significant in the QE should be the bulk of the photocathode, which dictates the optical properties, though impurities and defects in the bulk could be another dominant factor of the QE reduction.

\section{Conclusion}
A precise description of the PMT response requires formulation of the dependence of the QE on the angle and polarization of the incident photon, which was yet to be studied.
Hence I proposed the one-step model of photoelectron emission, or photoexcitation of an electron to an energy above a certain threshold for the emission, which is adequate for thin multi-alkali photocathodes and for visible light.
In this model the dependence of the QE on the angle and polarization is fully involved in the absorptance, and thus it can be rigorously described by the optics theory.
The theoretical function of the QE derived in this work accurately fit the measured dependence of the photocurrent on the angle and polarization at all the wavelengths from 320 to 680~nm simultaneously.
That demonstrated the picture of the photoelectron emission rendered by the one-step model.

The measurement furnished the new data of the refractive index of the NaKSbCs photocathode over the whole visible range.
It revealed the optical properties of the photocathode and the following new insights were obtained.
It was demonstrated for the first time that the dispersion of the refractive index of NaKSbCs can be described by the Lorentz model.
The gain in QE by reabsorbing the transmitted light through the photocathode is feeble because of the low transmittance of the efficient photocathode and the large reflection at the vacuum-photocathode interface.
The QE differences among the photocathodes of the same type are attributed dominantly to the thickness of the photocathode and $P_{\rm excite}$.
It implies that a key of the QE enhancement could be the condition of the photocathode surface on the vacuum or the degree of crystalline perfection of the bulk.

In terms of the methodology for deducing the refractive index and thickness of the photocathode, the measurement of the photocurrent has proven to be useful in this work.
It was also demonstrated for the first time that this method can be applied for the stratified antireflection coating and photocathode and that the refractive indices and thicknesses of both layers can be deduced at the same time.

To conclude, the expression of the QE derived in this work enables the accurate measurement of the optical properties for thin multi-alkali photocathodes in the visible range and hence the precise description of the PMT response.
It should be valid for the other types of thin photocathodes as long as the applicable range of the wavelength is reevaluated based on the threshold energy for the electron-electron scattering during the transportation, and if the Lorentz dispersion model is appropriate for those photocathodes.

\section*{Acknowledgment}
This work was supported by JSPS KAKENHI Grant-in-Aid for Challenging Exploratory Research (Grant Number JP26610068).
The MCP-PMTs measured in this work were produced for the Belle~II TOP counter with support from JSPS KAKENHI Grant-in-Aid for Scientific Research (S) (Grant Number JP26220706).
I would like to thank my colleagues at Nagoya University for discussion, in particular T.~Iijima and K.~Inami.
I have also benefited from discussion with H.~Watanabe, Y.~Hasegawa, H.~Nishizawa and H.~Yamaguchi (HAMAMATSU PHOTONICS K.K.).


%


\end{document}